\documentclass[fleqn,10pt]{wlscirep}
\usepackage[utf8]{inputenc}
\usepackage[T1]{fontenc}
\usepackage{bm}
\usepackage{lscape}
\usepackage{longtable}
\usepackage{array,makecell,multirow,threeparttable}
\usepackage{rotating}
\usepackage{graphicx}
\usepackage{booktabs}
\usepackage{tabularx}
\usepackage{amssymb}
\usepackage{ragged2e}
\usepackage{enumitem}

\newcolumntype{Y}{>{\centering\arraybackslash}X}

\title{Large language models in medical time series analysis}

\author[1,2]{Yu Han}
\author[1,3]{Cigdem Beyan}
\author[4]{Xiang Zhang}
\author[5]{Xiaofeng Liu}
\author[6]{Nan Liu}
\author[7,8,9]{Jimeng Sun}
\author[10,11,*]{Shenda Hong}
\author[12,*]{Cheng Ding}
\author[1,3]{Vittorio Murino}
\affil[1]{AI for Good (AIGO), Istituto Italiano di Tecnologia, Genoa, Liguria, Italy}
\affil[2]{DITEN, University of Genoa, Genoa, Liguria, Italy}
\affil[3]{Department of Computer Science, University of Verona, Verona, Veneto, Italy}
\affil[4]{Department of Computer Science, University of North Carolina at Charlotte, Charlotte, North Carolina, USA}
\affil[5]{Department of Biomedical Informatics and Data Science, Yale University, New Haven, Connecticut, USA}
\affil[6]{NUS Artificial Intelligence Institute, National University of Singapore, Singapore, Singapore}
\affil[7]{Siebel School of Computing and Data Science, University of Illinois at Urbana-Champaign, Urbana, Illinois, USA}
\affil[8]{Carle Illinois College of Medicine, University of Illinois at Urbana-Champaign, Urbana, Illinois, USA}
\affil[9]{Keiji AI, Seattle, Washington, USA}
\affil[10]{National Institute of Health Data Science, Peking University,  Beijing, China}
\affil[11]{Institute for Artificial Intelligence, Peking University, Beijing, China}
\affil[12]{Nanjing University of Aeronautics and Astronautics, College of Artificial Intelligence, Nanjing, Jiangsu, China}

\begin{abstract}
Medical time series (MedTS), including electrocardiograms (ECG), electroencephalograms (EEG), photoplethysmography (PPG), and vital-sign recordings, are central to clinical diagnosis and health monitoring. As large language models (LLMs) have advanced, a growing body of work has examined how their reasoning, generation, and knowledge-integration capabilities can support MedTS analysis. Yet existing studies remain scattered, and the field still lacks a clear view of how these models should be designed, integrated into clinical workflows, and evaluated. This review synthesizes recent work on large language models for medical time series analysis (MedTSLLMs), covering both methodological progress and issues related to real-world deployment. We review model architectures, data resources, and processing pipelines, and prompt design strategies adapted for diverse clinical scenarios. We further organize existing MedTS applications, ranging from diagnostic interpretation and report generation to longitudinal health monitoring and physiological signal synthesis, highlighting task-specific design choices, common evaluation protocols, and empirical findings reported across studies. By bringing together current practices and open challenges, this review aims to provide a clearer foundation for developing, evaluating, and deploying MedTSLLMs responsibly in healthcare. We also maintain a regularly updated list of MedTSLLM studies and resources at: \textcolor{blue}{https://github.com/hy727/MedTSLLM-Review.}

\end{abstract}
\begin{document}

\flushbottom
\maketitle

\thispagestyle{empty}

\section*{Introduction}

Medical time series (MedTS), such as electrocardiograms (ECG), electroencephalograms (EEG), photoplethysmography (PPG), and vital-sign recordings, are essential sources of physiological information in modern healthcare~\cite{abbaspourazad2023large,thapa2024sleepfm,ding2025ai}. They support a broad range of clinical tasks, including disease screening~\cite{ding2025ai}, diagnosis~\cite{wang2023contrast,wang2024medformer}, patient monitoring~\cite{hogan2025scaling,fraser2025integration}, risk assessment~\cite{pedroso2025leveraging}, and treatment~\cite{sun2025automated,partamian2025machine}. Compared with medical images or clinical text, MedTS data are more continuous, dynamic, and generally multi-channel, capturing subtle temporal and spatial patterns that reflect underlying physiological behaviors~\cite{mathew2024foundation,thapa2024sleepfm}. However, the interpretation in those studies remains challenging because clinically meaningful information is generally embedded in complex waveform morphology, temporal dependencies, cross-channel interactions, and patient-specific variations~\cite{hogan2025scaling,sun2025automated}. 

Recent advances in large language models (LLMs), such as GPT-5~\cite{singh2025openai}, Gemma-3~\cite{team2025gemma}, and DeepSeek-v4~\cite{xu2026deepseek}, have introduced emerging opportunities for the medicine domain. LLMs have demonstrated strong capabilities in language understanding, instruction following, knowledge integration, and text generation. These capabilities are particularly attractive for MedTS analysis, where clinical interpretation often requires not only signal pattern recognition but also contextual reasoning, medical knowledge integration, and communication with clinicians or patients. Inspired by these successes, a growing number of studies have begun to extend LLMs to MedTS analysis (MedTSLLMs), such as ECG-Chat~\cite{zhao2025ecg}, Thought2Text~\cite{mishra2025thought2text}, and SensorLM~\cite{zhang2025sensorlm}. These models aim to connect physiological signals with patient information and clinical knowledge, and support applications such as ECG interpretation, EEG decoding, wearable-based health assessment, clinical report generation, signal-based medical question answering (QA), and physiological signal synthesis.

Despite the promising progress of MedTSLLMs, the rapid growth of this field has made it increasingly difficult for clinicians and biosignal researchers to obtain a coherent view of its technical landscape and clinical relevance. This motivates the need for a timely review that can help domain experts understand not only what MedTSLLMs have achieved, but also how MedTS can be effectively connected to the LLM ecosystem. Unlike text and images, MedTS generally exhibits complex temporal structure, high variability, strong physiological dependence, and hierarchical clinical events. However, current studies often directly apply LLMs designed for text or images without fully accounting for these differences. To accommodate the characteristics of MedTS, several fundamental questions need to be addressed: (1) \textit{how MedTS should be tokenized and aligned with LLMs}, (2) \textit{how prompt templates should be designed for MedTS interpretation}, (3) \textit{how patient context and clinical knowledge should be incorporated with MedTS}, and (4) \textit{how model outputs can be evaluated and grounded in physiological evidence}. These questions have been partially discussed by recent reviews on LLMs in medicine and biosignal analysis~\cite{babu2025large,ansari2025survey,zhou2023survey,arava2025large,ding2024survey,khan2026ecg,liu2026survey}, but they have not yet been integrated into a unified MedTSLLM perspective. As summarized in Table~1, general medical LLM reviews have largely focused on text-based clinical applications~\cite{zhou2023survey}, whereas biosignal-oriented reviews have typically concentrated on individual modalities, including ECG~\cite{ansari2025survey,khan2026ecg}, EEG~\cite{babu2025large}, and wearable sensors~\cite{arava2025large}. More recent MedTS-centered reviews~\cite{ding2024survey,liu2026survey} have broadened the technical scope, but offer comparatively limited systematic discussion of the functional roles of LLMs, prompt design, and clinically oriented guidance across MedTS applications. As a result, existing reviews provide limited guidance on how to systematically develop, deploy, and evaluate MedTSLLMs. This review is therefore not intended to simply summarize existing methods, but to organize recent advances around the key question of how to better bridge MedTS and LLMs. By further discussing current limitations and open challenges, we aim to stimulate further research on more reliable, clinically grounded, and practically useful MedTSLLMs.

\begin{table}[t]
\centering
\caption{Comparison of our review with existing review papers in LLM for biosignals. Abbreviation: MedTS: medical time series. D: diagnosis. G: generation. H: health monitoring. S: medical time-series synthesis.}
\label{tab:survey_comparison}
\footnotesize
\setlength{\tabcolsep}{2.3pt}
\renewcommand{\arraystretch}{1.12}

\begin{tabular*}{\textwidth}{@{\extracolsep{\fill}}l l @{\hspace{0.25cm}} c c c c c c c c@{}}
\toprule
\textbf{Review} 
& \textbf{Focus} 
& \makecell{\textbf{Multimodal}\\\textbf{Integration}} 
& \textbf{Architecture} 
& \textbf{Pipeline} 
& \makecell{\textbf{LLM}\\\textbf{Role}} 
& \makecell{\textbf{Prompt}\\\textbf{Design}} 
& \makecell{\textbf{Evaluation}\\\textbf{Protocol}} 
& \makecell{\textbf{Clinical}\\\textbf{Guidance}} 
& \makecell{\textbf{Task}\\\textbf{Scope}} \\
\midrule[0.4pt]

Zhou et al.~\cite{zhou2023survey}   & Medicine & $\checkmark$ & $\checkmark$ & $\checkmark$ & $\times$      & $\times$      & $\checkmark$ & $\checkmark$ & D+G+H \\
Ansari et al.~\cite{ansari2025survey} & ECG      & $\checkmark$ & $\checkmark$ & $\times$      & $\times$      & $\times$      & $\checkmark$ & $\times$      & D+G \\
Khan et al.~\cite{khan2026ecg}   & ECG      & $\checkmark$ & $\checkmark$ & $\times$      & $\checkmark$ & $\times$      & $\times$      & $\checkmark$ & D+G \\
Babu et al.~\cite{babu2025large}   & EEG      & $\checkmark$ & $\checkmark$ & $\times$      & $\checkmark$ & $\times$      & $\times$      & $\times$      & D+G+S \\
Arava et al.~\cite{arava2025large}  & Wearable & $\checkmark$ & $\times$      & $\checkmark$ & $\checkmark$ & $\times$      & $\checkmark$ & $\checkmark$ & G+H \\
Ding et al.~\cite{ding2024survey}  & MedTS    & $\times$      & $\checkmark$ & $\times$      & $\times$      & $\times$      & $\times$      & $\times$      & D+G+H+S \\
Liu et al.~\cite{liu2026survey}  & MedTS    & $\checkmark$      & $\checkmark$ & $\checkmark$      & $\times$      & $\times$      & $\checkmark$      & $\times$      & D+G+H \\

\midrule[0.4pt]
\textbf{Our review} 
& MedTS 
& $\checkmark$ 
& $\checkmark$ 
& $\checkmark$ 
& $\checkmark$ 
& $\checkmark$ 
& $\checkmark$ 
& $\checkmark$ 
& D+G+H+S \\
\bottomrule
\end{tabular*}

\end{table}

As shown in Fig~1, this review provides a structured and comprehensive synthesis of MedTSLLMs, systematically covering their foundational modeling workflows, prompting strategies, and clinical applications. To give a clear guideline and fill the interdisciplinary gap for clinicians and practitioners, we first introduce the background of MedTSLLMs by explaining fundamental LLM concepts and the general workflow for adapting LLMs to MedTS data, including tokenization, pre-training, and fine-tuning. We then summarize prompting strategies in MedTSLLMs, covering instruction prompting, structured reasoning prompting, clinical-guided prompting, and soft prompting, followed by a detailed discussion of prompt template designs. To give a clear view of the practical deployment of MedTSLLMs in clinical settings, we then explore through the development of guidelines tailored for six distinct clinical application scenarios, with attention to task formulation, the functional roles of LLMs, and commonly used evaluation protocols. Finally, we discuss key challenges in current MedTSLLM studies from both methodological and clinical perspectives, and outline promising opportunities for future work, encouraging further exploration of LLMs for MedTS analysis.

In summary, the main contributions of this review are as follows:
\begin{itemize}
    \item \textbf{A structured taxonomy of MedTSLLMs.} We provide a comprehensive overview of recent MedTSLLM studies across representative physiological modalities, including ECG, EEG, PPG, vital signs, and wearable sensor data. Rather than only summarizing individual models, we organize existing methods according to their core design choices, including MedTS tokenization, model architectures, data resources, and processing pipelines. This taxonomy clarifies how MedTS data can be connected to the LLM ecosystem and distinguishes MedTSLLMs from general LLMs and task-specific MedTS models.

    \item \textbf{A systematic synthesis of prompting and evaluation strategies.} We provide a comprehensive analysis of prompting in MedTSLLMs, covering instruction prompting, structured reasoning, clinical-guided prompting, and soft prompting, together with six commonly used prompt-template designs. We further examine how patient context, clinical knowledge, retrieval mechanisms, and multimodal inputs are incorporated into model reasoning. In addition, we review current evaluation practices and highlight the need to assess not only predictive accuracy, but also physiological grounding, temporal consistency, clinical robustness, and evidence-based reasoning.
    
    \item \textbf{A clinically-oriented guideline for MedTSLLM applications, challenges, and future directions.} We review MedTSLLMs across six major clinical application scenarios: disease diagnosis, clinical report generation, medical QA, neuro-signal translation, health assessment, and physiological signal synthesis. For each scenario, we summarize task formulations, model roles, design considerations, and evaluation protocols, providing practical guidance for clinicians and researchers when selecting, developing, and assessing MedTSLLMs. We also identify key challenges in data quality, hallucination, temporal and mechanistic reasoning, evaluation standardization, safety, regulation, and clinical deployment, and outline corresponding opportunities for future research.

\end{itemize}

The rest of the article is organized as follows. Section 2 illustrates the background of MedTSLLMs. Section 3 outlines the prompting strategies in MedTSLLMs. Section 4 presents how MedTSLLMs are applied in different clinical scenarios. Section 5 discusses the existing challenges and promising opportunities of MedTSLLMs, and Section 6 concludes the review.

\begin{figure}[!htbp]
\centering
\includegraphics[width=1.0\linewidth]{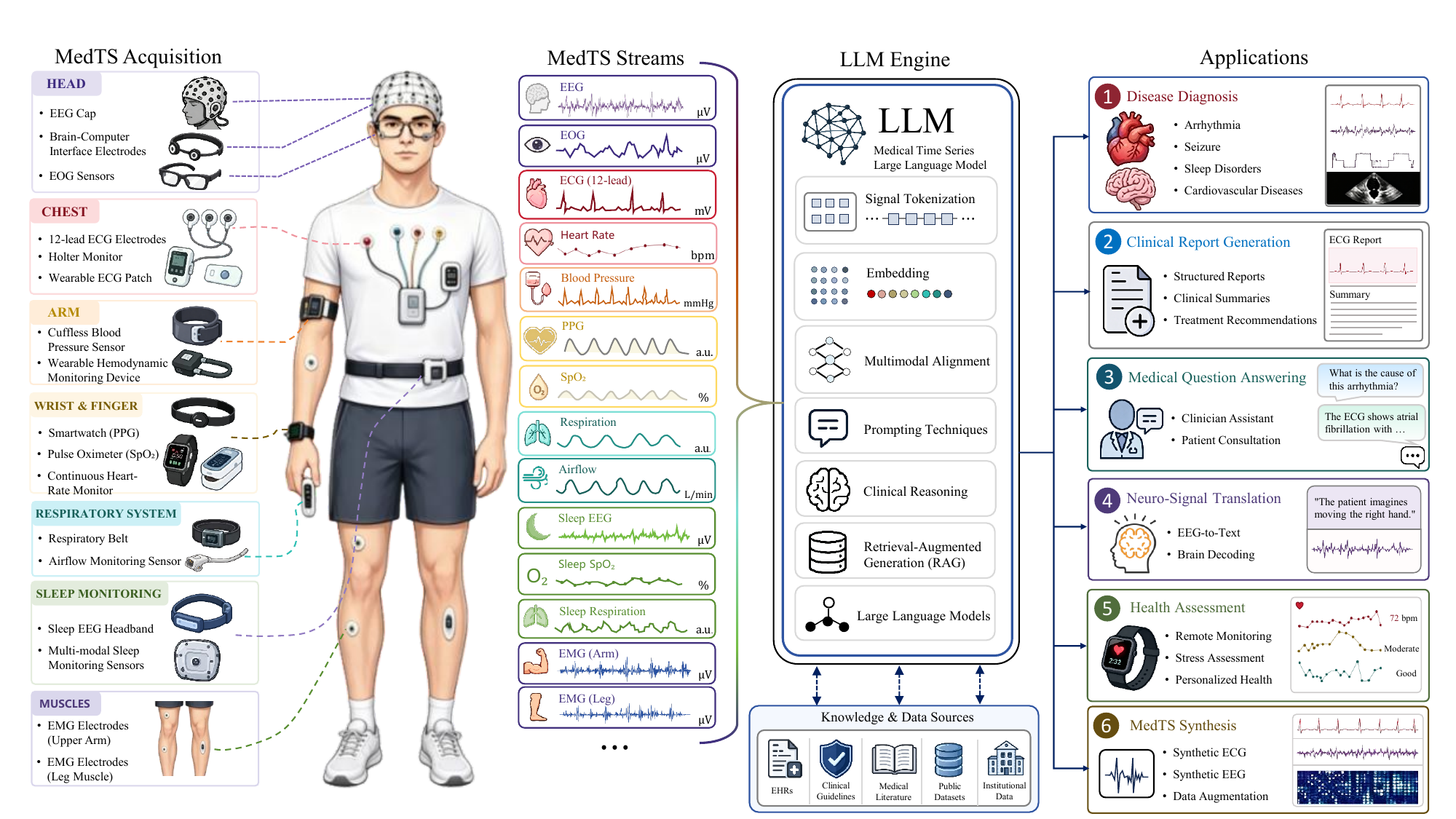}
\caption{\textbf{Overview of MedTSLLM workflow.} Here we outline the workflow of MedTSLLMs from multi-source MedTS acquisition and signal streams to LLM-based pipeline, knowledge integration, and downstream clinical applications.}
\label{fig:Prompt Template}
\end{figure}

\section*{Background of MedTSLLMs}
In this section, we introduce the background of MedTSLLMs from the perspective of LLM-style modeling. We first briefly review the fundamental concepts behind modern LLMs, including language models, the Transformer architecture, and scaling laws, to explain why LLMs can learn powerful representations from large-scale sequential data. We then introduce several basic terms commonly used in the LLM domain, such as token, embedding, and prompt, to make the discussion more accessible to readers from clinical and biomedical backgrounds. Building on these concepts, we finally outline a general workflow for MedTSLLM construction around three key steps: tokenization, pre-training, and fine-tuning, to provide a clear view of how raw physiological signals can be transformed into tokens, learned through large-scale pre-training, and further adapted to downstream medical tasks.

\subsection*{Fundamental Concepts}

To improve readability, particularly for clinicians and biomedical researchers, we briefly introduce several key terms that appear frequently throughout this review, including token, embedding, and prompt. These concepts provide a foundation for understanding the general workflow of MedTSLLMs.

\paragraph{Token} A token is the basic unit processed by an LLM. In natural language modeling, tokenization converts raw text into discrete units, such as words, subwords, characters, or punctuation marks, which can then be mapped into numerical representations and processed by neural networks~\cite{kaplan2025tokens,jin2024time}. In MedTSLLMs, this concept is extended from text to physiological time-series data. Since MedTS signals are continuous and often multi-channel, tokens are commonly constructed from temporal patches, channel-wise patches, discretized waveform units, or learned signal representations~\cite{luo2026toward,jin2024time}. In multimodal settings, tokens may also include image patches from signal visualizations, textual clinical descriptions, or metadata~\cite{liu2026teaching,makarov2025large,narayanswamy2025scaling}. By converting MedTS data into token-like units, MedTSLLMs can process physiological signals in a sequence-based manner and model relationships across time, channels, and modalities.

\paragraph{Embedding} An embedding is a continuous vector representation of a token in a numerical space~\cite{yu2026scaling}. In LLMs, discrete tokens are projected into dense vectors, often combined with positional information to preserve sequence order~\cite{zhang2024found}. These embeddings enable the model to encode token-level information, contextual relationships, and semantic patterns within a sequence. In MedTSLLMs, embeddings play the same fundamental role, but their construction is more diverse as physiological signals are continuous and semantically implicit~\cite{abbaspourazad2023large,zhou2025diagnosis}. Depending on the representation form and model architecture, MedTS embeddings can capture waveform dynamics, temporal order, channel-specific information, or spectral patterns. When multimodal data are used, modality-specific encoders are often adopted to obtain corresponding embeddings, such as using BioClinicalBERT~\cite{peng2019transfer} to encode clinical reports. These embeddings are then fed into the MedTSLLM to support contextual modeling and downstream medical tasks.

\paragraph{Prompt} A prompt is the input provided to an LLM to guide its generation or prediction. It may contain instructions, questions, examples, or contextual information that specify the task without changing the model architecture or updating model parameters~\cite{wan2024teach,wu2024prompt}. Prompt design can strongly influence model behavior because it determines how the task is framed, what information is provided, and what type of output is expected. In MedTSLLMs, prompts are extended to clinical and physiological settings. A prompt may include task instructions, MedTS representations, patient context, and relevant clinical knowledge~\cite{zhang2025leveraging}. For example, ECG waveform images can be provided together with an instruction such as ``\textit{Provide possible diagnostic suggestions based on these ECG images.}'' In this case, the waveform data provide physiological evidence, while the textual instruction specifies the clinical task. By organizing these components within a prompt, MedTSLLMs can condition their outputs on both the intended task and the available clinical information.

\paragraph{Large Language Models} A language model is a probabilistic model that captures the joint probability distribution of tokens, such as words, subwords, or other discrete units, in a text sequence~\cite{bengio2003neural}. By estimating how tokens are likely to appear given their preceding context, a language model can predict the next token, evaluate the likelihood of a sequence, or generate new sequences~\cite{sundermeyer2015feedforward}. LLMs extend this idea by training large-scale neural networks on massive corpora, enabling them to acquire broad linguistic knowledge, contextual understanding, and increasingly complex reasoning and generation capabilities~\cite{thirunavukarasu2023large}. The success of modern LLMs is mainly driven by two key factors: the Transformer architecture~\cite{vaswani2017attention} and scaling laws~\cite{kaplan2020scaling}.

\paragraph{The Transformer Architecture} The Transformer architecture provides an effective framework for modeling both local and long-range dependencies in sequential data. Its core component is the attention mechanism, which computes pairwise relevance scores among tokens in parallel~\cite{devlin2019bert}. These scores indicate how strongly each token should attend to other tokens when constructing its contextual representation. Unlike traditional language models like recurrent neural networks (RNNs)~\cite{mikolov2010recurrent,graves2012long}, which process tokens sequentially, Transformers allow all tokens in a sequence to interact simultaneously, greatly improving computational efficiency and scalability~\cite{vaswani2017attention}. By adaptively focusing on informative parts of the input sequence, the attention mechanism enables the model to capture complex contextual relationships and learn rich token representations.

\paragraph{Scaling Laws} Scaling laws describe the empirical relationship between model performance and key scaling factors, including the number of parameters, model depth, training data size, and computational budget~\cite{kaplan2020scaling}. In general, when these factors increase in a coordinated manner, model performance tends to improve predictably. Modern LLMs can therefore be viewed as scaled-up Transformer-based models with substantially more layers, parameters, and pre-training data~\cite{hoffmann2022training}. As model scale increases, LLMs often exhibit stronger generalization and contextual understanding ability. For example, GPT-3 with 175B parameters has shown much stronger few-shot learning and text generation abilities than GPT-2 with only 1.5B parameters, and the latest larger-scale GPT-4 even expands the model to multimodal reasoning and automatic coding capabilities~\cite{radford2019language,achiam2023gpt,floridi2020gpt}.

\subsection*{From General LLMs to MedTSLLMs} 

\paragraph{Scope of MedTSLLMs}  In this review, we use MedTSLLM as a broad term for LLM-based approaches to MedTS analysis, rather than referring to a single model architecture. MedTSLLMs may integrate language models with signal encoders, multimodal alignment modules, or other task-specific components to process physiological data. Accordingly, our scope covers different forms of LLM involvement in MedTS modeling, including representation learning, signal-language alignment, reasoning, generation, and agent-based analysis. This broader definition reflects the diverse ways in which current studies connect MedTS with the LLM framework.

\paragraph{LLM Architectures} As summarized in Table~2, existing MedTSLLM studies employ a range of LLM backbones that can be broadly categorized into \textit{encoder-only}, \textit{decoder-only}, and \textit{encoder-decoder} architectures. These architectures differ in how they encode contextual information and generate outputs, making them suitable for different MedTS scenarios and downstream tasks.

\textit{Encoder-only LLMs} consist of Transformer encoder layers with bidirectional self-attention, allowing each token to capture contextual information from the entire input sequence. This architecture provides strong bidirectional contextual modeling and efficient feature representation learning, making it particularly suitable for discriminative MedTS tasks, like disease detection~\cite{vaid2023foundational,choi2023ecgbert} and sleep stage classification~\cite{wang2023large}. However, since encoder-only LLMs are not designed for autoregressive generation, they usually require task-specific prediction heads and are less effective for open-ended report generation, QA, or instruction-following scenarios without additional generative modules. Representative examples include BERT-series models~\cite{devlin2019bert,peng2019transfer,yasunaga2022linkbert} and multimodal encoders such as ViLT~\cite{kim2021vilt}.

\textit{Decoder-only LLMs} utilize Transformer decoder layers trained with causal language modeling, where each token is predicted based on preceding tokens. This autoregressive design enables strong language generation and in-context learning, making them the dominant backbone for generation-oriented medical tasks such as medical QA~\cite{safranek2024automated,liu2024teach} and report generation~\cite{yang2025diagecg}. However, since decoder-only LLMs do not directly model full bidirectional signal context, they could be less efficient for representation learning and can suffer from hallucination and high computational cost when applied to clinically grounded MedTS reasoning. Representative models include closed-source LLMs such as GPT-series~\cite{radford2019language,ouyang2022training,achiam2023gpt,hurst2024gpt}, Gemini-series~\cite{team2023gemini,team2024gemini,comanici2025gemini}, and Claude-series~\cite{anthropic2024claude35sonnet}, as well as open-source models such as LLaMA-series~\cite{touvron2023llama1,touvron2023llama2,grattafiori2024llama}, Qwen-series~\cite{bai2023qwen,wang2024qwen2,hui2024qwen2}, and DeepSeek-series~\cite{liu2024deepseek,guo2025deepseek}. In addition, LLaMA-derived models, including Alpaca~\cite{taori2023stanford}, Vicuna~\cite{chiang2023vicuna}, and LeoLM~\cite{laion2023leolm}, further demonstrate how instruction-tuning and dialogue fine-tuning can adapt general LLMs to specific downstream scenarios.

\textit{Encoder-Decoder LLMs} combine a bidirectional encoder for input representation with an autoregressive decoder for output generation. The encoder captures contextual information from the input sequence, while the decoder generates the target output conditioned on the encoded representation~\cite{zeng2022glm,joseph2023multilingual,zhang2020pegasus}. This structure supports conditional generation and sequence-to-sequence modeling, making it well-suited to tasks where the input and output differ in modality or format, such as neuro signal translation~\cite{wang2024enhancing,zhou2024belt}. Nevertheless, encoder-decoder LLMs are architecturally more complex and typically involve a higher inference cost than encoder-only models, while being less dominant than decoder-only models in recent LLM development. Representative examples include BART~\cite{lewis2020bart}, Flan-T5~\cite{chung2024scaling}, CoCa~\cite{yu2022coca}, and GLM-series~\cite{zeng2022glm,glm2024chatglm}. 

\paragraph{Why Do We Need MedTSLLMs?}
Inspired by the success of LLMs in natural language processing, recent studies have begun to explore LLM-style modeling for MedTS analysis~\cite{brown2020language,moor2023foundation,chan2024medtsllm}. The motivation for MedTSLLMs is not simply to replace domain-specific MedTS foundation models, but to extend physiological signal modeling toward language-based understanding and reasoning. These foundation models can learn useful representations from physiological signals, but their outputs are generally constrained by predefined task formats, such as classification labels or risk scores~\cite{mathew2024foundation,mckeen2025ecg}. This makes them less suitable for clinical scenarios where physicians need not only predictions, but also interpretable descriptions and context-aware reasoning.

In contrast, LLMs offer a different set of capabilities as they are trained on large-scale text and multimodal corpora that span diverse domains. Therefore, MedTSLLMs provide a promising framework for connecting physiological signals with clinical language and medical knowledge. By converting MedTS data into LLM-compatible tokens, they can model complex temporal dynamics, channel dependencies, and clinically meaningful patterns while supporting flexible downstream medical tasks~\cite{liu2026teaching}. More importantly, their language interface makes MedTS analysis more human-friendly: clinicians can interact with the model through dialogue, ask follow-up questions, retrieve relevant information, generate structured diagnostic reports, and summarize patient conditions, which may improve clinical efficiency and decision support~\cite{zhao2025ecg,wan2025meit}. 

Nevertheless, MedTSLLMs also introduce important risks and practical barriers. They may generate hallucinated or unsupported outputs, show sensitivity to prompts, and produce inconsistent results for similar questions~\cite{hager2024evaluation,omar2025multi}. In clinical practice, these issues are further compounded by limited explainability, uncertain clinician trust, regulatory approval requirements, and the difficulty of integrating LLM-based systems into existing diagnostic workflows~\cite{thirunavukarasu2023large}. Additionally, the computational and deployment costs of MedTSLLMs may also be higher than those of task-specific models, making their cost-effectiveness uncertain in some clinical settings~\cite{samsi2023words,wang2024model}. Therefore, the central question is how to adapt LLMs to MedTS analysis in a reliable and clinically grounded manner. With their broad knowledge and strong capabilities in language understanding, reasoning, and generation, LLMs are likely to become an important component of future MedTS systems. Future work should further align physiological signals with medical knowledge, enabling these capabilities to be translated into trustworthy clinical support.

\begin{table}[!htbp]
\centering
\caption{Summary of the LLMs used in existing MedTS studies, including their underlying structures, numbers of parameters, pre-training data scales, representative applications, and key strengths and limitations. Column ``\# Params'' denotes the number of parameters. Abbreviations: K: thousand; M: million; B: billion; T: trillion; GB: gigabyte; IFT: instruction fine-tuning data.}
\footnotesize
\setlength{\tabcolsep}{1.3pt}
\renewcommand{\arraystretch}{1.3}

\begin{tabular}{@{}p{2.2cm}p{2.4cm}p{2.4cm}p{3.0cm}p{3.6cm}p{3.6cm}@{}}
\hline
Model Structures
& Models
& \# Params
& Pre-training Data Scale
& Representative Applications
& Key Strengths \& Limitations \\
\hline

\multirow{6}{2.2cm}{\raggedright Encoder-only}
& BERT~\cite{devlin2019bert}
& 110M/340M
& 3.3B tokens
& \multirow{6}{4cm}{\raggedright
Arrhythmia diagnosis;\newline
Sleep apnea detection;\newline
Sentiment analysis;\newline
Reading comprehension}
& \multirow{6}{3.6cm}{\justifying
\textbf{Strengths:} Bidirectional contextual modeling and efficient representation learning.\newline
\textbf{Limitations:} Limited generation and instruction-following capabilities; task-specific heads are often required.} \\

& ClinicalBERT~\cite{peng2019transfer}
& $\sim$110M
& 880M words
& & \\

& BiolinkBERT~\cite{yasunaga2022linkbert}
& $\sim$110M/333M
& 21GB text
& & \\

& BEiT~\cite{bao2021beit}
& 86M/304M
& 14M images
& & \\

& Ada-v2~\cite{openai2022embedding}
& -
& -
& & \\

& ViLT~\cite{kim2021vilt}
& $\sim$87M
& 5.1M image-text pairs
& & \\
\hline

\multirow{32}{2.2cm}{\raggedright Decoder-only}
& PaLM~\cite{chowdhery2023palm}
& 8B/62B/540B
& 780B tokens
& \multirow{32}{3.5cm}{\raggedright
Medical QA;\newline
Sleep monitoring;\newline
Text-to-ECG synthesis;\newline
EEG-to-text translation;\newline
Clinical report generation}

& \multirow{32}{3.5cm}{\justifying
\textbf{Strengths:} Strong generation, instruction following, in-context learning, and scalability.\newline
\textbf{Limitations:} Hallucination risk, high computational cost, and limited representation efficiency.} \\

& Vicuna~\cite{chiang2023vicuna}
& 7B/13B
& LLaMA+70K dialogues
& & \\

& Alpaca~\cite{taori2023stanford}
& 7B/13B
& LLaMA+52K IFT
& & \\

& Mistral~\cite{jiang2023mistral7b}
& 7B
& -
& & \\

& Mistral-v0.3~\cite{jiang2023mistral7b}
& 7.3B
& -
& & \\

& Mistral-v2 Large~\cite{mistralai2024large2}
& 123B
& -
& & \\

& OPT~\cite{zhang2022opt}
& 125M-175B
& 180B tokens
& & \\

& BLOOM~\cite{workshop2022bloom}
& 176B
& 366B tokens
& & \\

& Google Bard~\cite{thoppilan2022lamda}
& $\sim$137B
& 1.56T words
& & \\

& InternLM-2.5~\cite{cai2024internlm2}
& 1.8B/7B/20B
& $\sim$2.6T tokens
& & \\

& LeoLM~\cite{laion2023leolm}
& 7B/13B/70B
& LLaMA-2+65B tokens
& & \\

& GPT-2~\cite{radford2019language}
& 1.5B
& 40GB text
& & \\

& GPT-3.5~\cite{ouyang2022training}
& -
& -
& & \\

& GPT-3.5 Turbo~\cite{ouyang2022training}
& -
& -
& & \\

& GPT-4~\cite{achiam2023gpt}
& -
& -
& & \\

& GPT-4o~\cite{hurst2024gpt}
& -
& -
& & \\

& Gemini Pro~\cite{team2023gemini}
& -
& -
& & \\

& Gemini-1.5 Flash~\cite{team2024gemini}
& -
& -
& & \\

& Gemini-2.5 Flash~\cite{comanici2025gemini}
& -
& -
& & \\

& Gemini-Ultra~\cite{team2023gemini}
& -
& -
& & \\

& LLaMA~\cite{touvron2023llama1}
& 7B/13B/33B/65B
& 1T-1.4T tokens
& & \\

& LLaMA-2~\cite{touvron2023llama2}
& 7B/13B/34B/70B
& 2T tokens
& & \\

& LLaMA-3~\cite{grattafiori2024llama}
& 8B/70B
& $\sim$15T tokens
& & \\

& LLaVA~\cite{liu2023visual}
& 7B/13B
& LLaMA+595K\newline image-text pairs
& & \\

& LLaVA-v1.6~\cite{liu2024llavanext}
& 7B/13B/34B
& -
& & \\

& Qwen~\cite{bai2023qwen}
& 1.8B/7B/14B/72B
& 2.4T-3T tokens
& & \\

& Qwen-2~\cite{wang2024qwen2}
& 0.5B/\allowbreak1.5B/\allowbreak7B/\allowbreak\newline 57B/\allowbreak72B
& 7T-12T tokens
& & \\

& Qwen-2.5~\cite{hui2024qwen2}
& 0.5B/\allowbreak1.5B/\allowbreak3B/\allowbreak
7B/\allowbreak\newline 14B/\allowbreak32B/\allowbreak72B
& 18T tokens
& & \\

& DeepSeek-v3~\cite{liu2024deepseek}
& 671B total/\allowbreak\newline 37B activated
& 14.8T tokens
& & \\

& DeepSeek-r1~\cite{guo2025deepseek}
& 671B total/\allowbreak\newline 37B activated
& -
& & \\

& MiniMind-2~\cite{gong2025minimind2}
& 25.8M/\allowbreak26M/\allowbreak\newline 104M/\allowbreak145M
& 3B tokens
& & \\

& Claude-3.5~\cite{anthropic2024claude35sonnet}
& -
& -
& & \\
\hline

\multirow{7}{2.2cm}{\raggedright Encoder-Decoder}
& BART~\cite{lewis2020bart}
& 139M/406M
& 160GB text
& \multirow{7}{3.5cm}{\raggedright
Vitals calculation;\newline
EEG-to-text translation;\newline
Sentiment classification;\newline
Arrhythmia diagnosis;\newline
Ground ECG understanding}

& \multirow{7}{3.5cm}{\justifying
\textbf{Strengths:} Strong conditional generation and flexible sequence-to-sequence mapping.\newline
\textbf{Limitations:} Greater architectural complexity and computational cost; less commonly used in recent MedTSLLMs.} \\

& T5~\cite{joseph2023multilingual}
& 3B/11B
& 1T tokens
& & \\

& FLAN-T5~\cite{chung2024scaling}
& 3B/11B
& T5+180B tokens
& & \\

& PEGASUS~\cite{zhang2020pegasus}
& 223M/568M
& $\sim$1.5B documents
& & \\

& CoCa~\cite{yu2022coca}
& -
& -
& & \\

& GLM~\cite{zeng2022glm}
& 130B
& 400B tokens
& & \\

& GLM-2~\cite{glm2024chatglm}
& -
& -
& & \\
\hline

\end{tabular}
\end{table}

\subsection*{Tokenization}

Tokenization is a fundamental step for enabling LLMs to process MedTS data. Unlike natural language, MedTS data, such as ECG, EEG, and PPG, are continuous numerical sequences that cannot be directly handled by standard text tokenizers~\cite{jin2024time}. Therefore, existing MedTSLLMs generally transform raw physiological signals into LLM-compatible tokens before feeding them into the model. One common strategy is patch-based tokenization~\cite{goswami2024moment,abbaspourazad2023large}, which segments a signal into local patches and maps each patch into the embedding space through a signal encoder, such as a linear projection layer or ResNet-1D. Another strategy is discretization-based tokenization, which converts continuous signals into symbolic tokens, codebook indices, or numerical token sequences~\cite{yang2026heartllm}. These discrete representations are generally embedded into the prompts as textual numeric contexts, allowing the LLM to process physiological information together with natural language instructions.

Beyond signal tokenization, multimodal MedTSLLMs further construct tokens from multiple related modalities~\cite{liu2026teaching}. Textual information, such as clinical reports, task instructions, and patient metadata, is typically tokenized by an LLM tokenizer (e.g., BERT or BioClinicalBERT). Visual inputs, such as ECG plots or time-frequency images, are divided into image patches and encoded by a vision encoder (e.g., ViT or CLIP-ViT). These multimodal tokens are then projected into a shared embedding space, enabling the LLM to jointly process physiological signals, clinical text, and visual evidence~\cite{zhou2025diagnosis}.

Tokenization provides a unified interface for representing heterogeneous medical data in LLM-compatible formats~\cite{yang2026heartllm,jin2024time}. By converting continuous physiological signals, textual descriptions, visual representations, and structured clinical information into token sequences, MedTSLLMs can integrate different modalities within a shared modeling framework. This design reduces the complexity of long and high-frequency signals while preserving clinically meaningful local patterns, such as waveform morphology and rhythm changes. Moreover, multimodal tokenization facilitates cross-modal alignment between signal evidence, textual knowledge, visual cues, and patient-level context, enabling the model to integrate complementary information for diagnosis, question answering, report generation, and clinical reasoning~\cite{liu2026teaching}.

\subsection*{Pre-training}

Pre-training refers to the stage in which a model learns general representations from large-scale data before being adapted to specific downstream tasks~\cite{zoph2020rethinking}. In LLMs, this stage is usually conducted on massive text corpora with self-supervised objectives, such as next-token prediction or masked token reconstruction. By learning to predict or recover tokens from context, the model gradually captures semantic relationships and contextual dependencies without relying on task-specific annotations~\cite{du2024stacking}. For MedTS analysis, pre-training shifts the learning source from purely textual corpora~\cite{alsentzer2019publicly,peng2023study} to physiological signals~\cite{liu2024teach,qiang2025biecg} and clinically related information. MedTSLLMs are exposed to tokenized signal patches to learn reusable physiological patterns, such as local morphology, temporal dependencies, and rhythm-related variations. This builds a signal-level foundation for later task adaptation. Beyond single-modality input data, MedTSLLMs may further use multimodal data, like signal-text pairs or patient metadata, to connect physiological patterns with clinical semantics.~\cite{hu2024exploring,zhou2023survey} For example, ECG waveforms can be aligned with diagnostic reports so that the model learns not only the signal structure but also its clinical meaning. Such multimodal pre-training helps connect physiological signal patterns with clinical interpretation, while reducing the dependence on large task-specific annotations.

Current MedTSLLM pre-training strategies can be broadly summarized according to their learning objectives. Masked modeling, inspired by BERT-style biomedical language models (e.g., BioBERT~\cite{lee2020biobert}, BioClinicalBERT~\cite{peng2019transfer}, PubMedBERT~\cite{gu2021domain}), masks part of the input and trains the model to recover the missing content from the surrounding context. In MedTSLLMs, this idea is usually adapted by masking signal tokens or patches and reconstructing the missing physiological segments, encouraging the model to capture contextual dependencies within MedTS sequences. Generative pre-training follows the autoregressive paradigm used by GPT-series (e.g., BioGPT~\cite{luo2022biogpt}, BioMedGPT~\cite{luo2023biomedgpt}) and LLaMA-series biomedical models (e.g., PMC-LLaMA~\cite{wu2024pmc}, and MEDITRON~\cite{chen2023meditron}). For MedTSLLMs, this objective can be used to predict future signal segments or generate textual interpretations conditioned on physiological inputs. Contrastive pre-training learns by aligning related signal views or paired signal-text data while separating unrelated samples. This strategy is closely related to CLIP-style multimodal LLMs (e.g., BioMedCLIP~\cite{zhang2023biomedclip}, PMC-CLIP~\cite{lin2023pmc}), which align images and text in a shared representation space. In MedTSLLMs, similar contrastive objectives can align physiological signals with diagnostic reports, clinical descriptions, or augmented images, helping the model associate signal patterns with clinical semantics. After pre-training, MedTSLLMs obtain reusable representations that can be further adapted to downstream medical tasks through fine-tuning or prompting.

\subsection*{Fine-tuning}

Although large-scale pre-training enables MedTSLLMs to learn general-purpose physiological representations, the direct application of pre-trained models to clinical scenarios often remains insufficient because downstream tasks, signal modalities, patient populations, and annotation formats are highly heterogeneous~\cite{zheng2025learning}. Fine-tuning provides a crucial adaptation stage that transfers the broad knowledge acquired during pre-training to specific MedTS applications, such as arrhythmia diagnosis, seizure detection, sleep staging, and health status assessment. Compared with training MedTSLLMs from scratch, fine-tuning is more computationally feasible and can better exploit existing pre-trained encoders or LLM backbones while incorporating domain-specific supervision~\cite{toma2023clinical,zheng2025learning}. Existing fine-tuning strategies can be characterized along two complementary dimensions: the supervision format and the parameter-update scope. From the perspective of supervision format, existing methods mainly include task-specific supervised fine-tuning (SFT) and instruction fine-tuning (IFT). From the perspective of parameter updates, both SFT and IFT can be implemented through either full-parameter fine-tuning or parameter-efficient fine-tuning (PEFT), which updates only a small subset of model parameters. As summarized in Table 3, we discuss SFT, IFT, and PEFT as three widely adopted adaptation paradigms, noting that PEFT concerns parameter updating rather than supervision format.

\begin{table}[!htbp]
\centering
\caption{Comparison of common fine-tuning paradigms for MedTSLLMs in terms of
trainable parameter size, computational cost, data requirements, key strengths and limitations,
and representative applications.}

\footnotesize
\setlength{\tabcolsep}{2.5pt}
\renewcommand{\arraystretch}{1.35}

\begin{tabular}{@{}
>{\raggedright\arraybackslash}m{1.15cm}
>{\raggedright\arraybackslash}m{1.85cm}
>{\raggedright\arraybackslash}m{2.45cm}
>{\raggedright\arraybackslash}m{2.85cm}
>{\raggedright\arraybackslash}m{4.35cm}
>{\raggedright\arraybackslash}m{3.45cm}
@{}}
\hline

Paradigm &
Parameter size &
Computational Cost &
Data Requirements &
Key Strengths \& Limitations &
Representative Applications \\
\hline

SFT &
Large &
High &
Task-specific labeled signal-text pairs &
\textbf{Strengths:} Strong task specialization.\newline
\textbf{Limitations:} Data-intensive and prone to overfitting on limited datasets. &
Seizure classification;\newline
Arrhythmia diagnosis;\newline
ECG interpretation \\

IFT &
Large &
High &
High-quality instruction-\newline response pairs &
\textbf{Strengths:} Strong instruction following and controllable outputs.\newline
\textbf{Limitations:} Sensitive to instruction quality and coverage. &
Medical QA;\newline
Health assessment;\newline
Clinical report generation \\

PEFT &
Small &
Low &
SFT/IFT data &
\textbf{Strengths:} Compute- and memory-efficient.\newline
\textbf{Limitations:} Limited adaptation capacity and configuration-sensitive. &
Medical QA;\newline
EEG-to-text translation;\newline
Clinical report generation \\

\hline
\end{tabular}

\label{tab:finetuning_comparison}
\end{table}

\paragraph{Supervised Fine-Tuning (SFT)} adapts a pre-trained MedTSLLM to labeled downstream tasks by optimizing the model on task-specific signal-label pairs~\cite{ding2025ai,shi2024instruction}. In MedTS analysis, supervision may be derived from disease categories, event annotations, or clinical outcomes. For example, an ECG-based MedTSLLM can be fine-tuned to classify arrhythmia types from multi-lead ECG recordings.

The flexibility of SFT enables the development of MedTSLLMs with diverse task-specific capabilities by fine-tuning LLMs on specialized MedTS supervision data. For example, GEM~\cite{lan2025gem} performs SFT on fine-grained ECG instruction-response pairs, where each response is grounded in heartbeat-level physiological evidence. This design enables the model to generate clinically grounded ECG interpretations from time series, image, and text inputs. ECG-LM~\cite{yang2025ecg} conducts SFT using public clinical conversation datasets and a hospital-based ECG dataset, allowing the model to answer ECG-specific medical questions. EEG-MedRAG~\cite{wang2025eeg} applies SFT to BART using EEG-text pairs, enabling natural language decoding from EEG sequences.

SFT is particularly useful when the downstream task has reliable annotations and a clearly defined evaluation protocol. However, the performance of SFT depends heavily on the quality, scale, and distribution of labeled datasets~\cite{zhang2026instruction}. As most MedTS datasets are small, imbalanced, and collected from limited institutions or devices, models fine-tuned in a purely supervised manner may overfit to dataset-specific artifacts, subject-specific patterns, or acquisition conditions, rather than learning robust disease-related features.

\paragraph{Instruction Fine-Tuning (IFT)} adapts LLMs using instruction-formatted datasets, which typically consist of instruction-question-answer pairs~\cite{he2025survey,zhang2026instruction}. For MedTSLLMs, IFT aims to improve the ability of the model to interpret and follow diverse task-specific instructions, align its responses with medical requirements, and generate outputs suitable for MedTS analysis~\cite{liu2023large}. In this way, IFT serves as a critical step in transforming a general-purpose LLM into a specialized MedTSLLM. Compared with SFT, IFT places more emphasis on instruction-following behavior and response alignment. SFT is mainly used to incorporate domain-specific MedTS knowledge into a pre-trained LLM, thereby enhancing its understanding of medical narratives and physiological signal descriptions. In contrast, IFT focuses on guiding the model to produce responses that are consistent with given instructions, expected output formats, and clinical contexts, rather than merely improving next-token prediction. Consequently, while SFT often depends on the scale of domain-specific training corpora, IFT is more sensitive to the quality, diversity, and coverage of instruction data~\cite{zhou2023survey}. High-quality IFT for MedTSLLMs therefore requires instruction datasets that cover diverse MedTS modalities, task types, and realistic medical scenarios.

Recent MedTSLLMs have applied IFT to align model responses with task-specific medical instructions. For example, Health-LLM~\cite{kim2024health} undergoes IFT on instruction-response data, enabling the model to follow health-related instructions for consumer health assessment. PULSE~\cite{liu2024teach} performs IFT on a large-scale ECG image instruction dataset, enabling the model to understand ECG images, answer ECG-related questions, and generate diagnostic reports. These IFT-based approaches help bridge physiological signals and natural language instructions, allowing MedTSLLMs to better follow task-specific requirements and generate clinically relevant responses.

However, the effectiveness of IFT depends heavily on the quality of the training data. As MedTS signals are continuous, noisy, and often difficult to interpret without expert knowledge, weakly annotated or automatically generated responses may lead the model to learn superficial textual regularities rather than reliable physiological reasoning. Therefore, MedTS-oriented IFT requires carefully curated signal-response pairs, expert-validated annotations, clear task definitions, and explicit correspondence between physiological evidence and target outputs.

\paragraph{Parameter-Efficient Fine-Tuning (PEFT)} provides an efficient alternative to full fine-tuning by adapting LLMs with only a limited number of trainable parameters~\cite{dettmers2023qlora}. Instead of updating all parameters of a pre-trained LLM, PEFT generally freezes the backbone model and optimizes a small set of newly introduced or selected parameters. This design largely reduces computational cost and storage requirements, making PEFT particularly suitable for domain adaptation in resource-constrained medical scenarios~\cite{liu2024dora,gao2025enhancing}. Common PEFT methods include Low-Rank Adaptation (LoRA)~\cite{hu2022lora}, Prompt Tuning~\cite{jia2022visual,lester2021power}, Prefix Tuning~\cite{li2021prefix}, and Adapter Tuning~\cite{liu2022p,Liu2021PTuningVP}.

\textbf{LoRA} is one of the most widely used PEFT strategies. Rather than directly updating the original full-rank weight matrices, LoRA keeps the pre-trained weights fixed and injects trainable low-rank decomposition matrices into specific modules, commonly the query and value projection matrices in Transformer self-attention layers~\cite{hu2022lora}. During fine-tuning, only these low-rank matrices are optimized, while the original model parameters remain unchanged. In this way, LoRA substantially reduces the number of trainable parameters while still allowing the model to acquire task-specific and domain-specific knowledge.

\textbf{Prompt Tuning} and \textbf{Prefix Tuning} adapt LLMs from the perspectives of input conditioning and hidden-context conditioning. Prompt Tuning learns a small number of continuous soft prompt vectors that are prepended to the input embeddings, guiding the model toward specific downstream tasks without modifying the backbone parameters ~\cite{lester2021power}. Prefix Tuning further extends this idea by introducing trainable continuous prefixes into each Transformer layer, typically as additional key-value vectors in the attention mechanism~\cite{li2021prefix}. These learned prefixes act as task-specific contextual signals that steer generation behavior while preserving the original pre-trained weights.

\textbf{Adapter Tuning} introduces lightweight neural modules, called adapters, into Transformer layers. Typically, an adapter consists of a down-projection layer, a non-linear transformation, and an up-projection layer, forming a bottleneck structure~\cite{houlsby2019parameter}. During fine-tuning, only the adapter parameters are updated, while the original LLM remains frozen. This design enables different adapters to be trained for different medical tasks or data modalities and inserted into the same backbone model when needed. Therefore, adapter-based tuning provides a flexible and modular way to adapt LLMs to different MedTS tasks.

In general, PEFT is valuable for developing LLMs that meet the specific requirements of medical domains as it reduces computational demands while maintaining model performance. For example, ECG-Chat~\cite{zhao2025ecg} and DiagECG~\cite{yang2025diagecg} adopt LoRA-based PEFT for ECG report generation and open-ended ECG QA. MedualTime~\cite{ye2024medualtime} introduces dual multimodal adapters with learnable adaptation tokens injected into the top layers of the LLM, enabling efficient signal-text fusion with only a small number of trainable parameters. However, PEFT also has inherent limitations. Since the backbone LLM is largely frozen, PEFT may not fully adapt the model to highly specialized MedTS domains that require deep physiological understanding or complex clinical reasoning. Its performance is also sensitive to the choice of trainable modules, target layers, and parameter scale, which may lead to underfitting if the adaptation capacity is insufficient. In addition, maintaining multiple LoRA modules, prompts, prefixes, or adapters for different tasks and modalities can increase deployment complexity. Furthermore, PEFT reduces computational cost but does not address data quality issues. Noisy or weakly grounded fine-tuning data may still cause unreliable or clinically inconsistent responses.

\section*{Prompting Strategies in MedTSLLMs}

Prompt design plays a central role in enabling LLMs to effectively interpret MedTS. In this review, we categorize commonly used prompting techniques into four types: \textit{instruction prompting}, \textit{structured reasoning prompting}, \textit{clinical-guided prompting}, and \textit{soft prompting}. These strategies are typically implemented through diverse prompt templates that combine task instructions, signal-derived representations, contextual metadata, and domain-specific clinical knowledge. The overall objective is to reduce the representational gap between continuous physiological signals and the discrete, language-based reasoning mechanisms of LLMs. In this section, we systematically review the prompting techniques and corresponding prompt templates adopted in current MedTSLLM frameworks. By analyzing their respective strengths and limitations, we aim to provide practical guidance for researchers and clinicians in the design and selection of effective prompts for MedTSLLMs.

\subsection*{Prompting Techniques}

We organize existing prompting methods into four primary categories: \textit{instruction prompting}, \textit{structured reasoning prompting}, \textit{clinical-guided prompting}, and \textit{soft prompting}. This taxonomy is primarily organized according to prompt design and functionality, covering a broad spectrum from discrete, task-oriented prompts to learnable continuous prompt embeddings. Instruction prompting provides explicit task descriptions, structured reasoning prompting guides step-by-step inference, and clinical-guided prompting incorporates manually designed or retrieved domain knowledge into the LLM context. In contrast, soft prompting optimizes continuous prompt embeddings during model training or adaptation. These design differences are also generally associated with different stages of model use. The first three categories are primarily implemented through discrete prompts supplied at inference time, whereas soft prompting typically represents a training-time adaptation strategy. Nevertheless, the distinction is not absolute, as some approaches, particularly instruction prompting, may span both stages through IFT followed by instruction-based inference. We therefore include both inference-time prompting and prompting-related training mechanisms within the same taxonomy. This broader view reflects the close correspondence between inference-time prompts and the prompt-response pairs used during model adaptation, while enabling a coherent comparison of their controllability, interpretability, adaptability, and clinical usability across MedTS applications.

\begin{figure}[!htbp]
\centering
\includegraphics[width=1.0\linewidth]{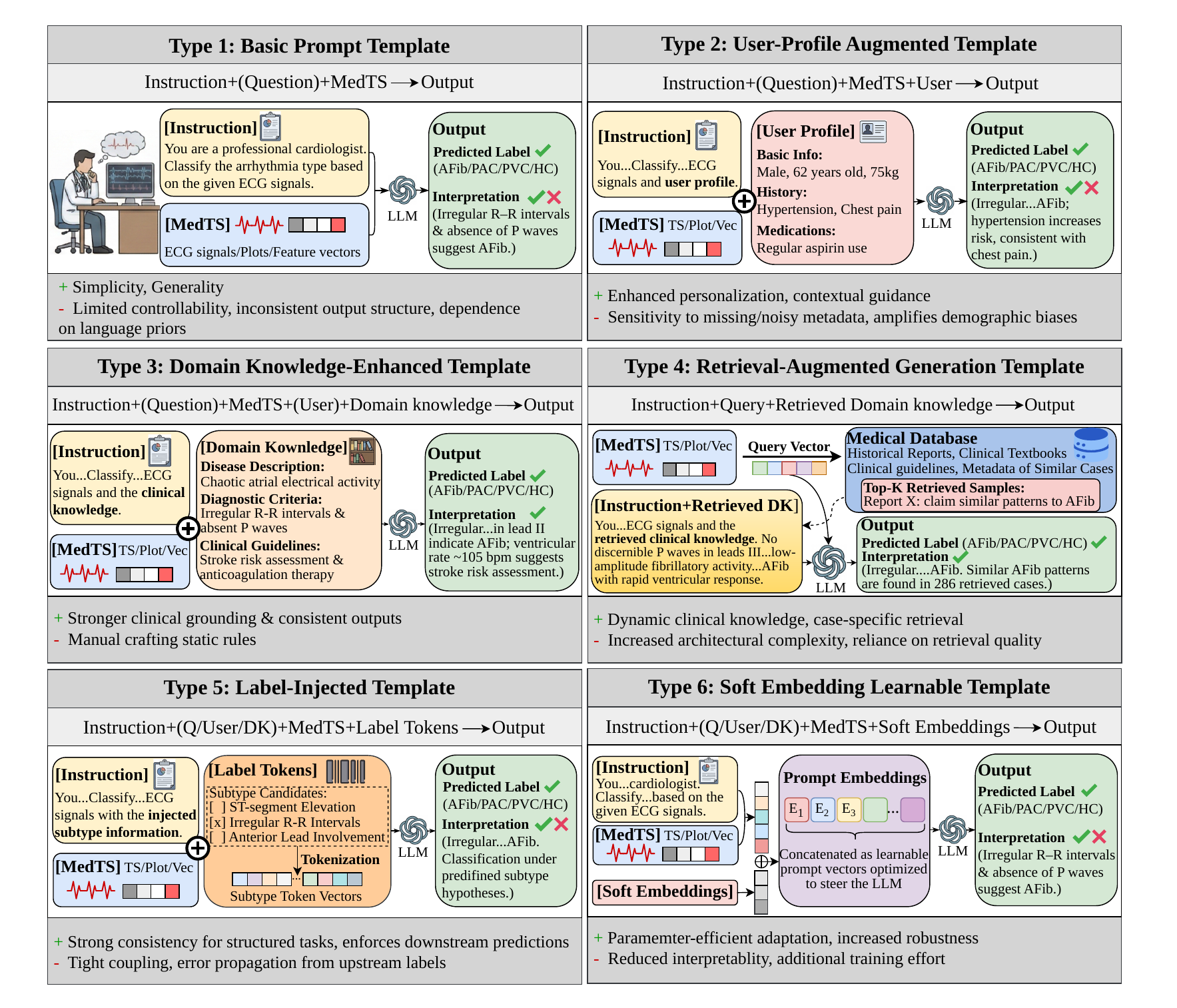}
\caption{\textbf{Diagram of the prompt templates used in MedTSLLM studies.} Here we take ECG-based arrhythmia diagnosis as an example to show how to design a prompt template under different prompt settings. Due to the space limitation, we use some abbreviations in this figure. Abbreviations: MedTS: medical time-series. TS: time-series. AFib: atrial fibrillation. PAC: premature atrial contractions. PVC: premature ventricular contractions. HC: healthy control. User: user profile. Q: question. DK: domain knowledge. To help readers assess the maturity and current adoption of different prompt designs, we report the number of representative studies using each template in the surveyed literature: Type 1: Basic Prompt Template ($n=10$), Type 2: User-Profile Augmented Template ($n=9$), Type 3: Domain Knowledge-Enhanced Template ($n=11$), Type 4: Retrieval-Augmented Generation Template ($n=3$), Type 5: Label-Injected Template ($n=5$), and Type 6: Soft Embedding Learnable Template ($n=2$).} 
\label{fig:Prompt Template}
\end{figure}

\subsubsection*{Instruction Prompting} 

Instruction prompting is one of the most widely adopted approaches in existing MedTSLLM studies. This strategy guides the model through explicit natural-language task descriptions provided at inference time. Typical prompts include statements such as ``\textit{You are a professional cardiologist. Interpret the provided ECG and identify key features and abnormalities in each lead.}'' or ``\textit{Analyze the provided EEG recordings to assess the individual’s stress level and sleep patterns, and generate a concise health assessment report summarizing the key findings.}'' In this paradigm, the prompt specifies the task objective, whereas the internal reasoning process is implicitly handled by the pretrained knowledge of the model.

Instruction prompting is generally implemented in either a zero-shot or few-shot configuration. In zero-shot prompting, the model receives task instructions together with raw medical inputs or their structured representations, without in-context examples. This configuration is common in MedTSLLM applications because large-scale labeled medical time-series data are scarce, and flexible deployment across heterogeneous clinical scenarios is often required. As a result, many studies adopt zero-shot instruction prompting to perform tasks such as ECG report generation~\cite{wan2025meit,yang2025diagecg}, abnormality classification~\cite{tang2026interpretable}, and clinical question answering (QA)~\cite{novak2023pulse,senkaiahliyan2023gpt}. For example, PSRT~\cite{jeong2024prediction} queries the model with a direct instruction to map physiological signal representations to stress predictions in a zero-shot manner. ECG-LM~\cite{yang2025ecg} similarly uses concise task instructions combined with ECG-derived textual descriptions and patient metadata to support ECG QA. By explicitly constraining the label space within the prompt template, ECG-LM encourages the production of precise and clinically consistent diagnostic outputs. In brain decoding applications, zero-shot instruction prompting has also been explored extensively. BSLA~\cite{liu2025llms} aligns autoencoder-derived EEG representations with the LLM embedding space and concatenates them with a task-specific instruction that asks the model to reconstruct the perceived linguistic content. Thought2Text~\cite{mishra2025thought2text} projects image sketches and object labels into a shared embedding space and concatenates them with role-based task instructions. In this framework, the prompt serves as a semantic query that guides the LLM to interpret EEG embeddings as language-like tokens, enabling the generation of coherent textual descriptions of visual stimuli without task-specific fine-tuning or in-context examples.

Although zero-shot instruction prompting is simple and flexible, its performance depends strongly on the capacity of the LLM to infer task intent from instructions alone. Consequently, outputs may be sensitive to prompt phrasing and may lack consistency. These limitations are particularly evident in MedTS tasks that require strict output formats, fine-grained clinical distinctions, or stable reporting styles. To address these issues, some studies adopt few-shot prompting, in which a small number of representative input-output examples are included to clarify task semantics and output structure. WDAI-LLM~\cite{bohi2024large}, for instance, incorporates examples that demonstrate how multimodal wearable time-series features, such as activity levels, heart rate dynamics, and sleep patterns, correspond to high-level health assessments across heterogeneous tasks. PH-LLM~\cite{cosentino2024towards} similarly provides expert-annotated input-output pairs that link summarized longitudinal personal health signals to clinically grounded interpretations, guiding the model to maintain consistent reasoning and output structure when evaluating unseen individuals. Overall, compared with purely zero-shot prompting, few-shot prompting offers a more reliable mechanism for clarifying task semantics and stabilizing output structure by implicitly demonstrating how MedTS data should be contextualized and interpreted when explicit instructions alone are insufficient.

\subsubsection*{Structured Reasoning Prompting} 

Structured reasoning prompting provides a systematic framework that encourages LLMs to perform multi-step and interpretable reasoning before producing a final output. In contrast to instruction prompting, which primarily defines what task should be executed, structured reasoning prompting additionally constrains how the reasoning process should unfold by decomposing complex MedTS tasks into explicit intermediate analytical steps. Within MedTSLLM systems, this paradigm includes widely adopted approaches such as Chain-of-Thought (CoT), Tree-of-Thought (ToT), and Reasoning and Acting (ReAct), all of which structure the reasoning process through explicit intermediate steps instead of relying solely on end-to-end output generation.

Among these approaches, CoT and ToT represent closely related structured reasoning strategies. CoT prompts the model to generate a single explicit reasoning trajectory prior to delivering the final answer. This mechanism has been shown to improve interpretability and performance in tasks that require the integration of multiple MedTS characteristics. For example, Zero-shot VQA~\cite{seki2025assessing} applies CoT to guide multimodal LLMs through stepwise analysis of ECG waveform morphology when addressing visual queries. ECG-ReGen~\cite{tang2025electrocardiogram} similarly adopts CoT to sequentially reason about salient cardiac rhythms and abnormal patterns grounded in retrieved ECG-report pairs. Health-LLM~\cite{kim2024health} also employs CoT to examine temporal dynamics and multiple physiological indicators before generating health assessments, and incorporates self-consistency at inference time by aggregating predictions derived from multiple independently sampled reasoning trajectories.

Despite these advantages, reliance on a single reasoning path makes CoT susceptible to early-stage reasoning errors, particularly in MedTS scenarios characterized by overlapping waveform patterns, signal noise, or subtle temporal and cross-channel variations~\cite{long2023large}. To mitigate this limitation, ToT extends the CoT framework by allowing exploration of multiple candidate reasoning paths, followed by evaluation and selection. This branching and selection mechanism improves robustness and flexibility, although it introduces additional computational overhead and larger prompt complexity. For instance, EEG-GPT~\cite{kim2024eeg} applies ToT to decompose EEG classification into multiple diagnostic branches aligned with distinct clinical perspectives, such as seizure activity and waveform abnormalities. Each branch independently evaluates outputs from domain-specific analysis modules, and intermediate findings are integrated to produce a final classification decision that mirrors a structured clinical workflow. Similarly, CHA-PPGHR~\cite{feli2025llm} leverages ToT to explore alternative strategies for preprocessing, peak detection, and signal quality assessment, ultimately selecting the most physiologically plausible heart-rate estimate under noisy PPG conditions.

ReAct combines explicit reasoning with action execution, enabling an LLM to iteratively alternate between internal deliberation, invocation of external tools or resources, and incorporation of observed outcomes into subsequent reasoning steps~\cite{yao2022react}. Unlike CoT and ToT, which generally operate within a static prompt context, ReAct is well-suited to agent-based settings that require iterative interaction with external systems, computational tools, or heterogeneous data sources. In MedTSLLM applications, ReAct is typically employed to support iterative analysis of MedTS data, adaptive invocation of domain-specific modules, such as signal processing or information retrieval components, and progressive refinement of decisions based on intermediate observations. For example, openCHA~\cite{abbasian2023conversational} proposes a personalized LLM-driven agent architecture that uses ReAct to decompose user queries into successive reasoning-action cycles. In this framework, the model alternates between task interpretation, interaction with external tools or knowledge bases, and updates of its reasoning state based on new evidence. This agent-oriented design demonstrates that ReAct-based orchestration can provide more robust, interpretable, and context-aware health analysis than direct single-pass prompting strategies.

\subsubsection*{Clinical-guided Prompting} 

Clinical-guided prompting refers to prompting strategies that explicitly encode clinical knowledge and decision logic into the prompt to regulate model interpretation and decision-making. In MedTSLLMs, this paradigm aligns model outputs with established clinical reasoning by embedding expert-defined diagnostic criteria, assessment workflows, or scoring frameworks within the prompt. By specifying how physiological signals and patient attributes should be evaluated and integrated, clinical-guided prompting constrains model behavior toward clinically meaningful interpretations, reduces reliance on general language priors that are not explicitly grounded in clinical rules, and improves controllability in safety-critical medical time-series applications~\cite{tang2025electrocardiogram}. According to the mechanism through which domain knowledge is incorporated, this paradigm can be divided into two main forms: manually crafted static prompts and dynamically retrieved prompts based on Retrieval-Augmented Generation (RAG).

In manually crafted prompts with static domain knowledge, clinical guidance is predefined and embedded directly into the prompt template as explicit rules, scoring systems, or physiological principles. Medical rubrics are translated into structured prompt instructions so that the MedTSLLM evaluates inputs according to consistent criteria. Representative studies illustrate different implementations of this static design. GEM~\cite{lan2025gem} employs grounding prompts derived from explicit ECG features and cardiology knowledge. Measurable signal attributes are transformed into structured prompt components through a cardiology-specific diagnosis guider, generating detailed instruction pairs offline. AHS~\cite{safranek2024automated} assesses the capability of MedTSLLMs to apply rule-based logic by embedding detailed scoring rubrics and point scales for patient history directly into the prompt. For wearable biosignals, CBPM-LLaMA~\cite{liu2024large} incorporates predefined physiological relationships, such as those among mean arterial pressure, cardiac output, and peripheral resistance, into the prompt template before requesting blood pressure prediction from extracted signal features.

In contrast, RAG-based prompting introduces clinical knowledge dynamically at inference time through active retrieval from external databases. In this setting, the query typically functions as a multimodal search key, such as raw physiological signals, extracted numerical features, or patient metadata, rather than a simple user question. The RAG system generally follows a two-stage process. First, the multimodal query is used to retrieve relevant knowledge from external sources, including textbooks or historical patient records. Second, the retrieved knowledge is incorporated into a flexible prompt template together with the original task instruction, forming the final input to the MedTSLLM. For example, Zero-shot RAG~\cite{yu2023zero} uses observed ECG abnormalities, such as ST segment elevation, as retrieval key vectors to obtain authoritative clinical guidance, which is then integrated into the prompt for LLM inference. EEG Emotion Copilot~\cite{chen2025eeg} retrieves information from a customized knowledge database while incorporating demographic attributes, facial features, and compressed EEG signals to infer emotional states. EEG-MedRAG~\cite{wang2025eeg} constructs a three-layer hypergraph to dynamically retrieve and integrate historical patient records, medical knowledge facts, and similar real-time EEG representations to inform current diagnostic reasoning. ECG-Chat~\cite{zhao2025ecg} employs a GraphRAG component constructed from cardiology textbooks to retrieve relevant symptom etiologies and clinical guidelines, grounding report generation in explicit diagnostic states and reducing hallucination.

Most clinical-guided prompting strategies are implemented in zero-shot settings, in which clinical knowledge is conveyed entirely through expert-designed or retrieval-augmented prompt templates rather than through example demonstrations. Clinical-guided prompting can therefore be regarded as a specialized extension of instruction prompting with stronger clinically grounded constraints. Notably, it should be distinguished from instruction prompting that merely incorporates patient-specific attributes. Although instruction prompts may include age, sex, symptoms, or medical history to condition outputs, they do not define how these factors should be systematically applied in diagnostic reasoning. In contrast, clinical-guided prompting encodes explicit diagnostic logic within the prompt, specifying how clinical variables and signal features should be interpreted, combined, or weighted to reach a conclusion. This logic is typically implemented through diagnostic criteria, scoring systems such as the HEART score~\cite{safranek2024automated}, rule-based definitions such as waveform thresholds or interval abnormalities, or clinician-style evaluation workflows that structure assessment of rhythm, morphology, and risk factors. In this configuration, patient attributes serve as inputs to predefined clinical rules rather than as passive contextual modifiers.

\subsubsection*{Soft Prompting}

Soft prompting refers to a class of prompting strategies in which task guidance is conveyed through learned continuous prompt representations rather than manually constructed natural-language instructions. Instead of designing textual prompts, this approach introduces a small set of trainable embeddings that are typically prepended to the model input or injected into intermediate layers~\cite{chang2024efficient}. These embeddings are optimized to regulate model behavior while keeping the backbone model largely fixed. This paradigm is particularly suitable for MedTS tasks, which often involve long temporal sequences, subtle dynamics, and heterogeneous signal modalities that are difficult to reliably control through textual prompt engineering alone~\cite{peng2024model}. In contrast to manually designed prompting strategies, which encode task instructions, clinical guidelines, diagnostic criteria, or reasoning procedures explicitly in text, soft prompting internalizes task-relevant knowledge within learned prompt vectors through optimization. As a result, soft prompting provides strong parameter efficiency and deployment flexibility, enabling effective task adaptation without extensive fine-tuning of LLMs. However, this efficiency is accompanied by reduced interpretability, because learned prompt embeddings are not human-readable and their functional contributions are difficult to analyze directly.

In current MedTSLLM research, soft prompting is commonly combined with frozen or lightly fine-tuned LLMs to adapt them to specific physiological signal understanding tasks. For example, Health-Learner~\cite{liu2023large} introduces a small set of continuous prompt embeddings that are integrated into the input space of a frozen LLM, enabling efficient few-shot adaptation for health and wearable data tasks while preserving the pretrained backbone parameters. JoLT~\cite{cai2024jolt} proposes a multimodal alignment framework in which learnable query tokens function as soft prompts to connect time-series encoders with a language decoder. These query embeddings interact with MedTS representations to condition the generation process of the LLM, allowing the production of textual summaries or interpretations of physiological signals. BELT-2~\cite{zhou2024belt2} further incorporates trainable prefix vectors into the attention layers of a BART model, where they act as soft prompts that modulate internal activations during decoding. This design supports multi-task EEG-to-language translation while retaining the majority of pretrained parameters.

\subsection*{Prompting Templates}

Beyond prompting techniques, MedTSLLM studies also vary in how prompts are structured to encode, contextualize, and present physiological signals alongside relevant clinical information to the LLM. Such structural design directly shapes task formulation, model interpretability, and clinical usability. As shown in Figure~2, this section first introduces a basic prompt template formulation and then progressively examines enriched and specialized templates adopted in existing MedTSLLM studies. Our goal is to clarify the design space of prompt templates and illustrate how different template components affect model capabilities and their suitability for specific clinical applications.

Across current MedTSLLM research, despite substantial variation in tasks and model architectures, prompt templates generally follow a shared foundational structure centered on explicit task instructions grounded in medical time-series data. At the most basic level, prompt templates can be abstracted as ``\textit{Instruction} + (\textit{Question}) + \textit{MedTS} $\rightarrow$ \textit{Output}'', where \textit{MedTS} denotes medical time-series inputs or their transformed representations, including raw signals, signal plots, textualized numerical summaries, or learned time-series or image token embeddings. The \textit{Output} varies according to the task, such as a clinician-style answer for question-answering, a diagnostic label for disease classification, a free-form interpretation for health assessment, or a structured clinical report for generation tasks. This formulation constitutes the core template underlying most MedTSLLM systems, with more advanced templates emerging through the inclusion of additional conditioning elements.

The most basic template strictly adheres to this structure without auxiliary context. In this configuration, the instruction, optionally expressed as a question, defines the task objective, and the MedTS input provides the sole evidential basis for reasoning. Such templates are frequently used in baseline and exploratory studies to assess whether LLMs can interpret physiological signals under minimal contextual assumptions. For example, EEG-GPT~\cite{kim2024eeg} conditions the LLM on EEG representations together with concise task instructions to perform classification and interpretation, evaluating model capability under limited constraints. ALPHA~\cite{tang2023alpha} combines wearable physiological measurements with health assessment instructions, prompting the LLM to infer anomalous states directly from signal inputs. MEIT~\cite{wan2025meit} pairs ECG embeddings with natural-language instructions to enable structured ECG report generation based solely on instructions and multimodal ECG data. The primary advantage of this template lies in its simplicity and generality, which facilitate adaptation across tasks and modalities. However, the absence of contextual or clinical constraints often reduces controllability, leads to inconsistent output structure, and increases dependence on pretrained language priors rather than explicit medical reasoning.

To enhance personalization and contextual relevance, many studies extend the basic template by incorporating patient-specific information, resulting in ``\textit{Instruction} + (\textit{Question}) + \textit{MedTS} + \textit{User} $\rightarrow$ \textit{Output}''. Here, \textit{User} typically includes demographic attributes, medical history, symptoms, or longitudinal context. In MedTS analysis, identical or similar signal characteristics may correspond to different clinical interpretations depending on individual factors such as age, baseline health status, or comorbidities~\cite{ahmad2023revolutionizing}. Explicitly encoding such information guides the LLM to condition its reasoning on patient-specific context rather than relying on generic assumptions learned during pretraining. ECG-LM~\cite{yang2025ecg}, for instance, integrates patient attributes, clinical background, and ECG descriptions to modulate diagnostic interpretation based on individualized risk factors. EEG Emotion Copilot~\cite{chen2025eeg} incorporates subject-level information together with EEG representations, enabling emotion inference and assisted medical record generation aligned with individual context. SignalGPT~\cite{liu2023biosignal} embeds user context within a report drafting workflow, where biomedical signals are interpreted in conjunction with background information to produce tailored narrative reports. Nevertheless, such augmentation introduces sensitivity to missing or noisy metadata and may amplify demographic biases if user attributes are overemphasized relative to signal evidence.

A further extension incorporates explicit domain knowledge, yielding templates of the form ``\textit{Instruction} + (\textit{Question}) + \textit{MedTS} + (\textit{User}) + \textit{Domain knowledge} $\rightarrow$ \textit{Output}''. In this setting, \textit{Domain knowledge} includes disease descriptions, diagnostic criteria, clinical guidelines, or scoring standards that define how signals and patient attributes should be interpreted. Compared with user-based augmentation, domain knowledge-enhanced templates provide stronger clinical grounding and more consistent outputs by directing the MedTSLLM to focus on clinically relevant signal attributes and to integrate them according to established criteria. For example, PULSE~\cite{liu2024teach} incorporates expert-annotated ECG descriptions and explanatory text as structured knowledge, enabling alignment between visual patterns and clinically meaningful narratives. In CHA-PPGHR~\cite{feli2025llm}, the agent framework iteratively augments PPG-based heart rate analysis with external textual summaries and intermediate observations generated by signal-processing tools, which are fed back into the prompt to refine subsequent reasoning.

In addition to hard-coded domain knowledge, Retrieval-Augmented Generation (RAG) introduces clinical knowledge dynamically at inference time. In this paradigm, the prompt can be abstracted as ``\textit{Instruction} + \textit{Query} + \textit{Retrieved domain knowledge} $\rightarrow$ \textit{Output}''. The \textit{Query} may include textual questions, MedTS token embeddings or feature vectors, patient case descriptions, or their combinations. \textit{Retrieved domain knowledge} may consist of relevant clinical reports, textbook passages, or metadata from physiologically similar cases. Such knowledge can be incorporated into the prompt in multiple representational forms, including natural-language text, structured graphs, or paired MedTS-diagnosis examples that share similar signal characteristics. For example, ECG-ReGen~\cite{tang2025electrocardiogram} encodes ECG signals to retrieve clinically similar reports, which are injected into the prompt as external evidence to guide report generation and QA. ECG-Bench~\cite{song2025retrieval} extracts multi-domain ECG features to query a dedicated database of historical signals and reports, injecting top-$k$ retrieved diagnostic reports into the prompt to support precise generation. Compared with predefined domain knowledge templates, RAG-based designs reduce the burden of manually crafting exhaustive static rules and allow case-specific retrieval of clinically similar patterns. However, retrieval-augmented systems increase architectural complexity and introduce dependence on retrieval accuracy and preprocessing quality.

Another specialized template explicitly injects labels or ground-truth tokens, expressed as ``\textit{Instruction} + (\textit{Question/User/Domain knowledge}) + \textit{MedTS} + \textit{Labels/Ground-truth tokens} $\rightarrow$ \textit{Output}''. Here, \textit{Labels/Ground-truth tokens} generally represent diagnostic hypotheses or semantic tokens used in signal-to-text tasks. For instance, MERL-CKEPE~\cite{liu2024zero} incorporates candidate diagnostic labels into the prompt as conditioning tokens, constraining classification outputs under predefined hypotheses. Thought2Text~\cite{mishra2025thought2text} injects semantic tokens derived from EEG signals to guide coherent text generation reflecting neural activity. EEG-ETB~\cite{zhang2023integrating} integrates word-level neural state labels and biometric annotations into the prompt, conditioning predictions on explicit cognitive-state information. Such designs enforce strong consistency between upstream predictions and downstream generation, which is beneficial for structured report generation and evaluation. However, tight coupling between labels and outputs could make the system highly sensitive to label quality, and errors or bias in upstream predictions are directly propagated to the generated results.

Finally, some studies replace or augment textual components with learned continuous representations, resulting in ``\textit{Instruction} + (\textit{Question/User/Domain knowledge}) + \textit{MedTS} + \textit{Soft embeddings} $\rightarrow$ \textit{Output}''. \textit{Soft embeddings} are obtained through soft prompting or prompt-tuning and concatenated with MedTS representations to condition a largely frozen LLM. Health-Learner~\cite{liu2023large} introduces trainable soft prompt embeddings prepended to the input sequence and optimized for few-shot health tasks, guiding interpretation of physiological and behavioral signals. JoLT~\cite{cai2023jolt} employs learnable query embeddings that attend to time-series representations and condition the language decoder, effectively functioning as soft prompts that align signals with natural-language outputs. Compared with manually specified prompt components, soft embeddings encode task- and signal-specific inductive biases through optimization, shaping internal attention and activation patterns. This approach enables parameter-efficient adaptation and improved robustness but reduces interpretability and requires additional training effort.

Overall, prompt templates in MedTSLLMs can be viewed as a structured progression from the basic ``\textit{Instruction} + (\textit{Question}) + \textit{MedTS} $\rightarrow$ \textit{Output}'' formulation toward increasingly enriched designs that incorporate user context, domain knowledge, labels, or learned embeddings. These extensions are driven primarily by task requirements, including personalization, clinical alignment, evidence grounding, and adaptation efficiency, rather than by prompting techniques alone. Prompt template design therefore represents a critical yet underexplored dimension in MedTSLLM research, complementing advances in prompting strategies and model architectures.

\begin{figure}[!htbp]
\centering
\includegraphics[width=1.0\linewidth]{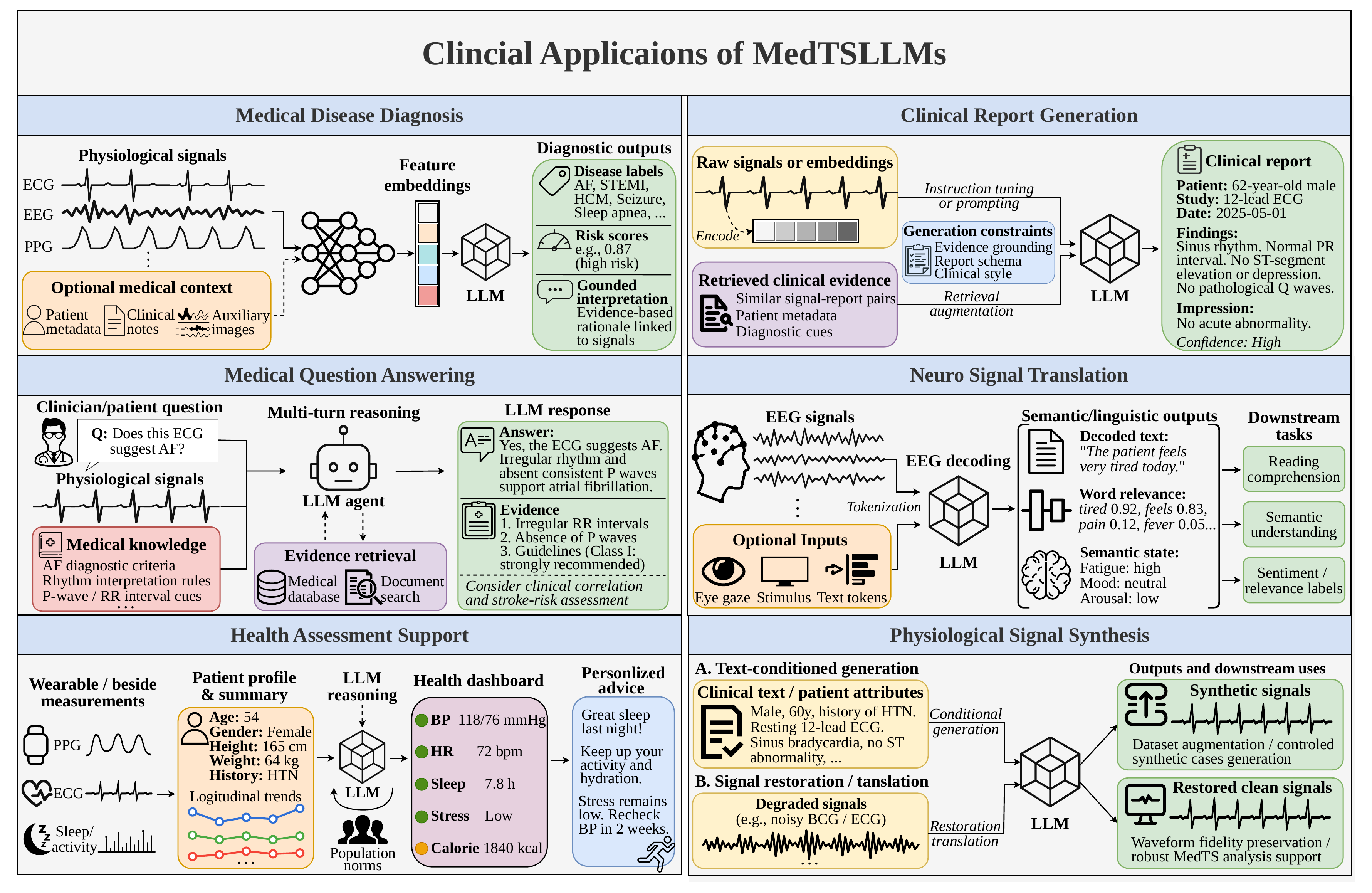}
\caption{\textbf{Integrated overview of potential applications of MedTSLLMs.} Here, we briefly present six main applications of MedTSLLMs and outline their frameworks and key components of MedTSLLMs in different clinical scenarios. Abbreviations: AF: atrial fibrillation. STEMI: ST-segment elevation myocardial infarction. HCM: hypertrophic cardiomyopathy. HTN: hypertension. BP: blood pressure. HR: heart rate.}
\label{fig:Prompt Template}
\end{figure}

\section*{Clinical Applications} 

In this section, we focus on six representative clinical application scenarios that are most clearly reflected in the current MedTSLLM literature, as illustrated in Figure~3. Each subsection analyzes a specific application scenario and describes how MedTSLLMs are formulated to address the corresponding task, with emphasis on input representation, functional role within the modeling pipeline, and evaluation methodology. Table~3 provides practical guidelines for the selection, design, and assessment of MedTSLLMs across distinct clinical use cases. Although large-scale prospective validation remains limited, the growing body of literature reviewed here reflects sustained efforts to evaluate the reliability, interpretability, and translational relevance of MedTSLLMs in real-world clinical settings.

\subsection*{Disease Diagnosis}

\paragraph{Guideline} 
To develop effective MedTSLLMs for disease diagnosis, existing studies generally follow two dominant paradigms that differ in how LLMs interact with physiological signals and diagnostic outputs~\cite{thirunavukarasu2023large}. In the first paradigm, an LLM, either pre-trained or trained from scratch, is used as a representation learner to encode MedTS features, which are then passed to a task-specific classification head for disease prediction. In the second paradigm, powerful pre-trained LLMs are adapted through the injection of domain-specific diagnostic knowledge, enabling direct generation of disease labels or diagnostic interpretations for individual patients. This section summarizes the principal strategies underlying these paradigms and reviews representative implementations.

Within the first paradigm, encoder-style language models, particularly BERT-based architectures, are widely employed as feature extractors for ECG signals. HeartBEiT~\cite{vaid2023foundational} leverages a BEiT model pre-trained on millions of ECG images using masked image modeling and reports consistent improvements across multiple diagnostic tasks, including hypertrophic cardiomyopathy detection, prediction of low left ventricular ejection fraction, and classification of ST-elevation myocardial infarction. ECGBERT~\cite{choi2023ecgbert} similarly pre-trains a BERT model on large-scale unlabeled ECG datasets, such as PTB-XL~\cite{wagner2020ptb}, MIT-BIH~\cite{moody2001impact}, and MIMIC-III~\cite{moody2020mimic}, using self-supervised learning to capture complex temporal dependencies and achieve strong performance on diverse cardiac classification tasks. Beyond unimodal representation learning, several studies incorporate multimodal ECG-text alignment to enhance diagnostic generalization and enable zero-shot learning. sEHR-ECG-Text~\cite{lalam2023ecg} introduces sEHR-BERT to encode structured electronic health records and demonstrates that joint pre-training with ECG signals and unstructured clinical notes improves the quality of ECG representations. ETP~\cite{liu2024etp} combines a BioClinicalBERT text encoder with a ResNet-18 signal encoder and aligns ECG signals with textual prompts to learn transferable representations that support linear probing and zero-shot classification on datasets such as PTB-XL~\cite{wagner2020ptb} and CPSC2018~\cite{liu2018open}. METS~\cite{li2024frozen} extends this multimodal alignment framework by incorporating LLM-driven test-time clinical knowledge. Specifically, complete sentence-level ECG report descriptions are fed into the language model to establish semantic correspondence between clinical narratives and signal features, thereby improving diagnostic robustness.

The second paradigm primarily relies on decoder-only LLMs, such as GPT- and LLaMA-style architectures, which directly generate diagnostic labels or textual interpretations from physiological signals. GPT-PPG~\cite{chen2025gpt} adapts the GPT architecture to continuous physiological data and pre-trains it on more than two hundred million PPG segments, demonstrating strong generalization to unseen distributions and robustness to noisy inputs. MERL-CKEPE~\cite{liu2024zero} jointly trains a ResNet-18 ECG encoder with a pre-trained Med-CPT~\cite{jin2023medcpt} model using multimodal contrastive learning to align ECG signals with clinical reports. During inference, GPT-4 generates knowledge-enriched prompts, resulting in improved diagnostic performance across multiple ECG benchmarks. CardioGPT~\cite{fu2024cardiogpt} trains a GPT model on over one million annotated ECG records to produce free-text diagnostic interpretations, enabling coherent and clinically meaningful narratives directly from ECG data. CQA-ESI~\cite{yu2024ecg} employs GPT-3.5 to retrieve expert knowledge from medical textbooks and generate detailed textual descriptions of ECG signals, whereas ECG-GPT~\cite{khunte2024automated} fine-tunes GPT-2 on UK-Biobank~\cite{bycroft2018uk} and CODE-15~\cite{ribeiro2020automatic} datasets to generate expert-level diagnostic predictions from ECG images without access to raw signals. Beyond purely language-based models, recent work integrates large vision-language models. GEM~\cite{lan2025gem} injects aligned ECG time-series representations into the language space of SFT-LLaVA, enabling the generation of grounded ECG interpretations that explicitly connect diagnostic conclusions to waveform-level evidence rather than producing unconstrained free-text descriptions.

Evaluation protocols for LLM-based ECG diagnostic systems vary according to task formulation but generally incorporate signal-level, text-level, and clinically oriented criteria. For ECG classification tasks, commonly reported metrics include AUC, accuracy, precision, recall, F1-score, and Hamming Loss, often evaluated across multiple disease categories and external cohorts. For free-text diagnostic generation, text similarity metrics, such as ROUGE, BLEU, and CIDEr~\cite{vedantam2015cider,khunte2024automated}, are frequently adopted. More recently, some studies have introduced LLM-based evaluators to assess diagnostic quality under predefined scoring frameworks. For example, GEM~\cite{lan2025gem} defines structured evaluation dimensions, including \textit{AnalysisRelevance}, \textit{LeadAssessmentCoverage}, and \textit{EvidenceBasedReasoning}, by prompting an LLM to systematically score the quality and clinical grounding of generated ECG interpretations.

\paragraph{Functional roles of LLMs} 
In disease diagnosis tasks, MedTSLLMs assume multiple functional roles within the diagnostic pipeline, depending on the mode of integration with MedTS representations. A MedTSLLM may serve as a backbone model that processes latent or tokenized MedTS features and supports end-to-end optimization for representation learning. In this configuration, the model functions as the principal computational module that extracts task-relevant semantic structure for downstream diagnosis.

MedTSLLMs may also operate as text or image encoders that transform heterogeneous inputs, including clinical narratives and medical images, into fixed-dimensional embeddings~\cite{lan2025gem}. These embeddings are subsequently aligned within multimodal architectures to enable diagnostic inference. In this role, the MedTSLLM contributes semantic abstraction and cross-modal alignment rather than directly producing diagnostic decisions.

In generative diagnostic settings, a MedTSLLM may act as a text decoder positioned at the final stage of the workflow. Here, modality-agnostic embeddings derived from physiological signals are mapped into the token space of the language model, which autoregressively generates diagnostic rationales or clinical interpretations~\cite{rajpurkar2022ai}. Alternatively, MedTSLLMs can function as classifiers that directly output diagnostic labels or probability distributions over disease categories through instruction-following or prompt-based inference, particularly in zero-shot or few-shot scenarios. In addition, MedTSLLMs may serve as prompt generators that construct or refine diagnostic prompts by incorporating task instructions, clinical context, and domain knowledge. This capability enhances inference flexibility and supports test-time adaptation without parameter updates~\cite{liu2024zero}.

\paragraph{Discussion}
A defining characteristic of MedTSLLM-based disease diagnosis is that LLMs do not operate directly on raw physiological time-series signals. Instead, reasoning is performed over intermediate representations, such as tokenized signals, projected embeddings, or text-aligned features. This design facilitates flexible diagnostic reasoning and language-based interpretation; however, it also renders diagnostic outcomes sensitive to the fidelity of upstream signal encoding. Clinically salient temporal or morphological patterns may be attenuated before reaching the language model.

When MedTSLLMs are deployed in generative or prompt-based configurations, particularly under zero-shot inference, decoder-style models may produce fluent and medically plausible conclusions that are not fully grounded in waveform-level evidence. These characteristics introduce substantial evaluation challenges. Conventional signal-level metrics assess predictive accuracy but do not determine whether generated diagnoses are physiologically justified, whereas text similarity metrics inadequately capture clinical correctness and evidential support. For example, ZETA~\cite{tang2026interpretable} aligns ECG representations with structured positive and negative clinical observations, demonstrating that accurate label prediction alone is insufficient without verifying that language-based reasoning is explicitly supported by signal-derived evidence. Overall, both model design and evaluation protocols must explicitly address the representational gap between medical time-series data and language-driven diagnostic reasoning in MedTSLLM-based systems.

\begin{landscape}
\scriptsize 
\setlength{\tabcolsep}{3pt} 
\renewcommand{\arraystretch}{1.4} 

\begin{longtable}{
    >{\raggedright\arraybackslash\hspace{0pt}}m{1.5cm}   
    >{\raggedright\arraybackslash\hspace{0pt}}m{2.2cm} 
    >{\raggedright\arraybackslash\hspace{0pt}}m{1.8cm} 
    >{\raggedright\arraybackslash\hspace{0pt}}m{2.2cm} 
    >{\raggedright\arraybackslash\hspace{0pt}}m{1.0cm}   
    >{\raggedright\arraybackslash\hspace{0pt}}m{1.3cm}   
    >{\raggedright\arraybackslash\hspace{0pt}}m{2.3cm}   
    >{\raggedright\arraybackslash\hspace{0pt}}m{2.0cm} 
    >{\raggedright\arraybackslash\hspace{0pt}}m{1.4cm} 
    >{\raggedright\arraybackslash\hspace{0pt}}m{1.6cm} 
    >{\raggedright\arraybackslash\hspace{0pt}}m{4.3cm} 
}

\caption{Summary of existing MedTSLLMs tailored to various clinical applications, in terms of their architecture, functional role, the number of parameters, the prompt techniques and templates, the source, scale, and detailed type of the pre-training/fine-tuning data, and the evaluation information, including the task and performance. Note that, in this review, arrhythmia diagnosis refers to rhythm-related disorders, whereas abnormality diagnosis denotes non-arrhythmic general abnormalities, such as myocardial infarction and ST/T changes. Specific cardiac diseases or functional conditions, such as LVEF and STEMI, are listed separately in the \textit{Evaluation} column when applicable. Abbreviations: M: million. B: billion. TS: time-series. QA: question-answering. CoT: chain-of-thought prompting. ToT: tree-of-thought prompting. Ins: instruction. Q: question. GT: ground truth. TN: textualized numeric context. DK: domain knowledge. Interp: Interpretation. User: user profile. CEN: clinical encounter notes. DTN: descriptive textual narratives. ImgEmb: image token embedding. TSEmb: time-series token embedding. TEXmb: text token embedding. ObjEmb: object-label token embedding. SE: soft embeddings. RR: retrieved reports. DxL: diagnosis labels. TF: time-frequency. BEL: brain electrode location. TW: target words. SHGC: structured hypergraph context. BP: blood pressure. HR: heart rate. CAS: Coronary atherosclerosis. CAM: Cardiac amyloidosis. PH: pulmonary hypertension. LVEF: left ventricular ejection fraction. HCM: hypertrophic cardiomyopathy. STEMI: ST-segment elevation myocardial infarction. AF: atrial fibrillation. IBI: interbeat interval. CBP: cuffless blood pressure} \label{tab:long_table_final} \\
\hline
\textbf{Application} & \textbf{Method} & \textbf{Architecture} & \textbf{Role of LLM} & \textbf{\# Params} & \textbf{Prompt Technique} & \textbf{Prompt Template} & \textbf{Dataset} & \textbf{Data Scale} & \textbf{Data Type} & \textbf{Evaluation (Task: Score)} \\ \hline
\endfirsthead

\caption[]{Cont.} \\
\hline
\textbf{Application} & \textbf{Method} & \textbf{Architecture} & \textbf{Role of LLM} & \textbf{\# Params} & \textbf{Prompt Technique} & \textbf{Prompt Template} & \textbf{Dataset} & \textbf{Data Scale} & \textbf{Data Type} & \textbf{Evaluation (Task: Score)} \\ \hline
\endhead

\hline
\endfoot

\hline
\endlastfoot


\multirow{28}{1.5cm}{Medical\newline disease\newline diagnosis} 
 & \!\footnotemark[1]HeartBEiT \cite{vaid2023foundational} & BEiT & Backbone & 86M & N/A & N/A & MSHS\newline PTB-XL & 8.5M samples\newline 2.1M patients & ECG & LVEF diagnosis: 0.86 AUROC\newline HCM diagnosis: 0.77 AUROC\newline STEMI diagnosis: 0.88 AUROC \\ \cline{2-11} 
 & sEHR-ECG-Text \cite{lalam2023ecg} & BERT & Text encoder & 15M & N/A & N/A & Mayo EHRs\newline PhysioNet2020\newline Chapman & 9M samples\newline 2.4M patients & ECG\newline EHRs\newline sEHRs & Arrhythmia diagnosis: 0.933 AUROC\newline CAS diagnosis: 0.892 AUROC\newline Myocarditis diagnosis: 0.905 AUROC\newline CAM diagnosis: 0.961 AUROC\newline PH diagnosis: 0.917 AUROC\newline LVEF diagnosis: 0.933 AUROC \\ \cline{2-11} 
 & ECGBERT \cite{choi2023ecgbert} & BERT & Backbone & N/A & N/A & N/A & MIMIC-III\newline PTB-XL\newline CPSC2018\newline Georgia\newline MIT-BIH\newline Apnea-ECG & 88K samples 29K patients & ECG & Arrhythmia diagnosis: 0.973 ACC\newline Sleep apnea detection: 0.725 ACC \\ \cline{2-11} 
 & METS \cite{li2024frozen} & Clinical\-BERT & Text encoder & N/A & N/A & N/A & MIMIC-III\newline PTB-XL\newline MIT-BIH & 22K samples & ECG\newline Clinical reports & Abnormality diagnosis: 0.842 ACC\newline Arrhythmia diagnosis: 0.746 ACC \\ \cline{2-11} 
 & \!\footnotemark[2]MERL-CKEPE \cite{liu2024zero} & GPT-4 & Prompt generator & 1.2B & Clinical-guided & Ins+Q+GT$\rightarrow$Prompts & MIMIC-ECG\newline PTB-XL\newline CPSC2018\newline CSN & 770K pairs & ECG\newline Clinical notes & Abnormality diagnosis: 0.906 AUC\newline Zero-shot ECG classification: 0.829 AUC \\ \cline{2-11} 
 & Zero-shot RAG \cite{yu2023zero} & LLaMA-2 GPT-3.5 & Text decoder & 7B\newline 13B\newline 175B & Clinical-guided & Ins+TN+DK$\rightarrow$\newline Labels+Interp & PTB-XL+\newline Apnea-ECG & 56K samples & ECG & Arrhythmia diagnosis: 0.669 F1\newline Sleep apnea detection: 0.801 F1 \\ \cline{2-11} 
 & ECG-GPT \cite{khunte2024automated} & GPT-2 & Text decoder & 2.39B & N/A & N/A & YNHHS\newline CODE15\newline UK Biobank\newline LRH\newline PTB-XL & 2.6M samples & ECG & Abnormality diagnosis: 4.69 CIDEr \\ \cline{2-11} 
 & CardioGPT \cite{fu2024cardiogpt} & CardioGPT & Classifier & N/A & N/A & N/A & GE \& Philips\newline NSH & 1.1M samples 750K patients & ECG\newline Clinical notes & Abnormality diagnosis: 0.91 F1 \\ \cline{2-11} 
 & \!\footnotemark[3]CQA-ESI \cite{yu2024ecg} & GPT-3.5 Biolink\-BERT & Text decoder\newline Text encoder & 26.8M 85.6M & Clinical-guided & Ins+Q+Plot$\rightarrow$\newline Metadata & PTB-XL\newline Chapman\newline ICBEB\newline MIMIC-IV-ECG & 660K pairs & ECG\newline Clinical reports & Arrhythmia diagnosis: 0.982 AUC \\ \cline{2-11} 
 & EEG-GPT \cite{kim2024eeg} & GPT-3 & Classifier & N/A & ToT & Ins+TN$\rightarrow$Labels & Temple-EEG & 2993 patients & EEG & Abnormality diagnosis: 0.86 AUROC \\ \cline{2-11} 
 & ETP \cite{liu2024etp} & Bio\-Clinical\-BERT & Text encoder & 1.2B & N/A & N/A & PTB-XL\newline CPSC2018 & 29K samples & ECG\newline Clinical reports & Seizure diagnosis: 0.861 AUC \\ \cline{2-11} 
 & GPT-PPG \cite{chen2025gpt} & GPT & Backbone & 19M-1B & N/A & N/A & Stanford\newline WESAD\newline DaLiA\newline IEEE\newline SBP\newline DBP\newline BIDMC & 480K samples & PPG & Arrhythmia diagnosis: 0.847 F1 \\ \cline{2-11} 
 & N/A \cite{zeljkovic2025beyond} & GPT-4 & Text decoder & N/A & N/A & N/A & CaRD & 150 patients & ECG\newline Clinical notes & Abnormality diagnosis: 0.837 F1 \\ \cline{2-11} 
 & \!\footnotemark[4]GEM \cite{lan2025gem} & GPT-4o ECG-CoCa SFT-LLaVA-7B & Text decoder Image+TS encoder & 7B & Clinical-guided & Ins+ImgEmb+\newline TSEmb+DK$\rightarrow$Interp & MIMIC-IV-ECG\newline PULSE\newline ECG-Bench & 30K pairs & ECG & Grounded ECG understanding: 0.863 ACC \\ \cline{2-11} 
 & \!\footnotemark[5]MedualTime \cite{ye2024medualtime} & GPT-2 & Backbone & 1M & N/A & N/A & PTB-XL\newline TUSZ & 50K samples & ECG\newline EEG\newline Clinical notes & Abnormality diagnosis: 0.73 F1 \\ \cline{2-11} 
 & \!\footnotemark[6]ZETA \cite{tang2026interpretable} & Flan-T5 Claude-3.5 LLaMA-3.1 & Text encoder Observation generator & 2.8B & Zero-shot & Ins+GT$\rightarrow$\newline Observations & PTB-XL\newline CPSC2018\newline CSN\newline CODE15 & 74K samples & ECG\newline Observations & Abnormality diagnosis: 0.956 AUC\newline Zero-shot ECG classification: 0.81 AUC \\ \hline

\multirow{18}{1.5cm}{Clinical\newline report\newline generation} 
 & JoLT \cite{cai2023jolt} & OPT-2.7B OPT-6.7B & Text decoder & 2.7B 6.7B & Soft prompt & (Ins/Q)+TSEmb+\newline SE$\rightarrow$Interp & PTB-XL & 14.6K pairs & ECG\newline Clinical notes & ECG interpretation: 0.414 METEOR \\ \cline{2-11} 
 & SignalGPT \cite{liu2023biosignal} & ChatGPT & Agent & N/A & Zero-shot & Ins+User+TN$\rightarrow$\newline Reports & TNMG & 827 samples & ECG\newline Clinical reports & Signal-to-text: 0.971 Recall \\ \cline{2-11} 
 & \!\footnotemark[7]MEIT \cite{wan2025meit} & GPT-4 GPT-Neo LLaMA-2 Mistral & Prompt generator\newline Text decoder & 3.45B-20B & Zero-shot & Ins+TSEmb$\rightarrow$Reports & PTB-XL\newline MIMIC-IV-ECG & 800K reports & ECG\newline Clinical reports & Report generation: 0.724 ROUGE-L \\ \cline{2-11} 
 & PhysioLLM \cite{fang2024physiollm} & GPT-4-turbo & Agent & N/A & Zero-shot & Ins+Q+User+TN+\newline Plot$\rightarrow$Reports & Fitbit & 24 patients & Fitbit\newline User biography & N/A \\ \cline{2-11} 
 & \!\footnotemark[8]ECG-Chat \cite{zhao2025ecg} & ECG-CoCa Vicuna-13B & Backbone\newline Report generator & 7B & Clinical-guided & Ins+User+TN$\rightarrow$\newline Reports & MIMIC-IV-ECG\newline CSN\newline SPH\newline PTB-XL\newline CPSC2018 & 240K pairs & ECG\newline Clinical reports & Report generation: 0.894 Recall\newline Report Retrieval: 0.93 Recall \\ \cline{2-11} 
 & ECG-ReGen \cite{tang2025electrocardiogram} & BERT GPT-4o-mini Gemini-Flash & Text encoder\newline Text decoder & 1.1B & Zero-shot CoT & Ins+Q+RR+DxL$\rightarrow$\newline Reports\&Answers & MIMIC-IV-ECG\newline PTB-XL\newline ECG-QA & 27K samples & ECG\newline Clinical reports & Report generation: 0.419 METEOR \\ \cline{2-11} 
 & \!\footnotemark[9]ECG-Bench \cite{song2025retrieval} & Llama-3.2-1B & Backbone & 1B & Clinical-guided & Ins+Query+RR+\newline TSEmb$\rightarrow$Reports & MIMIC-IV-ECG\newline PTB-XL\newline ECG-QA & 420K samples & ECG\newline Clinical reports & Report generation: 0.668 ROUGE-L \\ \cline{2-11} 
 & DiagECG \cite{yang2025diagecg} & LLaMA-3.2-3B & Backbone\newline Report generator & 3B & Zero-shot & Ins+User+(Q)+\newline TSEmb$\rightarrow$Reports & ECG-QA\newline PTB-XL\newline MIMIC-IV-ECG & 1.5M samples & ECG\newline Clinical reports & Report generation: 0.589 ROUGE-L\newline ECG QA: 0.558 ACC \\ \hline

\multirow{6}{1.5cm}{Medical\newline question-\newline answering} 
 & ECG-QA \cite{oh2023ecg} & GPT-4 GPT-3.5 davinci-003 & Answer generator & N/A & Zero-shot & Ins+Q+DK+TN$\rightarrow$\newline Answers & ECG-QA & 414K pairs & ECG\newline QA text& ECG QA: 0.758 AUROC \\ \cline{2-11} 
 & N/A \cite{senkaiahliyan2023gpt} & GPT-4V & Answer generator & N/A & Zero-shot & Ins+Image$\rightarrow$Interp & Wave-Maven\newline Eurorad\newline EEG-Atlas & 69 images & X-ray\newline CT\newline MRI\newline EEG\newline ECG\newline Clinical images & Image interpretation: 1.8/5 (GPT-4V) \\ \cline{2-11} 
 & N/A \cite{gunay2024comparison} & GPT-4 & Answer generator & N/A & N/A & N/A & 150 ECG Cases & 150 cases & ECG questions & ECG QA: 38/40 (GPT-4) \\ \cline{2-11} 
 & AutoHeart \cite{safranek2024automated} & GPT-3.5\newline GPT-4 & Answer generator & N/A & Clinical-guided & Ins+DK+CEN$\rightarrow$\newline HERAT score & N/A & 4 patients & Clinical notes & HEART score: 0.815/1 (GPT-3.5) \\ \cline{2-11} 
 & \!\footnotemark[10]openCHA \cite{abbasian2023conversational} & openCHA & Agent & N/A & ToT+ReAct & Ins+Q+DK+User+\newline DTN$\rightarrow$Interp & Wearable PPG\newline REM sleep & 16 patients & PPG\newline IMU\newline EHRs\newline Monitor images & N/A \\ \cline{2-11} 
 & ECG-LM \cite{yang2025ecg} & BioMedGPT GPT-3.5 & Answer generator & 7B & Zero-shot & Ins+User+TN$\rightarrow$\newline Answers & PTB-XL\newline PTB-XL+\newline ECG-QA & 232K pairs & ECG\newline Clinical reports\newline QA text & ECG QA: 0.577 ACC\newline Abnormality diagnosis: 0.647 F1\newline Zero-shot ECG classification: 0.57 F1 \\ \cline{2-11} 
 & \!\footnotemark[11]PULSE \cite{liu2024teach} & LLaVA-v1.6 GPT-4o LLaMA-3 & Performance evaluator\newline Answer generator & 7B & Clinical-guided & Ins+Q+Report+\newline Image$\rightarrow$Answers & ECGInstruct & 1.2M pairs & ECG plots\newline Clinical reports & ECG QA: 0.738 ACC \\ \cline{2-11} 
 & N/A \cite{novak2023pulse} & Google Bard GPT-4 & Answer generator & N/A & Zero-shot & Ins+Q+User+\newline TN$\rightarrow$Answers & ANA\newline ACO\newline USMLE\newline Guideline review & N/A & QA text & Cardiology reasoning: 0.84 ACC \\ \cline{2-11} 
 & \!\footnotemark[12]CHA-PPGHR \cite{feli2025llm} & GPT-3.5-turbo & Agent & N/A & ToT & Ins+User+Metadata$\rightarrow$\newline Answers & Wearable PPG & 2.5K samples & PPG & HR estimation: 2.83 MAE \\ \cline{2-11} 
 & Zero-shot VQA \cite{seki2025assessing} & Gemini Pro ChatGPT Plus & Answer generator & N/A & Zero-shot CoT & Ins+Q+Plot$\rightarrow$Interp & PTB-XL\newline COVID-19 ECG plots & 23K samples & ECG\newline ECG instructions & Zero-shot VQA: 0.231 F1 (95\% CI) \\ \cline{2-11} 
 & \!\footnotemark[13]ECG-Expert-QA \cite{wang2025ecg} & GPT-4o DeepSeek-V3 Qwen-2.5 & Answer generator & 25.8M & Zero-shot & Ins+Q+User+DK+\newline TN$\rightarrow$Answers & ECG-Expert-QA & 47K pairs & ECG\newline Clinical reports & ECG QA: 0.313 ROUGE-L \\ \cline{2-11} 
 & EEG Emotion Copilot \cite{chen2025eeg} & Qwen-2 InternLM-2.5 & Agent & 0.15B-7B & Clinical-guided & Ins+User+TN$\rightarrow$Interp & FACED & 3.4K samples & EEG & EEG QA: 0.45 F1 \\ \cline{2-11} 
 & \!\footnotemark[14]EEG-MedRAG \cite{wang2025eeg} & GPT-4o-mini Deepseek-r1 & Agent & N/A & Clinical-guided & Ins+Query+SHGC+\newline User$\rightarrow$Answers & CHB-MIT\newline OpenNeuro & 303 patients & EEG\newline Clinical reports & Report generation: 0.222 F1 \\ \hline

\multirow{16}{1.5cm}{Neuro\newline signal\newline translation} 
 & EEG-ETB \cite{zhang2023integrating} & GPT-3.5\newline GPT-4\newline LLaMA & Classifier & N/A & Zero-shot & Ins+Q+TW+GT$\rightarrow$\newline Labels & ZuCo & 12 patients & EEG\newline EyeGaze & Neural state classification: 0.795 ACC\newline Reading comprehension: 0.932 ACC \\ \cline{2-11} 
 & \!\footnotemark[15]DeWave \cite{duan2023dewave} & BART OPT-1.3B LLaMA-7B & Text decoder & 1.3B-7B & N/A & N/A & ZuCo & 13.6K pairs & EEG & EEG-to-text: 0.247 ROUGE-1 F1 \\ \cline{2-11} 
 & \!\footnotemark[16]MTAM \cite{qiu2023can} & BERT & Text encoder & 1.1B & N/A & N/A & ZuCo\newline K-EmoCon & N/A & EEG\newline Language context & Sentiment classification: 0.806 F1\newline Relation detection: 0.736 F1 \\ \cline{2-11} 
 & CFEHC \cite{mischler2024contextual} & Galactica-6.7B OPT-6.7B LLaMA-7B & Backbone & 6.7B-7B & N/A & N/A & ZuCo\newline SQuAD 2.0\newline BoolQ\newline OpenBookQA & N/A & iEEG\newline Language context & N/A \\ \cline{2-11} 
 & iEEG-GPT \cite{lee2024enhancing} & GPT-3.5-turbo & Text decoder & N/A & Zero-shot & Ins+Q+User+TN$\rightarrow$\newline Interp & N/A & N/A & iEEG\newline TF image\newline BEL map & N/A \\ \cline{2-11} 
 & \!\footnotemark[17]WERE \cite{zhang2024word} & BERT GPT-3.5 GPT-4 & Text encoder Classifier & 1.1B & Zero-shot & Ins+Q+TW+GT$\rightarrow$\newline Labels & ZuCo & 9 patients & EEG\newline EyeGaze & Neural state Classification: 0.974 ACC\newline Reading comprehension: 0.988 ACC \\ \cline{2-11} 
 & \!\footnotemark[18]Neuro-GPT \cite{cui2024neuro} & GPT-2 & Text decoder & 1B & N/A & N/A & TUH-EEG\newline BCI-2a & 20K samples & EEG & EEG interpretation: 0.645 ACC \\ \cline{2-11} 
 & BELT \cite{zhou2024belt} & BART & Text decoder & 1.4B & N/A & N/A & ZuCo & 8 patients & EEG & EEG-to-text: 0.326 ROUGE-1 F1\newline Zero-shot sentiment classification: 0.68 F1 \\ \cline{2-11} 
 & BELT-2 \cite{zhou2024belt2} & BART T5 LLaMA-2 & Text decoder & 4B-7.7B & Soft prompt & N/A & ZuCo & 20K samples & EEG & EEG-to-text: 0.376 ROUGE-1 F1\newline Sentiment classification: 0.733 F1 \\ \cline{2-11} 
 & CET-MAE \cite{wang2024enhancing} & BART & Text decoder & 4B & N/A & N/A & ZuCo & 23.5K samples & EEG & EEG-to-text: 0.518 ROUGE-1 F1 \\ \cline{2-11} 
 & BSLA \cite{liu2025llms} & Qwen-2.5 & Backbone Text decoder & 7B & Zero-shot & Ins+TSEmb$\rightarrow$Text & ChineseEEG & 10 patients & EEG\newline Reading stimulus & EEG interpretation: 0.733 F1 \\ \cline{2-11} 
 & \!\footnotemark[19]Thought2Text \cite{mishra2025thought2text} & GPT-4 LLaMA-3 Qwen-2.5 & Caption generator Performance evaluator & 7B\newline 8B & zero-shot & Ins+Img/TSEmb+\newline ObjEmb$\rightarrow$Text & CVPR2017 & 12K pairs & EEG\newline Image stimulus & EEG-to-text: 0.266 ROUGE-L \\ \hline

\multirow{22}{1.5cm}{Health\newline assessment\newline support} 
 & \!\footnotemark[20]ALPHA \cite{tang2023alpha} & GPT-3.5 GLM-2-Pro & Classifier & N/A & Zero-shot & Ins+Image$\rightarrow$Advice & Mmpd\newline Physbench & 312 patients & HR\newline SpO2\newline PPG & Vitals calculation: 0.27 MAE\newline Health-status classification: 0.87 ACC \\ \cline{2-11} 
 & Health-Learner \cite{liu2023large} & PaLM-24B & Predictor & 24B & Soft prompt & Ins+TN+SE$\rightarrow$Health values & MIT-BIH\newline MIMIC-III\newline PAMAP2 & 650 patients & Fitbit & IBI monitoring: 0.897 ACC\newline Activity recognition: 0.85 ACC\newline Calorie estimation: 48 MAE\newline Mental health estimation: 0.825 ACC \\ \cline{2-11} 
 & \!\footnotemark[21]AdaCT \cite{wang2023large} & BERT\newline GPT-2 & Backbone Classifier & 22M-3.55B & N/A & N/A & UCI HAR\newline Sleep-EDF & 61K samples & EEG & Epileptic seizure prediction: 0.979 F1\newline Sleep stage classification: 0.764 F1\newline Activity recognition: 0.995 F1 \\ \cline{2-11} 
 & \!\footnotemark[22]Health-LLM \cite{kim2024health} & HealthAlpaca & Classifier & 7B-70B & Few-shot CoT & Ins+(DK+User)+\newline TN$\rightarrow$Labels & PMData\newline LifeSnaps\newline GLOBEM & 278K samples & Fitbit & Stress monitoring: 0.21 MAE\newline Sleep disorder detection: 0.939 ACC\newline Depression monitoring: 0.24 MAE\newline Calorie estimation: 28.5 MAE \\ \cline{2-11} 
 & CBPM-LLaMA \cite{liu2024large} & LLaMA3-8B & Predictor & 6B-20B & Clinical-guided & Ins+(DK+User)+\newline TN$\rightarrow$BP values & CAS-BP & 12.6K pairs & ECG\newline PPG & CBP measurement: 7.08 MAE \\ \cline{2-11} 
 & WDAI-LLM \cite{bohi2024large} & LLaMA-2 & Predictor & 7B & Few-shot & Ins+(DK)+TN$\rightarrow$\newline Health values & Homekit2020 & 5K patients & Fitbit & HR estimation: 0.38 MAE\newline Calorie prediction: 572 MAE\newline Sleep monitoring: 0.53 MAE \\ \cline{2-11} 
 & PSRT \cite{jeong2024prediction} & GPT-4\newline GPT-3.5 & Classifier & N/A & Zero-shot & Ins+Q+TN$\rightarrow$Labels & SMC-HRV & 41 patients & PPG\newline ECG\newline EEG\newline Fitbit & Stress classification: 0.769 ACC \\ \cline{2-11} 
 & PH-LLM \cite{cosentino2024towards} & Gemini-Ultra-1.0 & Backbone & N/A & Few-shot & Ins+(DK+User)+\newline TN$\rightarrow$Advice & Case studies & 7K samples & Fitbit\newline Clinical reports & Sleep\&fitness QA: 0.79 ACC \\ \cline{2-11} 
 & SensorLM \cite{zhang2025sensorlm} & SensorLM & Text decoder & 15M-2.72B & Clinical-guided & Ins+DK+TN$\rightarrow$Interp & Activity\newline Metabolic & 242K samples & Fitbit\newline Captions & Zero-shot Classification: 0.475 F1\newline Sensor-to-text: 0.994 ACC\newline Text-to-sensor: 0.992 ACC \\ \hline

\multirow{7}{1.5cm}{Physiological\newline signal\newline synthesis} 
 & \!\footnotemark[23]Auto-TTE \cite{chung2023text} & Autoregressive\newline Transformer & Text decoder & N/A & N/A & N/A & PTB-XL\newline Sejong & 59K pairs & ECG\newline Clinical reports & Text-to-ECG: 0.857 AUROC \\ \cline{2-11} 
 & \!\footnotemark[24]ECG-LLM \cite{liu2024ecg} & LLaMA-2 LLaMA-3 & Backbone & 7B\newline 8B & N/A & N/A & PhysioNet2020 & 30K samples & ECG & ECG forecasting: 0.539 MAE \\ \cline{2-11} 
 & BCG2ECG \cite{zuo2024adapting} & Qwen2-1.5B & Backbone & 1.5B & N/A & N/A & Kansas & 40 patients & BCG\newline PPG\newline ECG & ECG reconstruction: 0.113 RMSE \\ \cline{2-11} 
 & \!\footnotemark[25]DiffuSETS \cite{lai2025diffusets} & Ada-v2 & Backbone & N/A & Zero-shot & Ins+Reports$\rightarrow$TXEmb & MIMIC-IV-ECG\newline PTB-XL & 794K samples & ECG\newline Clinical reports & Text-to-ECG: 0.841 F1 \\ \hline
\end{longtable}
\begin{tablenotes}  \tiny
    \item[1]1 \url{https://github.com/akhilvaid/HeartBEiT}
    \item[2]2 \url{https://github.com/cheliu-computation/MERL}
    \item[3]3 \url{https://github.com/comp-well-org/ESI}
    \item[4]4 \url{https://github.com/lanxiang1017/GEM}
    \item[5]5 \url{https://github.com/start2020/MedualTime}
    \item[6]6 \url{https://github.com/Tang-Jia-Lu/Zeta}
    \item[7]7 \url{https://github.com/AIoT-MLSys-Lab/MEIT}
    \item[8]8 \url{https://github.com/YubaoZhao/ECG-Chat}
    \item[9]9 \url{https://github.com/willxxy/ECG-Bench}
    \item[10]10 \url{https://github.com/Institute4FutureHealth/CHA}
    \item[11]11 \url{https://github.com/AIMedLab/PULSE}
    \item[12]12 \url{https://github.com/mohammadfeli/CHA-PPGHR}
    \item[13]13 \url{https://github.com/Zaozzz/ECG-Expert-QA}
    \item[14]14 \url{https://github.com/yi9206413-boop/EEG-MedRAG}
    \item[15]15 \url{https://github.com/duanyiqun/DeWave}
    \item[16]16 \url{https://github.com/Jielin-Qiu/EEG_Language_Alignment}
    \item[17]17 \url{https://github.com/Xemin0/ReadingEmbedding}
    \item[18]18 \url{https://github.com/wenhui0206/NeuroGPT}
    \item[19]19 \url{https://github.com/abhijitmishra/Thought2Text}
    \item[20]20 \url{https://github.com/McJackTang/LLM-HealthAssistant}
    \item[21]21 \url{https://github.com/wangbxj1234/AdaCE}
    \item[22]22 \url{https://github.com/mitmedialab/Health-LLM}
    \item[23]23 \url{https://github.com/TClife/text_to_ecg}
    \item[24]24 \url{https://github.com/dragonlfy/ECG-LLM}
    \item[25]25 \url{https://github.com/Raiiyf/DiffuSETS_Exp}

\end{tablenotes}
\end{landscape}

\subsection*{Clinical Report Generation}

\paragraph{Guideline}

When developing MedTSLLMs for clinical report generation, the central objective is to produce report-style clinical narratives rather than isolated diagnostic labels or short answers. Therefore, the key considerations can be characterized along two critical points: (1) \textit{how clinical information is organized into a coherent report}; and (2) \textit{how the generated report follows domain-specific reporting requirements}. The first point concerns the integration of clinical information, while the second concerns whether the generated report conforms to the expected structure, terminology, and clinical emphasis of a specific reporting scenario.

The first requirement is to integrate heterogeneous clinical information into a coherent report, which requires the model to present multiple levels of information in a structured narrative. A report may include diagnostic impressions, clinical findings, supporting descriptions, patient-specific context, and follow-up suggestions. For example, ECG report generation typically involves key waveform findings (e.g., rhythm patterns and ST-T changes), diagnostic implications, and recommendations for additional examinations or patient prognosis. ECG-Chat~\cite{zhao2025ecg} explicitly organizes patient information, medical history, ECG measurements, diagnostic results, and retrieved cardiology proof, and LLM interpretations into predefined report sections through a LaTeX-based reporting pipeline. DiagECG~\cite{yang2025diagecg} formulates report generation as an instruction-tuning task in which tabular patient features, ECG indicators, and discretized ECG tokens are jointly provided as the clinical context, enabling report generation conditioned on both signal evidence and contextual information. For wearable signals, reports generally describe abnormal events, longitudinal trends, and risk indicators. For example,  PhysioLLM~\cite{fang2024physiollm} summarizes longitudinal records into trends and correlations, converts them into personalized insights, and presents them as integrated health explanations with suggestions and follow-up guidance in the report.

Another important requirement is adherence to clinical reporting requirements. A useful report should not only mention possible abnormalities, but also express them using appropriate medical terminology, maintain a reasonable level of detail, and follow the style of expert-written clinical reports. For example, an ECG report should avoid simply stating ``abnormal ECG'' and should instead describe the relevant rhythm, interval, morphology, or ST-T findings that support the diagnostic impression. Current MedTSLLMs address this requirement through three main modeling paradigms. (1) Instruction-tuned MedTSLLMs, such as ECG-Chat~\cite{zhao2025ecg} and MEIT~\cite{wan2025meit}, learn reporting styles directly from curated signal-text instructions, enabling the model to produce reports that more closely resemble expert-written narratives. (2) Agent-style MedTSLLMs further extend report generation by coordinating multiple analytical modules before producing the final narrative. SignalGPT~\cite{liu2023biosignal} leverages a ChatGPT controller to invoke modality-specific biomedical signal processing engines, and a buffer module to automatically update user feedback and physician corrections to recall patient information. PhysioLLM~\cite{fang2024physiollm} uses a statistical analysis tool to identify key patterns, trends, and relationships in wearable data, supporting the LLM to generate more informative and personalized explanations in the report.
(3) Retrieval-based MedTSLLMs, including ECG-ReGen~\cite{tang2025electrocardiogram} and ECG-Bench~\cite{song2025retrieval}, retrieve similar signal-report pairs as reference cases, helping the model follow clinically plausible reporting patterns and improving the interpretability of generated reports.

Evaluation protocols commonly rely on large-scale ECG corpora (e.g., PTB-XL~\cite{wagner2020ptb}, MIMIC-IV-ECG~\cite{gow2023mimic}, ECGInstruct~\cite{liu2024teach}, and CODE-15). Reporting quality is typically assessed using natural language generation (NLG) metrics (e.g., BLEU, ROUGE, and METEOR) as well as semantic similarity metrics (e.g., BERTScore). Several studies additionally incorporate clinical efficacy metrics to evaluate whether generated reports accurately reflect expert-labeled medical conditions. Collectively, these findings indicate that reliable clinical report generation requires an appropriate signal-language interface, explicit evidence grounding through prompts, instruction tuning, or retrieval mechanisms, and evaluation protocols that jointly consider linguistic quality and clinical correctness.

\paragraph{Functional roles of LLMs} 
In clinical report generation, MedTSLLMs are commonly integrated into multi-stage signal-to-text pipelines. Their functional role depends on their position within the pipeline and the degree of responsibility assigned for signal interpretation, content planning, and clinical reasoning~\cite{van2023clinical}. These roles can therefore be conceptualized as local roles, in which the MedTSLLM operates as a modular component constrained by upstream design, and global roles, in which it governs the overall report generation process~\cite{consort2019reporting}.

At the local level, a MedTSLLM may function as a \textit{backbone model}. In this configuration, it directly models signal-conditioned representations in an autoregressive manner, tightly coupling signal understanding with language modeling~\cite{song2025retrieval}. As a result, downstream behavior is largely determined by internal representations learned during training rather than by external control modules. A related but more constrained local role is that of a \textit{text decoder}. In this setting, the MedTSLLM is placed downstream of specialized components, such as signal encoders, alignment modules, or retrieval systems, and focuses on token-level realization by transforming upstream representations into fluent and clinically appropriate text~\cite{cai2024jolt}. Decisions regarding content selection and prioritization are primarily made upstream, and the decoder executes linguistic rendering under these constraints. Another local role is that of a \textit{prompt generator}. Instead of directly producing the final report, the MedTSLLM constructs, augments, or restructures prompts that condition subsequent generation, for example by composing task instructions, incorporating retrieved evidence, injecting diagnostic cues, or enforcing reporting schemas. This role enables dynamic control and test-time adaptation without modifying model parameters.

At the global level, a MedTSLLM may act as an end-to-end \textit{report generator}, assuming responsibility for selecting, organizing, and synthesizing signal-derived findings into a coherent clinical narrative~\cite{yang2025diagecg,zhao2025ecg}. Compared with the text decoder role, this configuration entails higher-level content planning, including determining which abnormalities to emphasize, structuring report sections, integrating patient context, and maintaining consistency across sections. In addition, MedTSLLMs may operate as \textit{copilots} in interactive, human-in-the-loop workflows~\cite{liu2023biosignal}. Rather than autonomously finalizing reports, they assist clinicians by drafting text, refining intermediate outputs, responding to follow-up queries, and revising content based on feedback, thereby emphasizing responsiveness, interpretability, and iterative collaboration.

\paragraph{Discussion}
A central observation in recent MedTSLLM-based clinical report generation is that overall quality is often constrained less by linguistic fluency than by the preservation and alignment of clinically salient evidence prior to decoding~\cite{wan2025meit,bartels2022ecgcaptions}. Most pipelines rely on intermediate representations, such as projected embeddings, discretized tokens, or retrieved exemplars, to interface with an LLM. Although these interfaces enable flexible text generation, they inevitably compress long and high-resolution waveforms. Such compression may obscure subtle yet clinically meaningful variations, including morphological changes or interval deviations, that materially influence interpretation. This challenge is further exacerbated by the heterogeneity and limited standardization of free-text clinical reports used as supervision, which can hinder generalization across datasets~\cite{vanveen2024summarizationllm}.

Retrieval-augmented designs partially address these issues by grounding generation in authentic clinical phrasing and providing case-based context. However, they introduce dependencies on retrieval quality and corpus representativeness. In particular, similarity measured in embedding space does not always correspond to physiological similarity at the patient level~\cite{yu2023zero,hager2024limitationsllmclinical}. These properties also complicate evaluation. Clinical reports are typically concise and pattern-focused, and minor misalignments between encoded evidence and generated statements can lead to clinically significant distortions that are insufficiently penalized by conventional text-based metrics, such as BLEU or ROUGE. Consequently, several studies complement text metrics with clinically oriented, concept-level evaluation strategies, for example by extracting diagnostic statements or condition labels from generated reports to quantify clinical efficacy and faithfulness~\cite{johri2025craftmd}. Overall, effective MedTSLLM-driven report generation depends not only on architectural design and model scaling, but critically on robust signal-text alignment and evaluation protocols that jointly assess evidential grounding and clinical narrative accuracy.

\subsection*{Medical Question Answering}

\paragraph{Guideline}
Large language models can be systematically integrated into medical QA systems to support clinical reasoning and provide an interactive interface for querying physiological signals, clinical context, and diagnostic knowledge~\cite{ren2024healthcare,wang2024jmlr}. Compared with rule-based algorithms or systems trained on limited datasets, these approaches leverage the broad knowledge and reasoning capacity of LLMs to facilitate diagnostic dialogue and generate context-aware recommendations. Existing MedTSLLM studies on medical QA can be broadly categorized into two dominant design paradigms, which differ in the manner in which physiological signals are incorporated and in the functional role assigned to MedTSLLMs during clinical reasoning. The first paradigm formulates medical QA as a \textit{signal-grounded knowledge querying task}, in which the LLM answers predefined or benchmarked questions by reasoning over aligned physiological representations. The second paradigm conceptualizes QA as an \textit{interactive clinical reasoning process}, where LLMs operate as agents or copilots that integrate physiological signals, retrieved medical knowledge, and multi-turn dialogue to assist diagnostic decision making.

Within the first paradigm, MedTSLLMs are typically evaluated on structured or semi-structured QA benchmarks that explicitly associate physiological signals with clinical questions. To enable this setting, recent work has introduced comprehensive ECG-centered QA datasets, in which each signal is paired with clinically grounded questions and expert-verified answers spanning waveform morphology, rhythm analysis, abnormality identification, and diagnostic inference. For example, ECG-QA~\cite{oh2023ecg} evaluates whether models can correctly answer close-ended, diagnosis-oriented questions by reasoning over signal-derived features, whereas ECG-Expert-QA~\cite{wang2025ecg} introduces expert-level questions that require multi-step clinical inference beyond surface-level pattern recognition. Complementing these benchmarks, ECG-Instruct~\cite{liu2024teach} is proposed as a large-scale instruction tuning dataset designed to bridge structured ECG QA and instruction-driven medical reasoning. It adopts an instruction-following formulation that unifies diverse ECG understanding tasks, including QA, interpretation, and explanation, under natural language instructions. Building on these datasets, ECG-LM~\cite{yang2025ecg} integrates a frozen BioMedGPT~\cite{luo2023biomedgpt} into the QA pipeline by aligning ECG signals with clinical language during training, demonstrating that MedTSLLMs can extend beyond classification toward clinically meaningful QA. Similarly, PULSE~\cite{liu2024teach} extends this paradigm to ECG images and leverages multimodal LLMs to answer ECG-related questions directly from visual inputs under diverse QA formats.

The second paradigm emphasizes interactive and knowledge-augmented medical QA, in which MedTSLLMs function as reasoning agents that guide QA through multi-turn dialogue and contextual evidence integration. For instance, EEG-MedRAG~\cite{wang2025eeg} employs a hierarchical hypergraph to integrate EEG waveforms, clinical knowledge, and patient metadata for joint semantic-temporal retrieval in QA tasks. EEG Emotion Copilot~\cite{chen2025eeg} performs real-time, end-to-end emotion computation from brain activity by mapping wavelet-compressed EEG features to affective states and subsequently conducting multi-step reasoning to generate personalized treatment plans and structured medical records. openCHA~\cite{abbasian2023conversational} enables an LLM agent to access personal health data and recent medical literature to provide personalized and up-to-date responses. CHA-PPGHR~\cite{feli2025llm} follows this framework and leverages it as an intelligent agent that coordinates user queries with established analytical tools, such as heart rate estimation from wearable PPG data, to deliver reliable health insights. Collectively, these systems illustrate the potential of MedTSLLMs to support AI-driven medical consultation and assist clinicians in routine practice.

Evaluation of MedTSLLM-based medical QA remains heterogeneous but exhibits several common patterns that emphasize both clinical accuracy and response efficiency. Protocols frequently report metrics such as accuracy, precision, recall, F1-score, AUROC, and Hamming loss (HL), reflecting the focus on predefined questions grounded in MedTS. Several studies additionally adopt Exact Match (EM) to assess factual consistency and token-level alignment between generated answers and reference responses. Some benchmarks, including ECG-Expert-QA~\cite{wang2025ecg}, further incorporate natural language generation metrics, such as BLEU, ROUGE, and METEOR, to quantify lexical and semantic overlap. Model-to-Model Scoring is also introduced in certain settings, in which a strong LLM evaluates answer quality across dimensions such as clinical correctness and completeness. Beyond correctness, recent work evaluates the agentic viability of these systems by measuring average response time for real-time suitability and employing LLM-based evaluators to provide structured and scalable assessments of open-ended reasoning. However, multiple studies emphasize that linguistic fluency alone is insufficient. Effective evaluation must explicitly consider physiological grounding, adherence to clinical guidelines, and robustness to ambiguous or incomplete signal inputs.

\paragraph{Functional roles of LLMs} 
In medical QA, MedTSLLMs can assume multiple functional roles within the QA pipeline, reflecting different levels of control over reasoning and evidence integration. These roles include answer generators, LLM agents, and performance evaluators, each corresponding to a distinct mode of integration between language modeling, physiological evidence, and supervision signals.

In many benchmark-oriented settings, MedTSLLMs function as answer generators and serve as the terminal component of the QA workflow. In this role, the model receives question-conditioned representations derived from physiological signals, such as encoded features, symbolic summaries, or multimodal embeddings, and produces a natural language response. The MedTSLLM primarily operates as a conditional generation module that maps aligned signal-language representations to structured or semi-structured answer formats. Its contribution lies in semantic reasoning and linguistic abstraction, whereas the orchestration of the QA process is largely determined by upstream components~\cite{singhal2025toward}.

Beyond terminal answer generation, MedTSLLMs may operate as LLM agents with a more active role in medical QA. In this configuration, the model not only generates responses but also guides the reasoning process through multi-turn dialogue, contextual integration, and dynamic evidence retrieval~\cite{goodell2025large}. The MedTSLLM incorporates conversational history, intermediate reasoning states, and auxiliary information, such as retrieved signal segments or clinical metadata, to support iterative clarification and decision making. This agentic role shifts the emphasis from isolated answer production to interactive clinical reasoning that more closely resembles real-world diagnostic workflows~\cite{chen2025enhancing,hager2024evaluation}.

In addition, MedTSLLMs can function as performance evaluators that automatically assess QA outputs. Given the limited capacity of traditional metrics, such as accuracy or ROUGE, to capture nuanced clinical reasoning, an LLM may be employed to score or compare generated answers according to criteria including clinical correctness, completeness, coherence, and consistency with expert references~\cite{zhou2025automating}. This role supports scalable evaluation of open-ended or free-text QA tasks, complements conventional automatic metrics, and reduces reliance on exhaustive human annotation. The evaluator role highlights the dual utility of MedTSLLMs in both solving and assessing medical QA tasks~\cite{agrawal2025evaluation}.

\paragraph{Discussion}
Recent studies demonstrate the strong performance of MedTSLLMs on medical QA benchmarks. However, fluent and confident responses do not necessarily reflect correct physiological reasoning~\cite{singhal2023large}. Several ECG and EEG investigations report that LLM-generated answers can be linguistically plausible yet inconsistent with waveform-level evidence or established clinical guidelines, particularly when signal information is incomplete, ambiguous, or weakly represented~\cite{senkaiahliyan2023gpt,novak2023pulse}. These observations expose a fundamental tension between textual plausibility and physiological grounding.

Evaluation practices further illustrate this tension. Although benchmarks such as ECG-Expert-QA combine lexical overlap metrics with LLM-based comparative scoring for open-ended answers, clinician-centered evaluations indicate that automatic metrics alone frequently fail to identify clinically meaningful errors~\cite{fast2024autonomous, wang2025trustworthy}. This discrepancy between quantitative scores and clinical validity has been emphasized in recent medical AI assessments, motivating evaluation protocols that explicitly incorporate physiological evidence, guideline adherence, and robustness analysis rather than relying solely on surface-level textual similarity.

From a system design perspective, effective MedTSLLM frameworks for medical QA consistently emphasize robust signal-language alignment to ensure that responses are grounded in physiological evidence rather than prior textual knowledge. Incorporation of structured clinical context, including patient metadata, scoring criteria, and guideline constraints, further improves relevance and safety, particularly for decision-oriented queries. Moreover, interactive and multi-turn QA settings better approximate clinical practice, where clarification and progressive refinement are often required. Agent-based and retrieval-augmented architectures enhance reliability by decomposing complex queries into interpretable substeps and explicitly referencing intermediate evidence during answer generation.

\subsection*{Neuro Signal Translation}

\paragraph{Guideline}
When developing MedTSLLMs for neuro signal translation, particularly for EEG-to-text decoding, existing frameworks can be analyzed along three fundamental dimensions: (1) \textit{the supervision strategy that renders neural dynamics linguistically interpretable}, (2) \textit{the paradigm for cross-modal semantic alignment}, and (3) \textit{the mechanism for connecting and adapting MedTSLLMs to neural representations}.

Supervision and task formulation largely determine the type and granularity of semantic content that can be recovered from EEG. A representative line of research adopts reading-based, word-level supervision derived from reading corpora such as ZuCo~\cite{hollenstein2018zuco,hollenstein2020zuco}. In this setting, EEG signals, often synchronized with eye fixations, are segmented at word granularity and paired with linguistic targets obtained from the reading task. For example, studies on word-level neural state classification and reading comprehension employ LLMs to generate relevance annotations, such as high- versus low-relevance words, which serve as proxy semantic labels. EEG and eye-tracking encoders are then trained to predict these labels~\cite{zhang2024word,zhang2023integrating}. In this configuration, LLMs function primarily as semantic annotators rather than decoders, grounding neural activity in task-specific linguistic structure. An alternative supervision strategy emphasizes stimulus-based pseudo-captioning, which avoids explicit language processing during EEG acquisition. Thought2Text~\cite{mishra2025thought2text} trains a lightweight projection module to map EEG representations into the token embedding space of instruction-tuned LLMs for text generation. By using captions generated by GPT-4 as training targets, visual stimuli serve as an indirect semantic bridge between EEG and text.

Given the scarcity, noise, and subject variability of paired EEG-text data, many systems treat representation alignment and invariance as prerequisites for effective neuro translation. BELT~\cite{zhou2024belt} formulates EEG-language alignment as a contrastive learning problem, aligning EEG representations with subword-level language embeddings to support open-vocabulary translation. BELT-2~\cite{zhou2024belt2} strengthens this alignment by introducing a byte-pair encoding-level contrastive objective and query-based prompting over discretized EEG tokens, enabling multi-task EEG-to-language decoding within a unified alignment framework. Similarly, DeWave~\cite{duan2023dewave} promotes stable EEG-text correspondence by aligning discrete EEG tokens with textual embeddings, demonstrating how discretization facilitates cross-modal alignment under weak supervision. Beyond contrastive objectives, some studies treat alignment as an analyzable representational property. MTAM~\cite{qiu2023can} aligns EEG and language features using canonical correlation and Wasserstein distance, improving performance on sentiment analysis and semantic relation tasks while providing frequency- and topography-aware interpretability of alignment strength. In addition, self-supervised learning has emerged as a dominant strategy to mitigate data scarcity. CET-MAE~\cite{wang2024enhancing} and Neuro-GPT~\cite{cui2024neuro} employ masked signal modeling to reconstruct brain activity from corrupted inputs, thereby encouraging the learning of latent spatio-temporal structure. In visual-to-text scenarios, models such as Thought2Text~\cite{mishra2025thought2text} use visual stimuli as language-agnostic proxies and align EEG embeddings with semantically rich CLIP visual representations.

With respect to connection strategies between neural encoders and MedTSLLMs, existing frameworks span a spectrum from lightweight prompting interfaces to parameter-efficient fine-tuning, with a clear emphasis on minimizing adaptation under limited neural data. Some approaches treat decoder-only LLMs as neural interpreters by transforming preprocessed neural features into structured textual prompts. For example, iEEG-GPT~\cite{lee2024enhancing} fine-tunes GPT-3.5 Turbo with neuroscience knowledge and prompts it using region-specific spectral features derived from iEEG, enabling the generation of interpretable descriptions of brain states and cognitive processes. Other approaches introduce bridging mechanisms that connect neural encoders to frozen LLMs. BELT-2~\cite{zhou2024belt2} injects continuous EEG-derived virtual tokens into a frozen LLaMA2-7B via prefix tuning, allowing neural information to condition language generation without full end-to-end fine-tuning. Thought2Text~\cite{mishra2025thought2text} similarly freezes Qwen2.5-7B and trains a lightweight multilayer perceptron projector, originally designed for image-caption alignment, to support EEG decoding. In contrast, some studies integrate encoder-only or encoder-decoder LLMs as semantic backbones rather than generators. For instance, WERE~\cite{zhang2024word} fine-tunes a BERT model to produce contextualized word representations and uses GPT-4 as a supervisory agent to generate relevance labels that guide EEG-eye-tracking fusion models.

Evaluation protocols for MedTSLLMs in neuro signal translation are closely linked to task formulation and can be divided into generative and discriminative settings. For generative EEG-to-text tasks, evaluation primarily relies on standard natural language generation metrics, including ROUGE, BLEU, METEOR, and BERTScore, which measure lexical overlap and semantic similarity between generated text and reference descriptions. These metrics are commonly adopted in EEG translation and captioning-style tasks to assess surface fidelity and semantic consistency. In contrast, classification-oriented tasks, such as sentiment classification and semantic relation detection, use conventional discriminative metrics, including accuracy, precision, recall, and F1-score, often under intra-subject or inter-subject protocols to reflect generalization robustness. Some studies~\cite{mishra2025thought2text} additionally employ LLM-based tools, such as GPT-4, to evaluate fluency and semantic adequacy using predefined ordinal rating scales. Although such evaluations provide complementary qualitative insight, they remain sensitive to prompt configuration and evaluator bias, and therefore should be interpreted as auxiliary evidence rather than definitive performance indicators.

\paragraph{Functional roles of LLMs}
In neuro signal translation, MedTSLLMs are integrated at different stages of the EEG-to-language pipeline, ranging from representational anchoring to end-task decoding. Across existing studies, four functional roles appear most frequently: backbone, classifier, text decoder, and performance evaluator. Each role corresponds to a distinct point of integration and reflects a different balance between neural evidence and linguistic priors.

When serving as a backbone, a MedTSLLM provides a stable semantic space that structures learning from limited and noisy neural data. In this configuration, contextual embeddings of the LLM organize neural representations toward language without requiring explicit text generation. In some cases, the backbone may also refer to an EEG foundation model trained with self-supervised objectives, such as masked modeling, to capture generalizable spatio-temporal structure prior to language grounding. Certain studies~\cite{gao2025increasing,chen2025eeg} further treat MedTSLLMs as reference backbones by probing internal feature hierarchies to examine whether neural signals exhibit language-like representational properties.

For discriminative neuro-language tasks, MedTSLLMs often function as classification heads rather than generators. In this role, neural representations are mapped into discrete semantic states, such as sentiment categories, relation types, or word-level relevance labels. A common pattern is that MedTSLLMs provide task-relevant semantic structure through contextual embeddings or LLM-generated pseudo-labels, while neural encoders are trained to predict these targets. Consequently, performance in this setting reflects the adequacy of the semantic label space as well as the extent to which neural activity can support meaningful discrimination.

In generative EEG-to-text translation, MedTSLLMs operate as conditional decoders that transform EEG-conditioned representations into natural language. As neural signals rarely determine text uniquely, decoding quality depends heavily on how effectively the interface constrains linguistic priors. Existing systems therefore invest in bridging mechanisms, including learned projection modules, soft prompts or prefix tuning, and discretized neural tokens, to regulate extrapolation beyond neural evidence~\cite{croxford2025evaluating}.

MedTSLLMs are also increasingly employed as performance evaluators when conventional metrics are insufficient. In generative settings, overlap-based metrics capture only partial aspects of correctness, motivating the use of LLM-based scoring to assess fluency and semantic adequacy under rubric-based criteria. In classification and alignment tasks, LLM-based evaluation can serve as a qualitative check on whether decoded semantics are coherent and consistent with task context.

\paragraph{Discussion}
Recent research indicates that the primary contribution of LLMs to neuro signal translation lies in semantic structuring rather than generative capacity alone. In classification and reading-based settings, MedTSLLMs often function as semantic supervisors that provide linguistically meaningful guidance, shaping neural representations without explicit text generation. In contrast, generative EEG-to-text tasks rely more heavily on strong linguistic priors, which can produce fluent outputs but may obscure or distort underlying neural evidence if not carefully constrained~\cite{chen2023can}. This tension is amplified by cross-subject variability, as individual differences in neural dynamics frequently dominate task-relevant semantics and limit generalization.

At the system level, the combination of strong language priors and imperfect neural evidence increases the risk of misinterpretation. If generated content is influenced by uncurated or noisy knowledge sources, clinically misleading conclusions may arise. Therefore, neuro signal translation systems should prioritize curated and authoritative medical resources, including textbooks and peer-reviewed literature, to reduce the propagation of misinformation and improve the reliability of generated interpretations~\cite{karabacak2023advent}.

\subsection*{Health Assessment Support}

\paragraph{Guideline}
Health assessment support is a major application domain of MedTSLLMs. The objective is to quantify and interpret physiological states, such as stress, sleep quality, cardiovascular function, and energy expenditure, using MedTS collected from wearable or clinical sensors. In contrast to disease diagnosis, these tasks are often continuous, personalized, and measurement-centric, with an emphasis on numerical accuracy, longitudinal consistency, and physiological plausibility. Existing studies indicate that MedTSLLMs are typically not used as end-to-end signal predictors. Instead, MedTSLLMs are positioned as grounded reasoning and integration modules that connect physiological measurements to health-related interpretations. Under this perspective, current approaches can be organized into two dominant paradigms.

The first paradigm treats MedTSLLMs as numerically grounded health assessors, in which structured physiological measurements are embedded into prompts and interpreted through language-based reasoning. In this setting, raw ECG, PPG, or wearable signals are first converted into numerical descriptors, such as heartbeat intervals, heart rate, and sleep duration, and then embedded into textual prompts. For instance, Health-Learner~\cite{liu2023large} uses PaLM-24B as a universal health learner and applies few-shot prompt tuning to support assessment tasks, including atrial fibrillation classification, physical activity recognition, and metabolic calorie estimation. Health-LLM~\cite{kim2024health} develops HealthAlpaca and evaluates multiple LLMs with a context enhancement strategy that incorporates user profiles, health knowledge, and temporal sequences into prompts to improve grounding. WDAI-LLM~\cite{bohi2024large} uses digit-level tokenization and in-context learning with an off-the-shelf LLaMA-2 to perform zero-shot feature extraction from heart rate and step tuples for metabolic calorie prediction and sleep detection. AdaCT~\cite{wang2023large} uses plug-and-play adapters to convert raw EEG sequences into two-dimensional pseudo-images or text strings for seizure recognition and sleep stage classification.

The second paradigm augments numerical grounding with explicit domain knowledge, personalization, and physiological reasoning constraints, positioning MedTSLLMs as integrative engines for contextual health assessment. SensorLM~\cite{zhang2025sensorlm} builds a sensor-aware LLM pre-trained on large-scale heterogeneous wearable data aligned with semantic text descriptions, supporting zero-shot activity recognition and few-shot prediction of clinical conditions. PH-LLM~\cite{cosentino2024towards} fine-tunes Gemini Ultra-1.0 on a mixture of sleep and fitness case studies and uses a multimodal multilayer perceptron adapter to integrate numerical time-series sensor data as input tokens. It supports personal health by generating long-form coaching recommendations and predicting patient-reported sleep quality outcomes with performance comparable to human experts and specialized discriminative models. CBPM-LLaMA~\cite{liu2024large} leverages an instruction-tuned LLaMA-3 and uses context-enhanced prompts that combine ECG and PPG features with clinical knowledge and user demographics, achieving high accuracy in cuffless blood pressure assessment.

Evaluation of MedTSLLM-based health assessment systems requires task-specific protocols. For measurement tasks, such as heart rate estimation, blood pressure prediction, and calorie calculation, studies commonly report mean absolute error (MAE), root mean squared error (RMSE), and mean absolute percentage error (MAPE), often alongside clinically acceptable error margins or regulatory standards. In cuffless blood pressure estimation, error distributions are frequently reported to assess agreement with reference measurements. For categorical or ordinal assessments, including stress levels, activity recognition, and anomaly detection, standard classification metrics, including accuracy, precision, recall, F1-score, AUROC, and AUPRC, are widely used. Some works report the Exact Match Ratio (EMR) to evaluate whether predicted physiological feature changes exactly match reference values. In stress assessment, some studies compute respiratory irregularity, defined as the mean of the standard deviations of respiratory peaks and valleys, to quantify whether predicted stress states correspond to measurable breathing instability. In addition, some studies benchmark LLM performance against domain experts. In sleep assessment tasks, performance is reported as percentage accuracy relative to average human test takers, enabling direct comparison between MedTSLLMs and expert-level reasoning. This expert-aligned benchmarking provides an additional evaluation dimension beyond purely statistical metrics and reflects the practical competence of LLM-based health assessment systems in professional contexts.

\paragraph{Functional roles of LLMs}
In health assessment support, MedTSLLMs can serve multiple functional roles within the system pipeline, depending on the degree to which the model participates in physiological modeling, numerical inference, and semantic interpretation. Because health assessment requires coordination between structured measurements and contextual reasoning, existing studies commonly position MedTSLLMs as backbones, classifiers, predictors, or text decoders.

When a MedTSLLM is used as a backbone, the model encodes structured features derived from wearable sensors or behavioral summaries into a shared embedding space that is connected to lightweight task-specific heads~\cite{wang2023large}. In this configuration, the MedTSLLM does not directly output health states. Instead, it provides temporally and semantically enriched representations that support downstream stress assessment, sleep monitoring, and vital sign estimation.

MedTSLLMs also function as classifiers in categorical assessment tasks, such as stress grading, sleep stage categorization, and anomaly detection. In these settings, structured physiological measurements, often embedded into carefully designed prompts, are mapped to discrete health states. This role is particularly common in few-shot or instruction-based configurations, where classification is performed with minimal task-specific retraining.

For quantitative measurement tasks, MedTSLLMs often operate as predictors that produce continuous outputs, such as heart rate, blood pressure, and calorie expenditure. In this role, the model performs numerical reasoning over physiologically meaningful features and may use intermediate variables that reflect domain knowledge. This configuration is most effective when the output format is constrained to structured numerical forms, such as fixed-value outputs or bounded ranges, which reduces free-form generation and mitigates numerical instability.

When the objective includes natural-language interpretation, MedTSLLMs act as text decoders. After receiving structured physiological summaries or predicted indicators, the model generates contextual explanations, wellness feedback, or anomaly descriptions. This role is particularly relevant in personalized health assistant settings, where user-facing communication and interpretability complement numerical assessment~\cite{zhang2025sensorlm}.

\paragraph{Discussion}
MedTSLLM-based health assessment systems are commonly designed as feature-grounded reasoning modules that operate on structured physiological descriptors, including heart rate variability indices, pulse transit time, sleep summaries, and wearable-derived statistics. Core design strategies include explicit numerical formatting, instruction tuning, and incorporation of physiological knowledge, enabling stepwise reasoning over clinically meaningful variables for tasks such as stress prediction, sleep monitoring, and cuffless blood pressure estimation. Many systems constrain outputs to predefined formats, including fixed numerical values, categorical stress levels, and template-based assessments, to improve consistency and reduce physiologically implausible generations~\cite{merrill2026transforming}. Personalization and contextual calibration are frequently incorporated through demographic attributes, baseline statistics, and longitudinal summaries to align health state inference with individual variability~\cite{asgari2025framework}. However, performance remains sensitive to input representation and prompt structure, numerical reasoning capacity varies across models, and robustness under sensor noise or distribution shift is rarely evaluated over extended monitoring periods~\cite{weissman2025unregulated}. Furthermore, current evaluation practices remain fragmented and are generally insufficient to capture clinical failure modes~\cite{asgari2025framework,merrill2026transforming}. For example, a MedTSLLM may produce numerically accurate short-term estimates while violating physiological plausibility, such as generating mutually inconsistent heart rate and sleep-state interpretations, or giving recommendations that conflict with known physiological constraints. Similarly, models evaluated only on isolated windows may overlook temporal instability, including abrupt state changes across adjacent time segments, inconsistent longitudinal trends, or failure to maintain individualized baselines during prolonged monitoring~\cite{khasentino2025personal}. These limitations highlight the need for unified evaluation protocols that jointly assess measurement accuracy, physiological consistency, temporal stability, and robustness to sensor noise and clinical failure cases. Overall, MedTSLLMs can provide strong reasoning for health assessment support, but their reliability depends on structured feature design, explicit response constraints, careful physiological grounding, and clinically oriented evaluation standards.

\subsection*{Physiological Signal Synthesis}

\paragraph{Guideline}
MedTSLLM-based physiological signal synthesis is commonly formulated as conditional waveform generation, in which a model generates signals subject to explicit clinical constraints, including diagnostic descriptors, patient attributes, or paired physiological inputs. This conditional formulation is preferable to unconstrained generation, which can yield superficially plausible waveforms that lack clinically meaningful grounding. To be medically useful, synthetic signals must be morphologically realistic and semantically controllable. The resulting rhythm patterns, waveform morphology, and artifact characteristics should reflect the specified conditions rather than collapsing to generic ``realistic'' ECG patterns.

Existing work largely falls into two synthesis paradigms, distinguished by the source of conditioning information and by the manner in which the LLM module integrates heterogeneous modalities. The first paradigm is \textit{text-conditioned ECG generation}, in which clinical narratives and patient factors are encoded as semantic conditions that guide long-range waveform synthesis. Auto-TTE~\cite{chung2023text} maps clinical text reports, optionally augmented with demographic tokens, to discrete ECG token sequences using an autoregressive Transformer. This pipeline compresses raw 12-lead ECG into a discrete codebook via VQ-VAE and reconstructs waveforms with a HiFi-GAN-style decoder. In contrast, DiffuSETS~\cite{lai2025diffusets} adopts a diffusion pipeline in latent space with a VAE bridge. In this design, an LLM serves as a semantic embedding model that converts clinical reports into conditioning vectors, which are then combined with patient attributes, such as sex, age, and heart rate, to guide denoising-based ECG synthesis. DiffuSETS further argues that free-form clinical narratives provide richer conditioning signals than coarse disease labels, enabling more fine-grained control of synthesized ECG characteristics.

The second paradigm formulates synthesis as \textit{physiology-to-physiology translation}, with the objective of generating a clean waveform or a target-modality waveform from a degraded or alternative physiological input. This formulation is closely related to signal restoration. BCG2ECG~\cite{zuo2024adapting} exemplifies this paradigm by adapting a pre-trained LLM to model physiological sequences and improving reconstruction through multi-task learning, including auxiliary heart rate prediction and anomaly detection. This design illustrates that auxiliary supervision can regularize generation and preserve clinically salient morphology. In ECG restoration, ECG-LLM~\cite{liu2024ecg} uses autoregressive modeling to impute missing segments and restore low-quality 12-lead ECG, incorporates timestamp information, and improves efficiency by freezing Transformer layers while adapting lightweight embedding and projection modules.

Evaluation protocols for physiological signal synthesis have progressed toward multi-axis assessment that prioritizes clinical validity beyond waveform realism. This diagnosis-aware perspective typically considers three complementary dimensions: signal fidelity, semantic alignment, and downstream utility. For restoration and generation, fidelity is measured with error metrics such as MAE and MSE for reconstruction accuracy, as well as distributional similarity metrics, including FID and manifold-based precision and recall, to balance diversity and coverage while reducing sensitivity to phase shifts. To verify that generated signals remain consistent with the intended clinical context, studies report semantic alignment measures, including CLIP-style text-ECG alignment and feature-level checks, such as heart rate consistency with report descriptions. Clinical utility is assessed by the effect on downstream diagnostic performance, such as AUROC, and by expert preference studies or cardiologist-led Turing tests. Collectively, these criteria emphasize that synthesis quality is determined not by superficial smoothness, but by the extent to which clinically actionable information is restored or generated.

\paragraph{Functional roles of LLMs}
In physiological signal synthesis, MedTSLLMs are commonly integrated in two functional roles: \textit{generative backbones} and \textit{text decoders}. The distinction depends on whether the language model directly models the physiological sequence or converts clinical language into structured conditions that guide a separate generator.

As a generative backbone, a MedTSLLM serves as a long-context sequence model for synthesis and restoration. Continuous multi-lead signals are mapped into token-like representations through discretization or learned projections, enabling autoregressive completion, imputation, and cross-domain translation over physiological sequences. The primary contribution is the modeling of long-range temporal dependencies and inter-lead consistency, whereas lightweight adapters and output heads map hidden states back to waveform space~\cite{loni2025review}. This configuration is particularly common when synthesis is framed as reconstruction, including missing-segment imputation, artifact removal, and sensor-to-sensor translation.

When synthesis is controlled by clinical narratives, MedTSLLMs operate as text-side modules that inject semantic constraints into waveform generation. This role can be instantiated by a decoder-style Transformer that generates signal tokens conditioned on report text, or by a pre-trained language model that encodes free-form reports into dense embeddings used as conditioning vectors by downstream generators~\cite{zhou2025diagnosis}. In both cases, the functional contribution is the conversion of unstructured medical language into compact and controllable representations, enabling conditioning at the level of report semantics rather than coarse diagnostic labels.

\paragraph{Discussion}
Recent MedTSLLM-based synthesis systems primarily exploit long-context sequence modeling to bridge numerical waveforms and clinical context~\cite{yang2026heartllm}. A recurring design choice is explicit modality alignment through specialized encoders that translate signals into representations compatible with an LLM, including discrete tokenization via VQ-VAE or learned projections that map continuous temporal features into the latent space of the LLM. In addition, multi-task learning is frequently used to regularize generation by coupling reconstruction objectives with auxiliary supervision, such as heart rate prediction and anomaly detection, which encourages representations that preserve clinically salient structure.

Several challenges remain. First, accurate segment-level reconstruction is difficult when the mapping between conditioning inputs and electrical morphology is indirect. In BCG-to-ECG translation~\cite{zuo2024adapting}, for example, reconstruction of the QRS complex is often more reliable because it is strongly associated with mechanical contraction, whereas P and T waves can exhibit larger errors due to weaker and more variable correspondence. Second, non-invasive signals, such as BCG, are susceptible to motion artifacts and sensor noise, and available paired datasets remain limited, particularly for rare cardiac conditions. These constraints hinder the synthesis of diverse waveforms with clinically meaningful variability. Third, validation remains challenging because standard numerical losses, such as MSE, do not fully capture clinical fidelity. As a result, evaluation increasingly relies on complementary criteria, including semantic alignment measures and expert preference studies in which cardiologists judge whether generated signals are realistic and clinically plausible.

\section*{Challenges and Opportunities}

Although MedTSLLMs have demonstrated promising capabilities across diverse MedTS applications, their translation into reliable clinical tools remains challenging. To meet the demands of practical medicine, these models must be developed and evaluated under conditions that reflect the complexity, uncertainty, and safety requirements of real-world healthcare. In this section, we examine the key challenges that currently hinder the adoption of LLMs in MedTS applications and discuss potential opportunities for future research, with the aim of providing insights for researchers and practitioners on how these models can better serve clinicians, patients, and the broader public.

\subsection*{Hallucination}
Hallucination in MedTSLLMs refers to the generation of clinically plausible but unsupported outputs that are inconsistent with physiological signals, patient context, or validated medical knowledge~\cite{rawte2023survey,pal2023med}. Unlike hallucination in purely text-based LLMs, hallucination in MedTSLLMs often stems from the representational gap between continuous MedTS and language-based reasoning. Medical time series are typically converted into plots, textual numerical summaries, discrete tokens, or latent embeddings before they are processed by LLMs. During this conversion, clinically salient information, including subtle waveform morphology, temporal dynamics, cross-channel dependencies, and signal-quality artifacts, may be compressed or discarded. Consequently, MedTSLLMs may produce spurious abnormalities, overconfident diagnostic explanations, inaccurate clinical reports, or QA responses that are fluent but only weakly supported by the original signal evidence. This issue is particularly critical in disease diagnosis, clinical report generation, medical QA, neural signal translation, and health assessment, where hallucinated outputs may mislead clinicians or patients and undermine trust in MedTSLLM-based systems.

\paragraph{Opportunities} Mitigating hallucination in MedTSLLMs requires methods that strengthen physiological grounding across the entire modeling pipeline. First, representation-level grounding can be improved by designing signal encoders, tokenization strategies, and alignment objectives that preserve clinically meaningful temporal, morphological, and cross-channel information before it is passed to the LLM. Second, inference-time control can reduce unsupported generation through structured prompts, constrained output formats, uncertainty estimation, and an explicit distinction between signal observations, diagnostic hypotheses, and final conclusions ~\cite{dhuliawala2024chain}. Third, clinically guided and retrieval-augmented frameworks can incorporate external evidence from guidelines, textbooks, historical reports, or similar physiological cases, thereby enabling the model to generate outputs that are more traceable and evidence-based~\cite{croxford2025evaluating}. However, the retrieval process must itself be carefully validated, because irrelevant or physiologically mismatched cases may introduce additional hallucination risks. Finally, hallucination evaluation should go beyond general text-similarity metrics and incorporate clinically oriented criteria, including signal-output consistency, guideline adherence, expert review, and robustness under noisy, missing, or ambiguous MedTS inputs~\cite{manakul2023selfcheckgpt}.

\subsection*{Lack of High-Quality Domain Data}

A central challenge in developing MedTSLLMs is the limited availability of sufficient, high-quality, and clinically diverse domain data. Unlike general-domain LLMs, which benefit from massive web-scale corpora, MedTSLLMs require paired physiological signals, clinical annotations, textual reports, patient context, and, in some cases, longitudinal follow-up information~\cite{thapa2026multimodal,zhou2025diagnosis}. However, such data are difficult to collect because MedTS are generally sensitive, institution-dependent, device-specific, and costly to annotate by experts. As a result, many existing MedTSLLMs rely on a small number of public datasets or institution-specific cohorts, which limits the ability of these models to learn generalizable representations across diseases, populations, and clinical settings~\cite{he2023domain}. Data quality and distribution can also differ substantially across acquisition settings. 
In-hospital MedTS are typically acquired using clinical-grade equipment under relatively standardized protocols, 
whereas ambulatory and home monitoring often relies on consumer-grade wearables operating under less controlled, 
free-living conditions. Such data are more susceptible to motion artefacts, missingness, and device-dependent variability, 
potentially introducing cross-setting distribution shifts that compromise model generalizability~\cite{jamieson2025guide}.

This data limitation is reflected not only in the overall scarcity of high-quality MedTS data, but also in the uneven coverage of signal modalities and clinical conditions. Among the 62 reviewed studies, most focus on ECG data (34/62), followed by EEG data (19/62), whereas PPG (7/62) and wearable sensor data (8/62) remain much less explored. Therefore, the reported progress of MedTSLLMs is still largely shaped by ECG- and EEG-centered tasks, rather than by a comprehensive evaluation across the broader spectrum of MedTS data. Within each modality, existing datasets also tend to cover only a limited range of clinical scenarios. ECG datasets are mainly concentrated on common cardiac abnormalities, especially arrhythmias and frequently annotated ECG patterns, while manifestations of broader structural or rare diseases remain underrepresented~\cite{poterucha2025detecting}. EEG datasets are often centered on sleep staging, emotion recognition, reading comprehension, or a small set of seizure detection tasks, whereas high-quality real-world data for neurological disorders remain scarce~\cite{sun2024adaptive}. Similarly, PPG and wearable sensor datasets generally emphasize activity recognition or stress detection, with limited coverage of clinically complex disease trajectories~\cite{mahajan2025wearable}. This narrow data foundation may lead MedTSLLMs to perform well on open benchmarks with broad but shallow coverage, while limiting their reliability in practical scenarios that require differential diagnosis, rare disease recognition, longitudinal monitoring, or personalized treatment planning.

\paragraph{Opportunities} 

Addressing this data bottleneck requires the construction of clinically meaningful, multi-domain, and multi-institutional MedTS resources. Future datasets should cover not only common diagnostic labels but also rare diseases, comorbidities, ambiguous cases, disease progression, treatment responses, and patient-specific longitudinal trajectories~\cite{thapa2026multimodal}. For ECG, this requires moving beyond arrhythmia-centered datasets toward broader cardiovascular and systemic disease phenotypes. For EEG, greater efforts are needed to construct real-world datasets for brain disorders, with standardized annotations linked to clinical diagnoses, cognitive assessments, medication records, imaging findings, and follow-up outcomes~\cite{rossi2025sleepyland}. For PPG and wearable sensors, future resources should better reflect continuous and ambulatory monitoring scenarios by incorporating clinically relevant endpoints, temporal context, device information, motion artifacts, and patient-level metadata~\cite{pereira2020photoplethysmography}.

At the same time, data quality is as important as data scale. High-quality MedTSLLM datasets should include standardized acquisition protocols, device information, signal-quality indicators, expert-verified labels, structured reports, and patient context. Multi-center collaboration, federated learning, privacy-preserving data sharing, and carefully governed clinical data commons can help reduce privacy and institutional barriers. In addition, synthetic data and LLM-generated annotations may support data augmentation and instruction tuning, but they must be validated against expert annotations and real physiological evidence~\cite{zhou2025diagnosis}. Ultimately, the development of reliable MedTSLLMs depends on a shift from large but weakly curated datasets toward clinically grounded datasets that reflect the diversity, complexity, and uncertainty of real-world medicine.

\subsection*{Narrow Coverage of Downstream Clinical Tasks}

As illustrated in Figure~4, existing applications of MedTSLLMs can be broadly grouped into four major categories and 18 task subtypes: \textit{disease classification}, including arrhythmia diagnosis, abnormality diagnosis, specific cardiac disease diagnosis (e.g., LVEF, HCM, and STEMI), sleep disorder detection, and seizure detection; \textit{signal understanding}, including ECG interpretation, report generation, question answering, EEG-to-text generation, sentiment classification, and reading comprehension; \textit{health monitoring}, including stress monitoring, sleep monitoring, vital sign measurement (e.g., HR, CBP, and IBI monitoring), and fitness estimation; and \textit{signal synthesis}, including signal generation and signal restoration. Other less common applications are grouped into an additional \textit{others} category.

Our analysis shows that current MedTSLLM research is still concentrated on a relatively narrow set of downstream tasks. More than 28\% of studies focus on disease classification, primarily targeting heart-, brain-, or sleep-related conditions, while over 37\% evaluate signal understanding tasks, such as physiological signal interpretation, question answering, and report generation. By contrast, fewer than 18\% of studies investigate health monitoring applications, and less than 7\% address signal synthesis. This distribution suggests that most existing MedTSLLMs are still evaluated as diagnostic or signal-to-text systems, whereas their broader potential for longitudinal monitoring, personalized health assessment, signal restoration, and closed-loop clinical support remains insufficiently explored.

A similarly imbalanced pattern can be observed within each main task category. In disease classification, more than 68\% of studies focus on arrhythmia detection or broad ECG abnormality diagnosis, whereas only a small proportion, approximately 9\%, examine specific cardiac conditions, such as LVEF, HCM, or PH. Beyond cardiovascular diseases, sleep disorders, including sleep apnea, and neurological disorders, such as epilepsy, Alzheimer's disease, and Parkinson's disease, remain comparatively underrepresented, with only 5 out of 22 disease classification studies addressing these conditions. Likewise, signal understanding studies are dominated by question answering and report generation, while tasks involving EEG or PPG interpretation, EEG-to-text translation, and other modality-specific forms of physiological reasoning have received limited attention.

This limited task coverage constrains our ability to assess the generalizability of MedTSLLMs beyond a relatively small set of well-studied ECG-centric applications. Several key clinical tasks, such as patient risk stratification, prognosis, and outcome prediction, also remain comparatively underexplored, leaving model performance across physiological modalities, temporal scales, and clinical workflows insufficiently characterized~\cite{lan2025gem,yang2026heartllm}.

\begin{figure}[ht]
\centering
\includegraphics[width=0.9\linewidth]{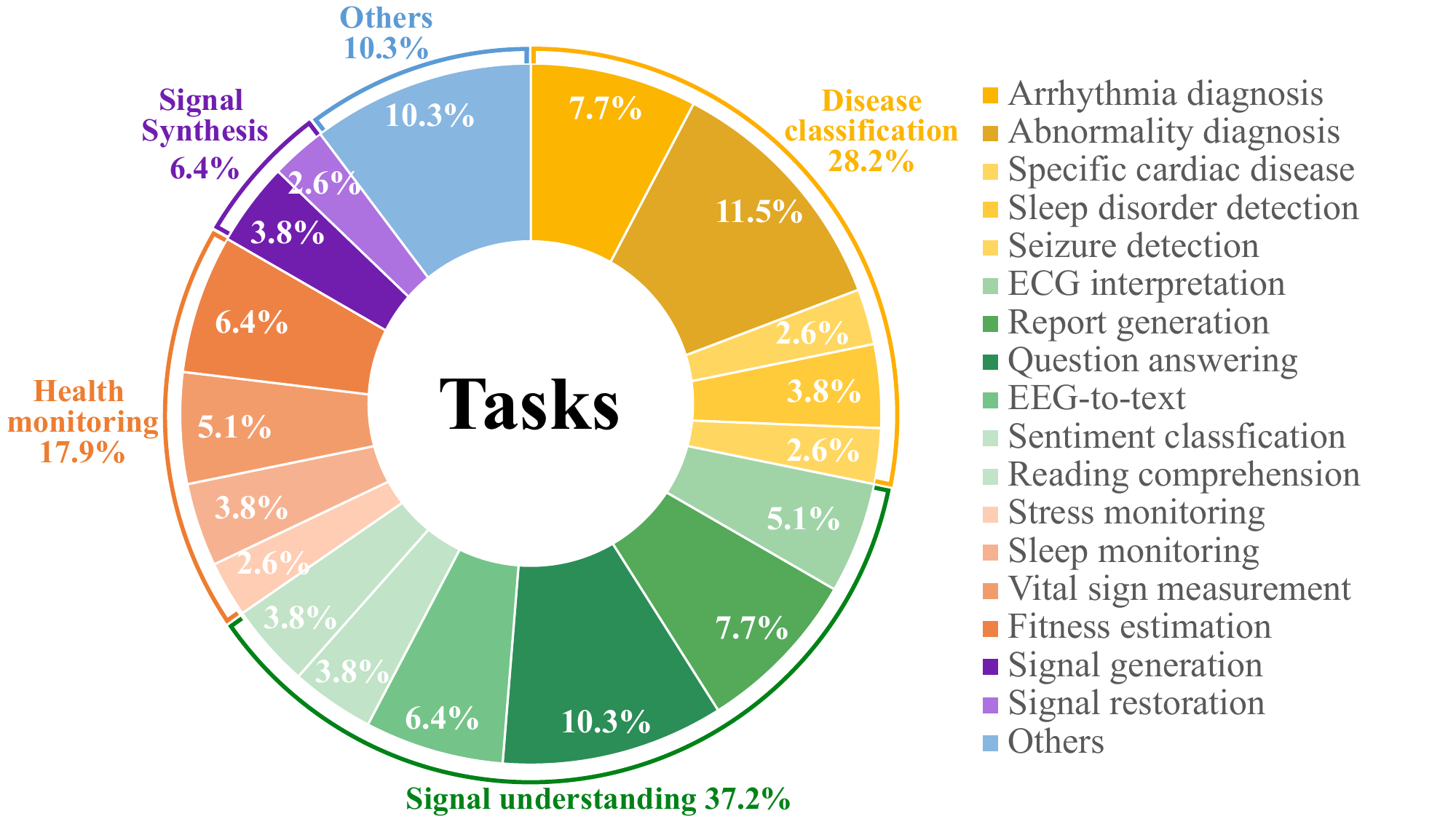}
\caption{\textbf{Categories of MedTSLLM applications in the reviewed papers.}}
\label{fig:categories}
\end{figure}

\paragraph{Opportunities}

Future research should expand the downstream evaluation of MedTSLLMs beyond the current focus on common diagnosis-oriented and signal-to-text tasks toward a broader range of clinically meaningful applications, including both underexplored disease areas~\cite{jiang2024neurolm,xie2025physllm,li2026anyecg} and less common task settings~\cite{kim2024health}. In disease diagnosis, greater attention should be paid to specific cardiac conditions, rare or less-studied cardiovascular diseases, sleep disorders, and neurological disorders. Beyond diagnosis, promising directions include longitudinal health monitoring, disease progression assessment, patient risk stratification and outcome prediction, signal quality assessment, signal restoration, and personalized patient support~\cite{zhang2025sensorlm,xie2025physllm}. Evaluating MedTSLLMs in such settings would provide a more realistic understanding of whether these models can capture long-term physiological patterns, adapt to heterogeneous patient conditions, and support decision-making beyond single-episode classification~\cite{li2026anyecg}. Beyond improving evaluation coverage, broader task design would help clarify the distinctive value of LLM-based frameworks, including their ability to integrate physiological signals with patient context, generate clinically grounded explanations, reason over multimodal information, and interact with users through flexible prompts.

\subsection*{Lack of Standard Evaluation Benchmarks and Metrics}

A major challenge in MedTSLLM research is the absence of standardized benchmarks and evaluation metrics~\cite{wu2025towards,agrawal2025evaluation}. Existing studies typically rely on task-specific metrics inherited from conventional machine learning or natural language generation, such as accuracy, F1-score, AUROC, Hamming loss, BLEU, ROUGE, METEOR, BERTScore, MAE, or RMSE. Although these metrics are useful for individual tasks, they provide only partial views of model performance and cannot fully determine whether a MedTSLLM is clinically reliable. For example, a model may achieve high diagnostic accuracy while generating explanations that are only weakly grounded in waveform evidence, or it may obtain favorable text-similarity scores while omitting clinically important findings~\cite{zhou2025automating}. This limitation is especially problematic because MedTSLLMs vary substantially in signal modality, input representation, prompt template, retrieval strategy, patient context, and functional role, which makes direct comparison across studies difficult.

More importantly, many current benchmarks do not systematically assess key clinical requirements, including physiological grounding, evidence-based reasoning, guideline adherence, robustness to noisy or missing signals, uncertainty awareness, and fairness across patient subgroups. Only a few recent studies have attempted more comprehensive evaluation. For example, GEM~\cite{lan2025gem} introduces the grounded ECG understanding task, which evaluates not only diagnostic accuracy but also analysis completeness, relevance, lead assessment coverage and accuracy, ECG feature grounding, evidence-based reasoning, and clinical diagnostic fidelity. Such multidimensional evaluation better reflects whether a MedTSLLM can produce clinically meaningful and trustworthy interpretations~\cite{khandekar2024medcalc}.

\paragraph{Opportunities}

Future evaluation of MedTSLLMs should move toward standardized and multidimensional benchmark systems~\cite{wu2024medjourney,karargyris2023federated}. A robust framework should jointly assess predictive correctness, linguistic quality, physiological faithfulness, clinical reasoning, robustness, and deployment readiness. In particular, benchmarks should evaluate whether model outputs are explicitly grounded in signal-derived evidence and whether diagnostic conclusions follow clinically accepted reasoning processes. For RAG-based or agentic MedTSLLMs, evaluation should further examine retrieval relevance, evidence consistency, tool-use reliability, and error propagation across reasoning steps.

Another important direction is to establish shared benchmark datasets and reporting protocols across ECG, EEG, PPG, and wearable sensor data. These protocols should include standardized data splits, preprocessing details, prompt templates, evaluation rubrics, and baseline models to improve reproducibility and comparability. Human expert evaluation remains essential for open-ended reports and high-risk decision-support outputs~\cite{croxford2025evaluating,vasey2022reporting}. Although LLM-based evaluators can support scalable assessment, their judgments should be calibrated against clinician annotations. Ultimately, evaluation of MedTSLLMs should measure not only whether a model provides the correct answer, but also whether the answer is derived from clinically trustworthy evidence and reasoning~\cite{hager2024evaluation}.

\subsection*{Limited Mechanism Reasoning}

A key limitation of current MedTSLLMs is that diagnosis, text generation, and health monitoring are still largely formulated as signal-to-label or signal-to-text mapping problems, rather than as mechanistically informed reasoning tasks~\cite{wu2025towards,chan2024leveraging,yang2025ecg}. Although these models can associate MedTS patterns with predictive labels, answers, or reports, they often provide limited insight into why such patterns arise from underlying physiological processes. In clinical practice, physicians are not only interested in whether a signal segment suggests a disease, but also in the underlying mechanism: how temporal changes, waveform morphology, patient context, and disease mechanisms jointly support a clinical interpretation. This gap reflects the broader distinction between correlation-based prediction and physiologically grounded reasoning.

Existing MedTSLLMs are not entirely devoid of explanatory ability. In tasks such as medical QA and clinical report generation, models often generate rationales together with their predictions~\cite{pham2025q,tang2025electrocardiogram}. For example, in ECG QA, an LLM may predict an arrhythmia-related label while referring to supporting patterns such as rhythm irregularities or abnormal R-R intervals, sometimes supplemented by retrieved clinical knowledge or similar cases. These designs improve the apparent interpretability of MedTSLLMs and make their outputs more consistent with clinical language. However, such explanations do not necessarily indicate a mechanistic understanding of disease processes. They are generally conditioned on predefined medical knowledge, retrieved cases, prompt templates, or statistical associations learned from biomedical text, rather than explicit modeling of how a physiological abnormality emerges, evolves, and leads to a clinical outcome. This limitation becomes more evident in complex scenarios involving comorbidities, medication effects, or rare diseases, where similar signal patterns can reflect different underlying mechanisms. In such cases, retrieved examples or textbook-like rationales may fail to capture patient-specific context, making it difficult for MedTSLLMs to determine whether an observed pattern reflects disease progression, treatment response, artifact, or an unrelated condition.

This challenge is closely related to temporal reasoning in MedTS. MedTS contains dynamic patterns across multiple temporal scales, from local waveform events and short-term episodes to long-term disease progression and continuous monitoring trajectories~\cite{wang2023contrast,yang2023biot}. However, existing MedTSLLMs have mainly focused on signal-to-text conversion and LLM adaptation, while less attention has been paid to long temporal context, streaming signals, temporal abstraction, and event hierarchies. As a result, they may struggle to model how local abnormalities accumulate into higher-level clinical states, how signal changes evolve before disease onset, or how transient patterns should be interpreted within a longer monitoring window. These issues are particularly important for longitudinal health assessment, early warning, prognosis, and personalized monitoring, where clinically meaningful reasoning depends not only on recognizing isolated signal patterns but also on understanding their temporal order, duration, recurrence, and interaction with patient context~\cite{merrill2026transforming,fang2024physiollm}.

\paragraph{Opportunities}
Future MedTSLLMs should move beyond correlation-based prediction and post-hoc textual rationalization toward temporally grounded and mechanistically informed reasoning. Promising directions include developing multi-scale temporal representations, modeling event-level hierarchies, incorporating physiological priors and clinical knowledge, and designing evaluation protocols that assess whether generated explanations are supported by both signal evidence and medical mechanisms~\cite{li2026mira,langer2025opentslm}. For continuous monitoring, models should also be able to process streaming inputs, update interpretations over time, and distinguish persistent trends from transient artifacts~\cite{li2026anyecg,merrill2026transforming}. These advances would help MedTSLLMs provide not only accurate predictions, but also clinically meaningful explanations that better align with real-world medical reasoning.

\subsection*{High Computational and Resource Demands}

The deployment of MedTSLLMs is constrained not only by model performance, but also by computational, memory, and energy requirements. These constraints are particularly relevant to wearable and continuous-monitoring settings, where signals such as ECG and PPG are collected continuously and may need to be processed with low latency over long periods~\cite{chen2024sparse}. In practice, MedTSLLMs generally rely on multi-stage pipelines that combine signal encoders, alignment modules, LLM backbones, and, in some cases, retrieval or agent-based components~\cite{yao2025efficient}. Such pipelines remain difficult to execute directly on wearable or other resource-constrained edge devices. Offloading inference to the cloud can reduce local hardware requirements, but introduces dependence on network connectivity, additional latency, and recurring computational costs. These issues become more pronounced in continuous monitoring, where repeated LLM inference may lead to substantial cumulative computation and energy use, raising concerns about both cost-effectiveness and environmental sustainability~\cite{leroux2025analog}.

\paragraph{Opportunities}
Improving deployment efficiency will require MedTSLLMs to be designed with resource constraints in mind rather than optimized solely for predictive or generative performance. Smaller language backbones, quantization, model compression, and knowledge distillation can reduce the computational and memory demands of inference, while parameter-efficient adaptation can lower the cost of task- or domain-specific model updates~\cite{woo2025synthetic}. Edge-cloud partitioning may further allow computationally intensive reasoning to be performed remotely while retaining time-sensitive signal processing on local devices~\cite{yao2025efficient}. For continuous monitoring, event-triggered or hierarchical designs offer another practical direction: lightweight models can process routine signal streams locally, while larger LLMs are invoked only when abnormal events, uncertain cases, or more complex reasoning require additional analysis~\cite{kasper2025wearable}. Accordingly, future studies should report deployment-related measures, such as inference latency, memory usage, energy consumption, and computational cost, alongside conventional performance metrics. Such evaluation is necessary to determine whether MedTSLLMs can operate reliably and efficiently in real-world wearable and clinical environments.

\subsection*{Regulatory Challenges}
MedTSLLMs introduce substantial regulatory challenges because they integrate physiological signal processing, language-based reasoning, and clinical decision support within highly heterogeneous pipelines. Unlike conventional medical AI systems that are designed for narrowly defined tasks, a single MedTSLLM framework may be adapted for disease diagnosis, report generation, medical QA, health assessment, or signal synthesis, with each use case involving distinct clinical risks~\cite{rajpurkar2022ai}. Moreover, model behavior is determined not only by the LLM but also by signal encoders, tokenization strategies, prompt templates, retrieval databases, and patient metadata. This multi-component structure complicates regulatory assessment because errors may arise from upstream signal compression, unreliable retrieval, prompt sensitivity, or hallucinated language outputs~\cite{mesko2023imperative,derraz2024new}. In addition, frequent updates through prompt modification, retrieval corpus expansion, or parameter-efficient tuning challenge traditional one-time approval mechanisms and create a need for continuous monitoring after deployment.

\paragraph{Opportunities} Future regulation of MedTSLLMs should adopt risk-adaptive and lifecycle-oriented frameworks. Low-risk applications, such as clinician-supervised report drafting, may require evidence standards that differ from those required for high-risk systems that directly influence diagnosis or treatment decisions. Regulatory assessment should therefore clearly specify the intended use, target user group, autonomy level, and potential clinical harm of each system~\cite{hacker2023regulating,derraz2024new}. A standardized evidence package could document signal acquisition procedures, preprocessing steps, tokenization strategies, prompt design, retrieval sources, model versions, and human oversight mechanisms. For adaptive systems, predefined change-control plans and post-deployment surveillance can support iterative improvement while maintaining safety. Prospective clinical validation, subgroup robustness analysis, and evaluation under noisy or incomplete signal conditions are also essential for ensuring reliable performance in real-world clinical settings~\cite{mesko2023imperative}.

\subsection*{Ethics and Safety}
Ethical and safety concerns in MedTSLLMs arise from the sensitivity of physiological signals and the high-stakes nature of medical interpretation. ECG, EEG, PPG, and wearable data can reveal cardiac, neurological, behavioral, and mental health information, especially when they are combined with demographic profiles or longitudinal records. This creates privacy risks during training, retrieval, and deployment. Safety risks are equally important. MedTSLLMs may generate fluent and clinically plausible explanations that are not fully grounded in physiological evidence~\cite{he2025survey}. Such hallucinations or evidence-output mismatches can lead to misleading diagnostic reports, incorrect QA responses, or unsafe health recommendations. Bias is another key concern, because models trained on limited datasets may perform unevenly across institutions, devices, patient groups, and rare disease populations~\cite{tian2024opportunities}.

\paragraph{Opportunities} Improving the ethical and safety profile of MedTSLLMs requires privacy-preserving data governance, evidence-grounded generation, and human-centered deployment. Sensitive physiological data should be protected through data minimization, de-identification, access control, and, where possible, federated or privacy-preserving learning~\cite{li2023multi}. To reduce hallucination, MedTSLLMs should produce structured outputs that clearly distinguish signal observations, diagnostic hypotheses, uncertainty estimates, and recommended next steps. Clinically guided prompting, constrained response formats, curated retrieval sources, and explicit grounding in signal-derived evidence can improve traceability~\cite{wei2023jailbroken}. Future benchmarks should also evaluate subgroup fairness, robustness to sensor noise, missing data, adversarial prompts, and retrieval failures. In clinical settings, MedTSLLMs should function as assistive tools that support expert decision-making rather than replace it.

\subsection*{The Implementation Gap in Real-World Clinical Settings}
Although MedTSLLMs have shown promising capabilities in disease diagnosis, clinical report generation, medical QA, health assessment, neural signal translation, and physiological signal synthesis, their current development remains largely exploratory. Most existing studies evaluate models on retrospective datasets or benchmark tasks, where inputs are preprocessed, labels are predefined, and clinical workflows are simplified~\cite{wornow2023shaky,agrawal2025evaluation}. This evaluation setting differs substantially from real-world healthcare environments, where MedTS data are generally noisy, incomplete, device-dependent, and embedded within complex patient trajectories~\cite{goetz2024generalization}. Moreover, MedTSLLMs are often tested as isolated models rather than as components of integrated clinical systems. Therefore, their practical value depends not only on model accuracy but also on latency, interoperability with electronic health records, robustness to sensor artifacts, clinician usability, regulatory compliance, and the ability to provide transparent evidence for each output~\cite{lavin2022technology}. The intended role of MedTSLLMs is also likely to differ across deployment settings. In hospital environments, where professional oversight and richer clinical context are available, 
MedTSLLMs can be positioned as assistive tools for diagnostic interpretation, clinical report generation, and decision support~\cite{bedi2026holistic}. In ambulatory and home settings, they may be better suited to longitudinal monitoring, risk screening, and escalation of suspicious findings for further clinical assessment, rather than autonomous diagnosis~\cite{pedroso2025leveraging,jamieson2025guide}.

At the current stage, MedTSLLMs are more likely to serve medicine as assistive systems rather than autonomous systems. Their near-term clinical utility may lie in drafting ECG or EEG reports, summarizing long-term wearable recordings, retrieving relevant clinical knowledge, highlighting suspicious patterns, and supporting clinicians in difficult or rare cases~\cite{al2023machine}. However, direct use for independent diagnosis or treatment recommendation requires stronger evidence. MedTSLLMs can move toward real clinical service only when they pass prospective multi-center validation, demonstrate stable performance across devices and patient subgroups, operate safely under noisy or missing data, and are embedded in human-in-the-loop workflows where clinicians can verify, correct, and override model outputs.

\paragraph{Opportunities}
Bridging this implementation gap requires a staged pathway from benchmark performance to clinically accountable deployment~\cite{you2025clinical}. First, MedTSLLMs should be evaluated in silent or shadow-mode clinical studies, where model outputs are generated in parallel with routine care but do not influence clinical decisions. Such studies can reveal real-world failure modes, workflow friction, and distribution shifts that are not visible in offline benchmarks~\cite{schrouff2022diagnosing}. Second, deployment should begin with low-risk and high-workload tasks, such as report drafting, signal summarization, documentation assistance, and retrospective case review. These tasks allow MedTSLLMs to reduce clinician burden while maintaining expert supervision.

For higher-risk applications, such as triage support, rare disease detection, continuous monitoring, and personalized health assessment, MedTSLLMs should be designed as evidence-grounded copilots. Rather than producing final decisions, they should present signal observations, candidate interpretations, uncertainty estimates, retrieved evidence, and recommended follow-up checks. In this form, MedTSLLMs can support clinicians by improving efficiency, expanding access to specialized knowledge, and identifying subtle patterns that may otherwise be overlooked~\cite{al2023machine,johnson2025artificial}. Ultimately, real clinical service will require not only stronger models but also standardized evaluation protocols, prospective validation, regulatory approval, clinician-centered interfaces, and continuous post-deployment monitoring.

\section*{Conclusions}

Recent advancements in LLMs are opening new possibilities for MedTS analysis by connecting physiological signals with language-based reasoning, multimodal understanding, and clinical decision support. In this review, we provide a practical and structured overview of MedTSLLMs tailored to various medical scenarios. We introduce the basic concepts and modeling workflow of MedTSLLMs, summarize representative tokenization and prompting strategies, and discuss how different prompt templates can support signal interpretation and clinical reasoning. Additionally, we review major clinical applications, including disease diagnosis, clinical report generation, medical question answering, neuro signal translation, health assessment, and physiological signal synthesis, while addressing current challenges and outlining future opportunities. As a narrative rather than systematic review, the literature selection was guided by the authors' domain knowledge and the scope of this review. Therefore, despite our efforts to provide broad coverage, some relevant or newly emerging studies may not have been included. To address the rapidly evolving nature of this field, we will maintain an accompanying GitHub repository to facilitate continuous updates to the collection of MedTSLLM studies. We believe that this review can help clinicians and researchers better understand the current landscape of MedTSLLMs and inspire further innovation in applying LLMs to MedTS analysis.

\section*{Author Contributions}
Yu Han conceived the study, conducted the literature review, organized and analyzed the included studies, developed the taxonomy, prepared the figures and tables, and drafted the manuscript. Cigdem Beyan contributed to the study design, supervised the review process, and revised the manuscript. Xiang Zhang, Xiaofeng Liu, Nan Liu, and Jimeng Sun contributed domain expertise, interpretation of the literature, and critical revision of the manuscript. Shenda Hong and Cheng Ding contributed to the study conception, methodological and clinical guidance, supervision, and critical revision of the manuscript. Vittorio Murino supervised and coordinated the study, contributed to its conceptual development, and revised the manuscript. All authors reviewed and approved the final manuscript.

\section*{Acknowledgment}
We sincerely thank Dr. Hongwei Ji for his insightful advice and constructive feedback, which contributed to the development and refinement of this work.

\section*{Funding}
This work was supported by the National Natural Science Foundation of China (Grant No. 32541017), the Opening Foundation of the State Key Laboratory of Transvascular Implantation Devices (Grant No. SKLTID2025102), and the Fundamental Research Funds for the Central Universities (Grant No. N22025028).

\section*{Conflict of Interest}
The authors declare that they have no conflicts of interest relevant to this work.

\section*{Declaration}
Generative AI tools were used only for language editing, grammatical correction, and improving clarity. All AI-assisted text was reviewed and revised by the authors, and all figures were created or manually redrawn by the authors.
\bibliography{main}

@article{choi2023ecgbert,
  title={Ecgbert: Understanding hidden language of ecgs with self-supervised representation learning},
  author={Choi, Seokmin and Mousavi, Sajad and Si, Phillip and Yhdego, Haben G and Khadem, Fatemeh and Afghah, Fatemeh},
  journal={arXiv preprint arXiv:2306.06340},
  year={2023}
}

@article{chen2025gpt,
  title={GPT-PPG: a GPT-based foundation model for photoplethysmography signals},
  author={Chen, Zhaoliang and Ding, Cheng and Kataria, Saurabh and Yan, Runze and Wang, Minxiao and Lee, Randall and Hu, Xiao},
  journal={Physiological Measurement},
  volume={46},
  number={5},
  pages={055004},
  year={2025},
  publisher={IOP Publishing}
}

@article{lan2025gem,
  title={Gem: Empowering mllm for grounded ecg understanding with time series and images},
  author={Lan, Xiang and Wu, Feng and He, Kai and Zhao, Qinghao and Hong, Shenda and Feng, Mengling},
  journal={arXiv preprint arXiv:2503.06073},
  year={2025}
}

@inproceedings{li2024frozen,
  title={Frozen language model helps ecg zero-shot learning},
  author={Li, Jun and Liu, Che and Cheng, Sibo and Arcucci, Rossella and Hong, Shenda},
  booktitle={Medical Imaging with Deep Learning},
  pages={402--415},
  year={2024},
  organization={PMLR}
}

@inproceedings{liu2024etp,
  title={Etp: Learning transferable ecg representations via ecg-text pre-training},
  author={Liu, Che and Wan, Zhongwei and Cheng, Sibo and Zhang, Mi and Arcucci, Rossella},
  booktitle={ICASSP 2024-2024 IEEE International Conference on Acoustics, Speech and Signal Processing (ICASSP)},
  pages={8230--8234},
  year={2024},
  organization={IEEE}
}

@article{yu2024ecg,
  title={Ecg semantic integrator (esi): A foundation ecg model pretrained with llm-enhanced cardiological text},
  author={Yu, Han and Guo, Peikun and Sano, Akane},
  journal={arXiv preprint arXiv:2405.19366},
  year={2024}
}

@inproceedings{devlin2019bert,
  title={Bert: Pre-training of deep bidirectional transformers for language understanding},
  author={Devlin, Jacob and Chang, Ming-Wei and Lee, Kenton and Toutanova, Kristina},
  booktitle={Proceedings of the 2019 conference of the North American chapter of the association for computational linguistics: human language technologies, volume 1 (long and short papers)},
  pages={4171--4186},
  year={2019}
}

@article{alsentzer2019publicly,
  title={Publicly available clinical BERT embeddings},
  author={Alsentzer, Emily and Murphy, John R and Boag, Willie and Weng, Wei-Hung and Jin, Di and Naumann, Tristan and McDermott, Matthew},
  journal={arXiv preprint arXiv:1904.03323},
  year={2019}
}

@article{achiam2023gpt,
  title={Gpt-4 technical report},
  author={Achiam, Josh and Adler, Steven and Agarwal, Sandhini and Ahmad, Lama and Akkaya, Ilge and Aleman, Florencia Leoni and Almeida, Diogo and Altenschmidt, Janko and Altman, Sam and Anadkat, Shyamal and others},
  journal={arXiv preprint arXiv:2303.08774},
  year={2023}
}

@article{kim2024eeg,
  title={EEG-GPT: exploring capabilities of large language models for EEG classification and interpretation},
  author={Kim, Jonathan W and Alaa, Ahmed and Bernardo, Danilo},
  journal={arXiv preprint arXiv:2401.18006},
  year={2024}
}

@inproceedings{yu2023zero,
  title={Zero-shot ECG diagnosis with large language models and retrieval-augmented generation},
  author={Yu, Han and Guo, Peikun and Sano, Akane},
  booktitle={Machine learning for health (ML4H)},
  pages={650--663},
  year={2023},
  organization={PMLR}
}

@article{seki2025assessing,
  title={Assessing the performance of zero-shot visual question answering in multimodal large language models for 12-lead ECG image interpretation},
  author={Seki, Tomohisa and Kawazoe, Yoshimasa and Ito, Hiromasa and Akagi, Yu and Takiguchi, Toru and Ohe, Kazuhiko},
  journal={Frontiers in cardiovascular medicine},
  volume={12},
  pages={1458289},
  year={2025},
  publisher={Frontiers Media SA}
}

@inproceedings{tang2025electrocardiogram,
  title={Electrocardiogram Report Generation and Question Answering via Retrieval-Augmented Self-Supervised Modeling},
  author={Tang, Jialu and Xia, Tong and Lu, Yuan and Mascolo, Cecilia and Saeed, Aaqib},
  booktitle={ICASSP 2025-2025 IEEE International Conference on Acoustics, Speech and Signal Processing (ICASSP)},
  pages={1--5},
  year={2025},
  organization={IEEE}
}

@article{kim2024health,
  title={Health-llm: Large language models for health prediction via wearable sensor data},
  author={Kim, Yubin and Xu, Xuhai and McDuff, Daniel and Breazeal, Cynthia and Park, Hae Won},
  journal={arXiv preprint arXiv:2401.06866},
  year={2024}
}

@article{feli2025llm,
  title={An LLM-Powered Agent for Physiological Data Analysis: A Case Study on PPG-based Heart Rate Estimation},
  author={Feli, Mohammad and Azimi, Iman and Liljeberg, Pasi and Rahmani, Amir M},
  journal={arXiv preprint arXiv:2502.12836},
  year={2025}
}

@article{abbasian2023conversational,
  title={Conversational health agents: A personalized llm-powered agent framework},
  author={Abbasian, Mahyar and Azimi, Iman and Rahmani, Amir M and Jain, Ramesh},
  journal={arXiv preprint arXiv:2310.02374},
  year={2023}
}

@article{jeong2024prediction,
  title={The Prediction of Stress in Radiation Therapy: Integrating Artificial Intelligence with Biological Signals},
  author={Jeong, Sangwoon and Pyo, Hongryull and Park, Won and Han, Youngyih},
  journal={Cancers},
  volume={16},
  number={11},
  pages={1964},
  year={2024},
  publisher={MDPI}
}

@inproceedings{mishra2025thought2text,
  title={Thought2Text: text generation from EEG signal using large language models (LLMs)},
  author={Mishra, Abhijit and Shukla, Shreya and Torres, Jose and Gwizdka, Jacek and Roychowdhury, Shounak},
  booktitle={Findings of the Association for Computational Linguistics: NAACL 2025},
  pages={3747--3759},
  year={2025}
}

@article{bohi2024large,
  title={Large language models for wearable data analysis and interpretation},
  author={B{\"o}hi, Simon and Gashi, Shkurta},
  journal={The Second Tiny Papers Track at ICLR 2024},
  year={2024},
  publisher={OpenReview}
}

@article{cosentino2024towards,
  title={Towards a personal health large language model},
  author={Cosentino, Justin and Belyaeva, Anastasiya and Liu, Xin and Furlotte, Nicholas A and Yang, Zhun and Lee, Chace and Schenck, Erik and Patel, Yojan and Cui, Jian and Schneider, Logan Douglas and others},
  journal={arXiv preprint arXiv:2406.06474},
  year={2024}
}

@article{chen2025eeg,
  title={EEG emotion copilot: Optimizing lightweight LLMS for emotional EEG interpretation with assisted medical record generation},
  author={Chen, Hongyu and Zeng, Weiming and Chen, Chengcheng and Cai, Luhui and Wang, Fei and Shi, Yuhu and Wang, Lei and Zhang, Wei and Li, Yueyang and Yan, Hongjie and others},
  journal={Neural Networks},
  pages={107848},
  year={2025},
  publisher={Elsevier}
}

@article{liu2023large,
  title={Large language models are few-shot health learners},
  author={Liu, Xin and McDuff, Daniel and Kovacs, Geza and Galatzer-Levy, Isaac and Sunshine, Jacob and Zhan, Jiening and Poh, Ming-Zher and Liao, Shun and Di Achille, Paolo and Patel, Shwetak},
  journal={arXiv preprint arXiv:2305.15525},
  year={2023}
}

@inproceedings{cai2024jolt,
  title={JoLT: jointly learned representations of language and time-series for clinical time-series interpretation (student abstract)},
  author={Cai, Yifu and Srinivasan, Arvind and Goswami, Mononito and Choudhry, Arjun and Dubrawski, Artur},
  booktitle={Proceedings of the AAAI Conference on Artificial Intelligence},
  volume={38},
  number={21},
  pages={23447--23448},
  year={2024}
}

@article{zhou2024belt2,
  title={Belt-2: Bootstrapping eeg-to-language representation alignment for multi-task brain decoding},
  author={Zhou, Jinzhao and Duan, Yiqun and Chang, Fred and Do, Thomas and Wang, Yu-Kai and Lin, Chin-Teng},
  journal={arXiv preprint arXiv:2409.00121},
  year={2024}
}

@article{liu2025llms,
  title={LLMs Help Alleviate the Cross-Subject Variability in Brain Signal and Language Alignment},
  author={Liu, Yifei and Ye, Hengwei and Li, Shuhang},
  journal={arXiv preprint arXiv:2501.02621},
  year={2025}
}

@article{tang2023alpha,
  title={Alpha: Anomalous physiological health assessment using large language models},
  author={Tang, Jiankai and Wang, Kegang and Hu, Hongming and Zhang, Xiyuxing and Wang, Peiyu and Liu, Xin and Wang, Yuntao},
  journal={arXiv preprint arXiv:2311.12524},
  year={2023}
}

@inproceedings{wan2025meit,
  title={MEIT: Multimodal Electrocardiogram Instruction Tuning on Large Language Models for Report Generation},
  author={Wan, Zhongwei and Liu, Che and Wang, Xin and Tao, Chaofan and Shen, Hui and Xiong, Jing and Arcucci, Rossella and Yao, Huaxiu and Zhang, Mi},
  booktitle={Findings of the Association for Computational Linguistics: ACL 2025},
  pages={14510--14527},
  year={2025}
}

@article{yang2025ecg,
  title={ECG-LM: Understanding Electrocardiogram with a Large Language Model},
  author={Yang, Kai and Hong, Massimo and Zhang, Jiahuan and Luo, Yizhen and Zhao, Suyuan and Zhang, Ou and Yu, Xiaomao and Zhou, Jiawen and Yang, Liuqing and Zhang, Ping and others},
  journal={Health Data Science},
  volume={5},
  pages={0221},
  year={2025},
  publisher={AAAS}
}

@article{liu2023biosignal,
  title={BioSignal Copilot: Leveraging the power of LLMs in drafting reports for biomedical signals},
  author={Liu, Chunyu and Ma, Yongpei and Kothur, Kavitha and Nikpour, Armin and Kavehei, Omid},
  journal={medRxiv},
  pages={2023--06},
  year={2023},
  publisher={Cold Spring Harbor Laboratory Press}
}

@article{oh2023ecg,
  title={Ecg-qa: A comprehensive question answering dataset combined with electrocardiogram},
  author={Oh, Jungwoo and Lee, Gyubok and Bae, Seongsu and Kwon, Joon-myoung and Choi, Edward},
  journal={Advances in Neural Information Processing Systems},
  volume={36},
  pages={66277--66288},
  year={2023}
}

@article{liu2024teach,
  title={Teach multimodal llms to comprehend electrocardiographic images},
  author={Liu, Ruoqi and Bai, Yuelin and Yue, Xiang and Zhang, Ping},
  journal={arXiv preprint arXiv:2410.19008},
  year={2024}
}

@article{liu2024zero,
  title={Zero-shot ecg classification with multimodal learning and test-time clinical knowledge enhancement},
  author={Liu, Che and Wan, Zhongwei and Ouyang, Cheng and Shah, Anand and Bai, Wenjia and Arcucci, Rossella},
  journal={arXiv preprint arXiv:2403.06659},
  year={2024}
}

@article{zhang2023integrating,
  title={Integrating llm, eeg, and eye-tracking biomarker analysis for word-level neural state classification in semantic inference reading comprehension},
  author={Zhang, Yuhong and Li, Qin and Nahata, Sujal and Jamal, Tasnia and Cheng, Shih-kuen and Cauwenberghs, Gert and Jung, Tzyy-Ping},
  journal={arXiv preprint arXiv:2309.15714},
  year={2023}
}

@inproceedings{cai2023jolt,
  title={Jolt: Jointly learned representations of language and time-series},
  author={Cai, Yifu and Goswami, Mononito and Choudhry, Arjun and Srinivasan, Arvind and Dubrawski, Artur},
  booktitle={Deep Generative Models for Health Workshop NeurIPS 2023},
  year={2023}
}

@article{vaid2023foundational,
  title={A foundational vision transformer improves diagnostic performance for electrocardiograms},
  author={Vaid, Akhil and Jiang, Joy and Sawant, Ashwin and Lerakis, Stamatios and Argulian, Edgar and Ahuja, Yuri and Lampert, Joshua and Charney, Alexander and Greenspan, Hayit and Narula, Jagat and others},
  journal={NPJ digital medicine},
  volume={6},
  number={1},
  pages={108},
  year={2023},
  publisher={Nature Publishing Group UK London}
}

@article{lalam2023ecg,
  title={Ecg representation learning with multi-modal ehr data},
  author={Lalam, Sravan Kumar and Kunderu, Hari Krishna and Ghosh, Shayan and Kumar, Harish and Awasthi, Samir and Prasad, Ashim and Lopez-Jimenez, Francisco and Attia, Zachi I and Asirvatham, Samuel and Friedman, Paul and others},
  journal={Transactions on Machine Learning Research},
  year={2023}
}

@article{fu2024cardiogpt,
  title={Cardiogpt: An ecg interpretation generation model},
  author={Fu, Guohua and Zheng, Jianwei and Abudayyeh, Islam and Ani, Chizobam and Rakovski, Cyril and Ehwerhemuepha, Louis and Lu, Hongxia and Guo, Yongjuan and Liu, Shenglin and Chu, Huimin and others},
  journal={IEEE Access},
  volume={12},
  pages={50254--50264},
  year={2024},
  publisher={IEEE}
}

@article{khunte2024automated,
  title={Automated diagnostic reports from images of electrocardiograms at the point-of-care},
  author={Khunte, Akshay and Sangha, Veer and Oikonomou, Evangelos K and Dhingra, Lovedeep S and Aminorroaya, Arya and Coppi, Andreas and Shankar, Sumukh Vasisht and Mortazavi, Bobak J and Bhatt, Deepak L and Krumholz, Harlan M and others},
  journal={medRxiv},
  year={2024}
}

@article{tang2026interpretable,
  title={Interpretable multimodal zero shot ECG diagnosis via structured clinical knowledge alignment},
  author={Tang, Jialu and Pham, Hung Manh and De Lathauwer, Ignace and Schipper, Henk S and Lu, Yuan and Ma, Dong and Saeed, Aaqib},
  journal={npj Cardiovascular Health},
  volume={3},
  number={1},
  pages={1},
  year={2026},
  publisher={Nature Publishing Group UK London}
}

@article{jin2023medcpt,
  title={Medcpt: Contrastive pre-trained transformers with large-scale pubmed search logs for zero-shot biomedical information retrieval},
  author={Jin, Qiao and Kim, Won and Chen, Qingyu and Comeau, Donald C and Yeganova, Lana and Wilbur, W John and Lu, Zhiyong},
  journal={Bioinformatics},
  volume={39},
  number={11},
  pages={btad651},
  year={2023},
  publisher={Oxford University Press}
}

@inproceedings{zhao2025ecg,
  title={Ecg-chat: A large ecg-language model for cardiac disease diagnosis},
  author={Zhao, Yubao and Kang, Jiaju and Zhang, Tian and Han, Puyu and Chen, Tong},
  booktitle={2025 IEEE International Conference on Multimedia and Expo (ICME)},
  pages={1--6},
  year={2025},
  organization={IEEE}
}

@article{yang2025diagecg,
  title={DiagECG: An LLM-Driven Framework for Diagnostic Reasoning via Discretized ECG Tokenization},
  author={Yang, Jinning and Shi, Wen},
  journal={arXiv preprint arXiv:2508.15338},
  year={2025}
}

@inproceedings{fang2024physiollm,
  title={Physiollm: Supporting personalized health insights with wearables and large language models},
  author={Fang, Cathy Mengying and Danry, Valdemar and Whitmore, Nathan and Bao, Andria and Hutchison, Andrew and Pierce, Cayden and Maes, Pattie},
  booktitle={2024 IEEE EMBS International Conference on Biomedical and Health Informatics (BHI)},
  pages={1--8},
  year={2024},
  organization={IEEE}
}

@article{song2025retrieval,
  title={Retrieval-Augmented Generation for Electrocardiogram-Language Models},
  author={Song, Xiaoyu and Han, William and Chen, Tony and Duan, Chaojing and Rosenberg, Michael A and Liu, Emerson and Zhao, Ding},
  journal={arXiv preprint arXiv:2510.00261},
  year={2025}
}

@inproceedings{bartels2022ecgcaptions,
  title={Learning to Automatically Generate Accurate ECG Captions},
  author={Bartels, Mathieu G. G. and Najdenkoska, Ivona and van de Leur, Rutger R. and Sammani, Arjan and Taha, Karim and Knigge, David M. and Doevendans, Pieter A. and Worring, Marcel and van Es, Ren{\'e}},
  booktitle={Proceedings of the 5th International Conference on Medical Imaging with Deep Learning},
  series={Proceedings of Machine Learning Research},
  volume={172},
  pages={86--102},
  year={2022},
  publisher={PMLR}
}

@article{hager2024limitationsllmclinical,
  title={Evaluation and mitigation of the limitations of large language models in clinical decision-making},
  author={Hager, Paul and Jungmann, Felix and Holland, Ryan and others},
  journal={Nature Medicine},
  volume={30},
  pages={2613--2622},
  year={2024},
  doi={10.1038/s41591-024-03097-1},
  publisher={Springer Nature}
}

@article{johri2025craftmd,
  title={An evaluation framework for clinical use of large language models in patient interaction tasks},
  author={Johri, S. and Jeong, J. and Tran, B. A. and others},
  journal={Nature Medicine},
  volume={31},
  pages={77--86},
  year={2025},
  doi={10.1038/s41591-024-03328-5},
  publisher={Springer Nature}
}

@article{vanveen2024summarizationllm,
  title={Adapted large language models can outperform medical experts in clinical text summarization},
  author={Van Veen, Dave and Van Uden, C. and Blankemeier, Louis and others},
  journal={Nature Medicine},
  volume={30},
  pages={1134--1142},
  year={2024},
  doi={10.1038/s41591-024-02855-5},
  publisher={Springer Nature}
}

@article{wang2025ecg,
  title={ECG-Expert-QA: A Benchmark for Evaluating Medical Large Language Models in Heart Disease Diagnosis},
  author={Wang, Xu and Kang, Jiaju and Han, Puyu and Zhao, Yubao and Liu, Qian and He, Liwenfei and Zhang, Lingqiong and Dai, Lingyun and Wang, Yongcheng and Tao, Jie},
  journal={arXiv preprint arXiv:2502.17475},
  year={2025}
}

@article{wang2025eeg,
  title={EEG-MedRAG: Enhancing EEG-based Clinical Decision-Making via Hierarchical Hypergraph Retrieval-Augmented Generation},
  author={Wang, Yi and Luo, Haoran and Meng, Lu and Jia, Ziyu and Zhou, Xinliang and Wen, Qingsong},
  journal={arXiv preprint arXiv:2508.13735},
  year={2025}
}

@article{fast2024autonomous,
  title={Autonomous medical evaluation for guideline adherence of large language models},
  author={Fast, Dennis and Adams, Lisa C and Busch, Felix and Fallon, Conor and Huppertz, Marc and Siepmann, Robert and Prucker, Philipp and Bayerl, Nadine and Truhn, Daniel and Makowski, Marcus and others},
  journal={NPJ Digital Medicine},
  volume={7},
  number={1},
  pages={358},
  year={2024},
  publisher={Nature Publishing Group UK London}
}

@inproceedings{wang2025trustworthy,
  title={Trustworthy medical question answering: An evaluation-centric survey},
  author={Wang, Yinuo and Wang, Baiyang and Mercer, Robert and Rudzicz, Frank and Roy, Sudipta Singha and Ren, Pengjie and Chen, Zhumin and Wang, Xindi},
  booktitle={Proceedings of the 2025 Conference on Empirical Methods in Natural Language Processing},
  pages={27477--27490},
  year={2025}
}

@article{senkaiahliyan2023gpt,
  title={GPT-4V (ision) unsuitable for clinical care and education: a clinician-evaluated assessment},
  author={Senkaiahliyan, Senthujan and Toma, Augustin and Ma, Jun and Chan, An-Wen and Ha, Andrew and An, Kevin R and Suresh, Hrishikesh and Rubin, Barry and Wang, Bo},
  journal={arXiv preprint arXiv:2403.12046},
  year={2023}
}

@article{singhal2023large,
  title={Large language models encode clinical knowledge},
  author={Singhal, Karan and Azizi, Shekoofeh and Tu, Tao and Mahdavi, S Sara and Wei, Jason and Chung, Hyung Won and Scales, Nathan and Tanwani, Ajay and Cole-Lewis, Heather and Pfohl, Stephen and others},
  journal={Nature},
  volume={620},
  number={7972},
  pages={172--180},
  year={2023},
  publisher={Nature Publishing Group}
}

@article{novak2023pulse,
  title={The pulse of artificial intelligence in cardiology: a comprehensive evaluation of state-of-the-art large language models for potential use in clinical cardiology},
  author={Novak, Andrej and Zeljkovi{\'c}, Ivan and Rode, Fran and Lisi{\v{c}}i{\'c}, Ante and Nola, Iskra A and Pavlovi{\'c}, Nikola and Manola, {\v{S}}ime},
  journal={medRxiv},
  pages={2023--08},
  year={2023},
  publisher={Cold Spring Harbor Laboratory Press}
}

@article{luo2023biomedgpt,
  title={Biomedgpt: Open multimodal generative pre-trained transformer for biomedicine},
  author={Luo, Yizhen and Zhang, Jiahuan and Fan, Siqi and Yang, Kai and Wu, Yushuai and Qiao, Mu and Nie, Zaiqing},
  journal={arXiv preprint arXiv:2308.09442},
  year={2023}
}

@article{wang2024enhancing,
  title={Enhancing eeg-to-text decoding through transferable representations from pre-trained contrastive eeg-text masked autoencoder},
  author={Wang, Jiaqi and Song, Zhenxi and Ma, Zhengyu and Qiu, Xipeng and Zhang, Min and Zhang, Zhiguo},
  journal={arXiv preprint arXiv:2402.17433},
  year={2024}
}

@inproceedings{qiu2023can,
  title={Can brain signals reveal inner alignment with human languages?},
  author={Qiu, Jielin and Han, William and Zhu, Jiacheng and Xu, Mengdi and Weber, Douglas and Li, Bo and Zhao, Ding},
  booktitle={Findings of the Association for Computational Linguistics: EMNLP 2023},
  pages={1789--1804},
  year={2023}
}

@inproceedings{zhang2024word,
  title={From word embedding to reading embedding using large language model, eeg and eye-tracking},
  author={Zhang, Yuhong and Yang, Shilai and Cauwenberghs, Gert and Jung, Tzyy-Ping},
  booktitle={2024 46th Annual International Conference of the IEEE Engineering in Medicine and Biology Society (EMBC)},
  pages={1--4},
  year={2024},
  organization={IEEE}
}

@article{mischler2024contextual,
  title={Contextual feature extraction hierarchies converge in large language models and the brain},
  author={Mischler, Gavin and Li, Yinghao Aaron and Bickel, Stephan and Mehta, Ashesh D and Mesgarani, Nima},
  journal={Nature Machine Intelligence},
  volume={6},
  number={12},
  pages={1467--1477},
  year={2024},
  publisher={Nature Publishing Group UK London}
}

@article{zhou2024belt,
  title={BELT: bootstrapped EEG-to-language training by natural language supervision},
  author={Zhou, Jinzhao and Duan, Yiqun and Chang, Yu-Cheng and Wang, Yu-Kai and Lin, Chin-Teng},
  journal={IEEE Transactions on Neural Systems and Rehabilitation Engineering},
  year={2024},
  publisher={IEEE}
}

@inproceedings{cui2024neuro,
  title={Neuro-gpt: Towards a foundation model for eeg},
  author={Cui, Wenhui and Jeong, Woojae and Th{\"o}lke, Philipp and Medani, Takfarinas and Jerbi, Karim and Joshi, Anand A and Leahy, Richard M},
  booktitle={2024 IEEE International Symposium on Biomedical Imaging (ISBI)},
  pages={1--5},
  year={2024},
  organization={IEEE}
}

@inproceedings{lee2024enhancing,
  title={Enhancing neural decoding with large language models: A GPT-based approach},
  author={Lee, Dong Hyeok and Chung, Chun Kee},
  booktitle={2024 12th International Winter Conference on Brain-Computer Interface (BCI)},
  pages={1--4},
  year={2024},
  organization={IEEE}
}

@article{duan2023dewave,
  title={Dewave: Discrete eeg waves encoding for brain dynamics to text translation},
  author={Duan, Yiqun and Zhou, Jinzhao and Wang, Zhen and Wang, Yu-Kai and Lin, Chin-Teng},
  journal={arXiv preprint arXiv:2309.14030},
  year={2023}
}

@inproceedings{chen2023can,
  title={Can large language models provide security \& privacy advice? measuring the ability of llms to refute misconceptions},
  author={Chen, Yufan and Arunasalam, Arjun and Celik, Z Berkay},
  booktitle={Proceedings of the 39th Annual Computer Security Applications Conference},
  pages={366--378},
  year={2023}
}

@article{karabacak2023advent,
  title={The advent of generative language models in medical education},
  author={Karabacak, Mert and Ozkara, Burak Berksu and Margetis, Konstantinos and Wintermark, Max and Bisdas, Sotirios},
  journal={JMIR Medical Education},
  volume={9},
  pages={e48163},
  year={2023},
  publisher={JMIR Publications Toronto, Canada}
}

@inproceedings{zuo2024adapting,
  title={Adapting LLMs for Ballistocardiographic Signals: A Multi-Task Learning Framework for BCG to ECG Reconstruction},
  author={Zuo, Boyang and Lei, Kun and Wang, Xingjun},
  booktitle={2024 6th International Conference on Robotics, Intelligent Control and Artificial Intelligence (RICAI)},
  pages={1060--1065},
  year={2024},
  organization={IEEE}
}

@inproceedings{liu2024ecg,
  title={ECG-LLM: Leveraging Large Language Models for Low-Quality ECG Signal Restoration},
  author={Liu, Longfei and Cui, Guosheng and Wan, Cheng and Wu, Dan and Li, Ye},
  booktitle={2024 IEEE International Conference on Bioinformatics and Biomedicine (BIBM)},
  pages={3537--3542},
  year={2024},
  organization={IEEE}
}

@article{lai2025diffusets,
  title={DiffuSETS: 12-Lead ECG generation conditioned on clinical text reports and patient-specific information},
  author={Lai, Yongfan and Chen, Jiabo and Zhao, Qinghao and Zhang, Deyun and Wang, Yue and Geng, Shijia and Li, Hongyan and Hong, Shenda},
  journal={Patterns},
  year={2025},
  publisher={Elsevier}
}

@inproceedings{chung2023text,
  title={Text-to-ecg: 12-lead electrocardiogram synthesis conditioned on clinical text reports},
  author={Chung, Hyunseung and Kim, Jiho and Kwon, Joon-Myoung and Jeon, Ki-Hyun and Lee, Min Sung and Choi, Edward},
  booktitle={ICASSP 2023-2023 IEEE International Conference on Acoustics, Speech and Signal Processing (ICASSP)},
  pages={1--5},
  year={2023},
  organization={IEEE}
}

@article{zhang2025sensorlm,
  title={SensorLM: Learning the Language of Wearable Sensors},
  author={Zhang, Yuwei and Ayush, Kumar and Qiao, Siyuan and Heydari, A Ali and Narayanswamy, Girish and Xu, Maxwell A and Metwally, Ahmed A and Xu, Shawn and Garrison, Jake and Xu, Xuhai and others},
  journal={arXiv preprint arXiv:2506.09108},
  year={2025}
}

@inproceedings{liu2024large,
  title={Large language models for cuffless blood pressure measurement from wearable biosignals},
  author={Liu, Zengding and Chen, Chen and Cao, Jiannong and Pan, Minglei and Liu, Jikui and Li, Nan and Miao, Fen and Li, Ye},
  booktitle={Proceedings of the 15th ACM International Conference on Bioinformatics, Computational Biology and Health Informatics},
  pages={1--11},
  year={2024}
}

@article{wang2023large,
  title={Large transformers are better eeg learners},
  author={Wang, Bingxin and Fu, Xiaowen and Lan, Yuan and Zhang, Luchan and Zheng, Wei and Xiang, Yang},
  journal={arXiv preprint arXiv:2308.11654},
  year={2023}
}

@article{merrill2026transforming,
  title={Transforming wearable data into personal health insights using large language model agents},
  author={Merrill, Mike A and Paruchuri, Akshay and Rezaei, Naghmeh and Kovacs, Geza and Perez, Javier and Liu, Yun and Schenck, Erik and Hammerquist, Nova and Sunshine, Jake and Tailor, Shyam and others},
  journal={Nature Communications},
  year={2026},
  publisher={Nature Publishing Group UK London}
}

@article{weissman2025unregulated,
  title={Unregulated large language models produce medical device-like output},
  author={Weissman, Gary E and Mankowitz, Toni and Kanter, Genevieve P},
  journal={NPJ Digital Medicine},
  volume={8},
  number={1},
  pages={148},
  year={2025},
  publisher={Nature Publishing Group UK London}
}

@article{asgari2025framework,
  title={A framework to assess clinical safety and hallucination rates of LLMs for medical text summarisation},
  author={Asgari, Elham and Monta{\~n}a-Brown, Nina and Dubois, Magda and Khalil, Saleh and Balloch, Jasmine and Yeung, Joshua Au and Pimenta, Dominic},
  journal={npj Digital Medicine},
  volume={8},
  number={1},
  pages={274},
  year={2025},
  publisher={Nature Publishing Group UK London}
}

@article{hollenstein2018zuco,
  title={ZuCo, a simultaneous EEG and eye-tracking resource for natural sentence reading},
  author={Hollenstein, Nora and Rotsztejn, Jonathan and Troendle, Marius and Pedroni, Andreas and Zhang, Ce and Langer, Nicolas},
  journal={Scientific data},
  volume={5},
  number={1},
  pages={180291},
  year={2018},
  publisher={Nature Publishing Group}
}

@inproceedings{hollenstein2020zuco,
  title={ZuCo 2.0: A dataset of physiological recordings during natural reading and annotation},
  author={Hollenstein, Nora and Troendle, Marius and Zhang, Ce and Langer, Nicolas},
  booktitle={Proceedings of the Twelfth Language Resources and Evaluation Conference},
  pages={138--146},
  year={2020}
}

@article{wagner2020ptb,
  title={PTB-XL, a large publicly available electrocardiography dataset},
  author={Wagner, Patrick and Strodthoff, Nils and Bousseljot, Ralf-Dieter and Kreiseler, Dieter and Lunze, Fatima I and Samek, Wojciech and Schaeffter, Tobias},
  journal={Scientific data},
  volume={7},
  number={1},
  pages={154},
  year={2020},
  publisher={Nature Publishing Group UK London}
}

@article{moody2020mimic,
  title={MIMIC-III waveform database (version 1.0)},
  author={Moody, Benjamin and Moody, George and Villarroel, Mauricio and Clifford, G and Silva, Ikaro},
  journal={PhysioNet},
  volume={3},
  year={2020}
}

@article{moody2001impact,
  title={The impact of the MIT-BIH arrhythmia database},
  author={Moody, George B and Mark, Roger G},
  journal={IEEE engineering in medicine and biology magazine},
  volume={20},
  number={3},
  pages={45--50},
  year={2001},
  publisher={IEEE}
}

@article{liu2018open,
  title={An open access database for evaluating the algorithms of electrocardiogram rhythm and morphology abnormality detection},
  author={Liu, Feifei and Liu, Chengyu and Zhao, Lina and Zhang, Xiangyu and Wu, Xiaoling and Xu, Xiaoyan and Liu, Yulin and Ma, Caiyun and Wei, Shoushui and He, Zhiqiang and others},
  journal={Journal of Medical Imaging and Health Informatics},
  volume={8},
  number={7},
  pages={1368--1373},
  year={2018},
  publisher={American Scientific Publishers}
}

@article{ribeiro2020automatic,
  title={Automatic diagnosis of the 12-lead ECG using a deep neural network},
  author={Ribeiro, Ant{\^o}nio H and Ribeiro, Manoel Horta and Paix{\~a}o, Gabriela MM and Oliveira, Derick M and Gomes, Paulo R and Canazart, J{\'e}ssica A and Ferreira, Milton PS and Andersson, Carl R and Macfarlane, Peter W and Meira Jr, Wagner and others},
  journal={Nature communications},
  volume={11},
  number={1},
  pages={1760},
  year={2020},
  publisher={Nature Publishing Group UK London}
}

@article{gow2023mimic,
  title={MIMIC-IV-ECG: Diagnostic Electrocardiogram Matched Subset},
  author={Gow, Brian and Pollard, Tom and Nathanson, Larry A and Johnson, Alistair and Moody, Benjamin and Fernandes, Chrystinne and Greenbaum, Nathaniel and Waks, Jonathan W and Eslami, Parastou and Carbonati, Tanner and others},
  journal={Type: dataset},
  year={2023}
}

@article{safranek2024automated,
  title={Automated HEART score determination via ChatGPT: Honing a framework for iterative prompt development},
  author={Safranek, Conrad W and Huang, Thomas and Wright, Donald S and Wright, Catherine X and Socrates, Vimig and Sangal, Rohit B and Iscoe, Mark and Chartash, David and Taylor, R Andrew},
  journal={Journal of the American College of Emergency Physicians Open},
  volume={5},
  number={2},
  pages={e13133},
  year={2024},
  publisher={Wiley Online Library}
}

@inproceedings{yao2022react,
  title={React: Synergizing reasoning and acting in language models},
  author={Yao, Shunyu and Zhao, Jeffrey and Yu, Dian and Du, Nan and Shafran, Izhak and Narasimhan, Karthik R and Cao, Yuan},
  booktitle={The eleventh international conference on learning representations},
  year={2022}
}

@article{long2023large,
  title={Large language model guided tree-of-thought},
  author={Long, Jieyi},
  journal={arXiv preprint arXiv:2305.08291},
  year={2023}
}

@article{peng2024model,
  title={Model tuning or prompt Tuning? a study of large language models for clinical concept and relation extraction},
  author={Peng, Cheng and Yang, XI and Smith, Kaleb E and Yu, Zehao and Chen, Aokun and Bian, Jiang and Wu, Yonghui},
  journal={Journal of biomedical informatics},
  volume={153},
  pages={104630},
  year={2024},
  publisher={Elsevier}
}

@article{chang2024efficient,
  title={Efficient prompting methods for large language models: A survey},
  author={Chang, Kaiyan and Xu, Songcheng and Wang, Chenglong and Luo, Yingfeng and Liu, Xiaoqian and Xiao, Tong and Zhu, Jingbo},
  journal={arXiv preprint arXiv:2404.01077},
  year={2024}
}

@article{ahmad2023revolutionizing,
  title={Revolutionizing healthcare: How deep learning is poised to change the landscape of medical diagnosis and treatment},
  author={Ahmad, Ahsan and Tariq, Aftab and Hussain, Hafiz Khawar and Gill, Ahmad Yousaf},
  journal={Journal of Computer Networks, Architecture and High Performance Computing},
  volume={5},
  number={2},
  pages={458--471},
  year={2023}
}

@article{zhou2023survey,
  title={A survey of large language models in medicine: Progress, application, and challenge},
  author={Zhou, Hongjian and Liu, Fenglin and Gu, Boyang and Zou, Xinyu and Huang, Jinfa and Wu, Jinge and Li, Yiru and Chen, Sam S and Zhou, Peilin and Liu, Junling and others},
  journal={arXiv preprint arXiv:2311.05112},
  year={2023}
}

@article{thirunavukarasu2023large,
  title={Large language models in medicine},
  author={Thirunavukarasu, Arun James and Ting, Darren Shu Jeng and Elangovan, Kabilan and Gutierrez, Laura and Tan, Ting Fang and Ting, Daniel Shu Wei},
  journal={Nature medicine},
  volume={29},
  number={8},
  pages={1930--1940},
  year={2023},
  publisher={Nature Publishing Group US New York}
}

@article{zeljkovic2025beyond,
  title={Beyond text: the impact of clinical context on GPT-4’s 12-lead electrocardiogram interpretation accuracy},
  author={Zeljkovic, Ivan and Novak, Andrej and Lisicic, Ante and Jordan, Ana and Serman, Ana and Jurin, Ivana and Pavlovic, Nikola and Manola, Sime},
  journal={Canadian journal of cardiology},
  volume={41},
  number={7},
  pages={1406--1414},
  year={2025},
  publisher={Elsevier}
}

@article{gunay2024comparison,
  title={Comparison of emergency medicine specialist, cardiologist, and chat-GPT in electrocardiography assessment},
  author={G{\"u}nay, Serkan and {\"O}zt{\"u}rk, Ahmet and {\"O}zerol, Hakan and Yi{\u{g}}it, Yavuz and Erenler, Ali Kemal},
  journal={The American journal of emergency medicine},
  volume={80},
  pages={51--60},
  year={2024},
  publisher={Elsevier}
}

@inproceedings{vedantam2015cider,
  title={Cider: Consensus-based image description evaluation},
  author={Vedantam, Ramakrishna and Lawrence Zitnick, C and Parikh, Devi},
  booktitle={Proceedings of the IEEE conference on computer vision and pattern recognition},
  pages={4566--4575},
  year={2015}
}

@article{ye2024medualtime,
  title={MedualTime: A dual-adapter language model for medical time series-text multimodal learning},
  author={Ye, Jiexia and Zhang, Weiqi and Li, Ziyue and Li, Jia and Zhao, Meng and Tsung, Fugee},
  journal={arXiv preprint arXiv:2406.06620},
  year={2024}
}

@article{lee2020biobert,
  title={BioBERT: a pre-trained biomedical language representation model for biomedical text mining},
  author={Lee, Jinhyuk and Yoon, Wonjin and Kim, Sungdong and Kim, Donghyeon and Kim, Sunkyu and So, Chan Ho and Kang, Jaewoo},
  journal={Bioinformatics},
  volume={36},
  number={4},
  pages={1234--1240},
  year={2020},
  publisher={Oxford University Press}
}

@article{gu2021domain,
  title={Domain-specific language model pretraining for biomedical natural language processing},
  author={Gu, Yu and Tinn, Robert and Cheng, Hao and Lucas, Michael and Usuyama, Naoto and Liu, Xiaodong and Naumann, Tristan and Gao, Jianfeng and Poon, Hoifung},
  journal={ACM Transactions on Computing for Healthcare (HEALTH)},
  volume={3},
  number={1},
  pages={1--23},
  year={2021},
  publisher={ACM New York, NY}
}

@inproceedings{peng2019transfer,
  title={Transfer learning in biomedical natural language processing: an evaluation of BERT and ELMo on ten benchmarking datasets},
  author={Peng, Yifan and Yan, Shankai and Lu, Zhiyong},
  booktitle={Proceedings of the 18th BioNLP workshop and shared task},
  pages={58--65},
  year={2019}
}

@article{luo2022biogpt,
  title={BioGPT: generative pre-trained transformer for biomedical text generation and mining},
  author={Luo, Renqian and Sun, Liai and Xia, Yingce and Qin, Tao and Zhang, Sheng and Poon, Hoifung and Liu, Tie-Yan},
  journal={Briefings in bioinformatics},
  volume={23},
  number={6},
  pages={bbac409},
  year={2022},
  publisher={Oxford University Press}
}

@article{chen2023meditron,
  title={Meditron-70b: Scaling medical pretraining for large language models, 2023},
  author={Chen, Zeming and Cano, Alejandro Hern{\'a}ndez and Romanou, Angelika and Bonnet, Antoine and Matoba, Kyle and Salvi, Francesco and Pagliardini, Matteo and Fan, Simin and K{\"o}pf, Andreas and Mohtashami, Amirkeivan and others},
  journal={URL https://arxiv. org/abs/2311.16079},
  year={2023}
}

@article{wu2024pmc,
  title={PMC-LLaMA: toward building open-source language models for medicine},
  author={Wu, Chaoyi and Lin, Weixiong and Zhang, Xiaoman and Zhang, Ya and Xie, Weidi and Wang, Yanfeng},
  journal={Journal of the American Medical Informatics Association},
  volume={31},
  number={9},
  pages={1833--1843},
  year={2024},
  publisher={Oxford University Press}
}

@inproceedings{lin2023pmc,
  title={Pmc-clip: Contrastive language-image pre-training using biomedical documents},
  author={Lin, Weixiong and Zhao, Ziheng and Zhang, Xiaoman and Wu, Chaoyi and Zhang, Ya and Wang, Yanfeng and Xie, Weidi},
  booktitle={International Conference on Medical Image Computing and Computer-Assisted Intervention},
  pages={525--536},
  year={2023},
  organization={Springer}
}

@article{zhang2023biomedclip,
  title={Biomedclip: a multimodal biomedical foundation model pretrained from fifteen million scientific image-text pairs},
  author={Zhang, Sheng and Xu, Yanbo and Usuyama, Naoto and Xu, Hanwen and Bagga, Jaspreet and Tinn, Robert and Preston, Sam and Rao, Rajesh and Wei, Mu and Valluri, Naveen and others},
  journal={arXiv preprint arXiv:2303.00915},
  year={2023}
}

@article{chowdhery2023palm,
  title={Palm: Scaling language modeling with pathways},
  author={Chowdhery, Aakanksha and Narang, Sharan and Devlin, Jacob and Bosma, Maarten and Mishra, Gaurav and Roberts, Adam and Barham, Paul and Chung, Hyung Won and Sutton, Charles and Gehrmann, Sebastian and others},
  journal={Journal of machine learning research},
  volume={24},
  number={240},
  pages={1--113},
  year={2023}
}

@article{singhal2025toward,
  title={Toward expert-level medical question answering with large language models},
  author={Singhal, Karan and Tu, Tao and Gottweis, Juraj and Sayres, Rory and Wulczyn, Ellery and Amin, Mohamed and Hou, Le and Clark, Kevin and Pfohl, Stephen R and Cole-Lewis, Heather and others},
  journal={Nature medicine},
  volume={31},
  number={3},
  pages={943--950},
  year={2025},
  publisher={Nature Publishing Group US New York}
}

@article{brown2020language,
  title={Language models are few-shot learners},
  author={Brown, Tom and Mann, Benjamin and Ryder, Nick and Subbiah, Melanie and Kaplan, Jared D and Dhariwal, Prafulla and Neelakantan, Arvind and Shyam, Pranav and Sastry, Girish and Askell, Amanda and others},
  journal={Advances in neural information processing systems},
  volume={33},
  pages={1877--1901},
  year={2020}
}

@article{touvron2023llama1,
  title={Llama: Open and efficient foundation language models},
  author={Touvron, Hugo and Lavril, Thibaut and Izacard, Gautier and Martinet, Xavier and Lachaux, Marie-Anne and Lacroix, Timoth{\'e}e and Rozi{\`e}re, Baptiste and Goyal, Naman and Hambro, Eric and Azhar, Faisal and others},
  journal={arXiv preprint arXiv:2302.13971},
  year={2023}
}

@article{touvron2023llama2,
  title={Llama 2: Open foundation and fine-tuned chat models},
  author={Touvron, Hugo and Martin, Louis and Stone, Kevin and Albert, Peter and Almahairi, Amjad and Babaei, Yasmine and Bashlykov, Nikolay and Batra, Soumya and Bhargava, Prajjwal and Bhosale, Shruti and others},
  journal={arXiv preprint arXiv:2307.09288},
  year={2023}
}

@article{babu2025large,
  title={Large language models for eeg: A comprehensive survey and taxonomy},
  author={Babu, Naseem and Mathew, Jimson and Vinod, AP},
  journal={arXiv preprint arXiv:2506.06353},
  year={2025}
}

@article{ansari2025survey,
  title={A survey of transformers and large language models for ECG diagnosis: advances, challenges, and future directions},
  author={Ansari, Mohammed Yusuf and Yaqoob, Mohammed and Ishaq, Mohammed and Flushing, Eduardo Feo and Mangalote, Iffa Afsa Changaai and Dakua, Sarada Prasad and Aboumarzouk, Omar and Righetti, Raffaella and Qaraqe, Marwa},
  journal={Artificial Intelligence Review},
  volume={58},
  number={9},
  pages={261},
  year={2025},
  publisher={Springer}
}

@article{hu2022lora,
  title={Lora: Low-rank adaptation of large language models.},
  author={Hu, Edward J and Shen, Yelong and Wallis, Phillip and Allen-Zhu, Zeyuan and Li, Yuanzhi and Wang, Shean and Wang, Liang and Chen, Weizhu and others},
  journal={Iclr},
  volume={1},
  number={2},
  pages={3},
  year={2022}
}

@inproceedings{li2021prefix,
  title={Prefix-tuning: Optimizing continuous prompts for generation},
  author={Li, Xiang Lisa and Liang, Percy},
  booktitle={Proceedings of the 59th Annual Meeting of the Association for Computational Linguistics and the 11th International Joint Conference on Natural Language Processing (Volume 1: Long Papers)},
  pages={4582--4597},
  year={2021}
}

@inproceedings{liu2022p,
  title={P-tuning: Prompt tuning can be comparable to fine-tuning across scales and tasks},
  author={Liu, Xiao and Ji, Kaixuan and Fu, Yicheng and Tam, Weng and Du, Zhengxiao and Yang, Zhilin and Tang, Jie},
  booktitle={Proceedings of the 60th Annual Meeting of the Association for Computational Linguistics (Volume 2: Short Papers)},
  pages={61--68},
  year={2022}
}

@article{Liu2021PTuningVP,
  title={P-Tuning v2: Prompt Tuning Can Be Comparable to Fine-tuning Universally Across Scales and Tasks},
  author={Xiao Liu and Kaixuan Ji and Yicheng Fu and Zhengxiao Du and Zhilin Yang and Jie Tang},
  journal={ArXiv},
  year={2021},
  volume={abs/2110.07602},
  url={https://api.semanticscholar.org/CorpusID:238857040}
}

@inproceedings{jia2022visual,
  title={Visual prompt tuning},
  author={Jia, Menglin and Tang, Luming and Chen, Bor-Chun and Cardie, Claire and Belongie, Serge and Hariharan, Bharath and Lim, Ser-Nam},
  booktitle={European conference on computer vision},
  pages={709--727},
  year={2022},
  organization={Springer}
}

@inproceedings{lester2021power,
  title={The power of scale for parameter-efficient prompt tuning},
  author={Lester, Brian and Al-Rfou, Rami and Constant, Noah},
  booktitle={Proceedings of the 2021 conference on empirical methods in natural language processing},
  pages={3045--3059},
  year={2021}
}

@inproceedings{houlsby2019parameter,
  title={Parameter-efficient transfer learning for NLP},
  author={Houlsby, Neil and Giurgiu, Andrei and Jastrzebski, Stanislaw and Morrone, Bruna and De Laroussilhe, Quentin and Gesmundo, Andrea and Attariyan, Mona and Gelly, Sylvain},
  booktitle={International conference on machine learning},
  pages={2790--2799},
  year={2019},
  organization={PMLR}
}

@article{rajpurkar2022ai,
  title={AI in health and medicine},
  author={Rajpurkar, Pranav and Chen, Emma and Banerjee, Oishi and Topol, Eric J},
  journal={Nature medicine},
  volume={28},
  number={1},
  pages={31--38},
  year={2022},
  publisher={Nature Publishing Group US New York}
}

@article{van2023clinical,
  title={Clinical text summarization: adapting large language models can outperform human experts},
  author={Van Veen, Dave and Van Uden, Cara and Blankemeier, Louis and Delbrouck, Jean-Benoit and Aali, Asad and Bluethgen, Christian and Pareek, Anuj and Polacin, Malgorzata and Reis, Eduardo Pontes and Seehofnerova, Anna and others},
  journal={Research square},
  pages={rs--3},
  year={2023}
}

@article{consort2019reporting,
  title={Reporting guidelines for clinical trials evaluating artificial intelligence interventions are needed},
  journal={Nature Medicine},
  volume={25},
  number={10},
  pages={1467--1468},
  year={2019},
  publisher={Nature Publishing Group US New York}
}

@article{ren2024healthcare,
  title={Healthcare copilot: Eliciting the power of general llms for medical consultation},
  author={Ren, Zhiyao and Zhan, Yibing and Yu, Baosheng and Ding, Liang and Tao, Dacheng},
  journal={arXiv preprint arXiv:2402.13408},
  year={2024}
}

@article{wang2024jmlr,
  title={Jmlr: Joint medical llm and retrieval training for enhancing reasoning and professional question answering capability},
  author={Wang, Junda and Yang, Zhichao and Yao, Zonghai and Yu, Hong},
  journal={arXiv preprint arXiv:2402.17887},
  year={2024}
}

@article{chen2025enhancing,
  title={Enhancing diagnostic capability with multi-agents conversational large language models},
  author={Chen, Xi and Yi, Huahui and You, Mingke and Liu, WeiZhi and Wang, Li and Li, Hairui and Zhang, Xue and Guo, Yingman and Fan, Lei and Chen, Gang and others},
  journal={NPJ digital medicine},
  volume={8},
  number={1},
  pages={159},
  year={2025},
  publisher={Nature Publishing Group UK London}
}

@article{hager2024evaluation,
  title={Evaluation and mitigation of the limitations of large language models in clinical decision-making},
  author={Hager, Paul and Jungmann, Friederike and Holland, Robbie and Bhagat, Kunal and Hubrecht, Inga and Knauer, Manuel and Vielhauer, Jakob and Makowski, Marcus and Braren, Rickmer and Kaissis, Georgios and others},
  journal={Nature medicine},
  volume={30},
  number={9},
  pages={2613--2622},
  year={2024},
  publisher={Nature Publishing Group US New York}
}

@article{goodell2025large,
  title={Large language model agents can use tools to perform clinical calculations},
  author={Goodell, Alex J and Chu, Simon N and Rouholiman, Dara and Chu, Larry F},
  journal={npj Digital Medicine},
  volume={8},
  number={1},
  pages={163},
  year={2025},
  publisher={Nature Publishing Group UK London}
}

@article{zhou2025automating,
  title={Automating expert-level medical reasoning evaluation of large language models},
  author={Zhou, Shuang and Xie, Wenya and Li, Jiaxi and Zhan, Zaifu and Song, Meijia and Yang, Han and Espinoza, Cheyenna and Welton, Lindsay and Mai, Xinnie and Jin, Yanwei and others},
  journal={npj Digital Medicine},
  year={2025},
  publisher={Nature Publishing Group UK London}
}

@article{agrawal2025evaluation,
  title={The evaluation illusion of large language models in medicine},
  author={Agrawal, Monica and Chen, Irene Y and Gulamali, Freya and Joshi, Shalmali},
  journal={npj Digital Medicine},
  volume={8},
  number={1},
  pages={600},
  year={2025},
  publisher={Nature Publishing Group UK London}
}

@article{gao2025increasing,
  title={Increasing alignment of large language models with language processing in the human brain},
  author={Gao, Changjiang and Ma, Zhengwu and Chen, Jiajun and Li, Ping and Huang, Shujian and Li, Jixing},
  journal={Nature computational science},
  volume={5},
  number={11},
  pages={1080--1090},
  year={2025},
  publisher={Nature Publishing Group US New York}
}

@article{croxford2025evaluating,
  title={Evaluating clinical AI summaries with large language models as judges},
  author={Croxford, Emma and Gao, Yanjun and First, Elliot and Pellegrino, Nicholas and Schnier, Miranda and Caskey, John and Oguss, Madeline and Wills, Graham and Chen, Guanhua and Dligach, Dmitriy and others},
  journal={npj Digital Medicine},
  volume={8},
  number={1},
  pages={640},
  year={2025},
  publisher={Nature Publishing Group UK London}
}

@article{loni2025review,
  title={A review on generative AI models for synthetic medical text, time series, and longitudinal data},
  author={Loni, Mohammad and Poursalim, Fatemeh and Asadi, Mehdi and Gharehbaghi, Arash},
  journal={npj Digital Medicine},
  volume={8},
  number={1},
  pages={281},
  year={2025},
  publisher={Nature Publishing Group UK London}
}

@article{zhou2025diagnosis,
  title={Diagnosis of cardiac conditions from 12-lead electrocardiogram through natural language supervision},
  author={Zhou, Xue and Li, Tianhui and Hayama, Hiromasa and Nakamura, Keijiro and Liu, Shing-Hong and Chen, Wenxi and Zhu, Xin},
  journal={npj Digital Medicine},
  volume={8},
  number={1},
  pages={697},
  year={2025},
  publisher={Nature Publishing Group UK London}
}

@inproceedings{yang2026heartllm,
  title={HeartLLM: Discretized ECG Tokenization for LLM-Based Diagnostic Reasoning},
  author={Yang, Jinning and Sun, Wenjie and Shi, Wen},
  booktitle={Proceedings of the AAAI Conference on Artificial Intelligence},
  volume={40},
  number={40},
  pages={34250--34258},
  year={2026}
}

@article{rawte2023survey,
  title={A survey of hallucination in large foundation models},
  author={Rawte, Vipula and Sheth, Amit and Das, Amitava},
  journal={arXiv preprint arXiv:2309.05922},
  year={2023}
}

@inproceedings{pal2023med,
  title={Med-halt: Medical domain hallucination test for large language models},
  author={Pal, Ankit and Umapathi, Logesh Kumar and Sankarasubbu, Malaikannan},
  booktitle={Proceedings of the 27th Conference on Computational Natural Language Learning (CoNLL)},
  pages={314--334},
  year={2023}
}

@inproceedings{manakul2023selfcheckgpt,
  title={Selfcheckgpt: Zero-resource black-box hallucination detection for generative large language models},
  author={Manakul, Potsawee and Liusie, Adian and Gales, Mark},
  booktitle={Proceedings of the 2023 conference on empirical methods in natural language processing},
  pages={9004--9017},
  year={2023}
}

@inproceedings{dhuliawala2024chain,
  title={Chain-of-verification reduces hallucination in large language models},
  author={Dhuliawala, Shehzaad and Komeili, Mojtaba and Xu, Jing and Raileanu, Roberta and Li, Xian and Celikyilmaz, Asli and Weston, Jason},
  booktitle={Findings of the association for computational linguistics: ACL 2024},
  pages={3563--3578},
  year={2024}
}

@article{mesko2023imperative,
  title={The imperative for regulatory oversight of large language models (or generative AI) in healthcare},
  author={Mesk{\'o}, Bertalan and Topol, Eric J},
  journal={NPJ digital medicine},
  volume={6},
  number={1},
  pages={120},
  year={2023},
  publisher={Nature Publishing Group UK London}
}

@article{derraz2024new,
  title={New regulatory thinking is needed for AI-based personalised drug and cell therapies in precision oncology},
  author={Derraz, Bouchra and Breda, Gabriele and Kaempf, Christoph and Baenke, Franziska and Cotte, Fabienne and Reiche, Kristin and K{\"o}hl, Ulrike and Kather, Jakob Nikolas and Eskenazy, Deborah and Gilbert, Stephen},
  journal={NPJ precision oncology},
  volume={8},
  number={1},
  pages={23},
  year={2024},
  publisher={Nature Publishing Group UK London}
}

@inproceedings{hacker2023regulating,
  title={Regulating ChatGPT and other large generative AI models},
  author={Hacker, Philipp and Engel, Andreas and Mauer, Marco},
  booktitle={Proceedings of the 2023 ACM conference on fairness, accountability, and transparency},
  pages={1112--1123},
  year={2023}
}

@article{he2025survey,
  title={A survey of large language models for healthcare: from data, technology, and applications to accountability and ethics},
  author={He, Kai and Mao, Rui and Lin, Qika and Ruan, Yucheng and Lan, Xiang and Feng, Mengling and Cambria, Erik},
  journal={Information Fusion},
  volume={118},
  pages={102963},
  year={2025},
  publisher={Elsevier}
}

@article{tian2024opportunities,
  title={Opportunities and challenges for ChatGPT and large language models in biomedicine and health},
  author={Tian, Shubo and Jin, Qiao and Yeganova, Lana and Lai, Po-Ting and Zhu, Qingqing and Chen, Xiuying and Yang, Yifan and Chen, Qingyu and Kim, Won and Comeau, Donald C and others},
  journal={Briefings in Bioinformatics},
  volume={25},
  number={1},
  pages={bbad493},
  year={2024},
  publisher={Oxford University Press}
}

@inproceedings{li2023multi,
  title={Multi-step jailbreaking privacy attacks on chatgpt},
  author={Li, Haoran and Guo, Dadi and Fan, Wei and Xu, Mingshi and Huang, Jie and Meng, Fanpu and Song, Yangqiu},
  booktitle={Findings of the Association for Computational Linguistics: EMNLP 2023},
  pages={4138--4153},
  year={2023}
}

@article{wei2023jailbroken,
  title={Jailbroken: How does llm safety training fail?},
  author={Wei, Alexander and Haghtalab, Nika and Steinhardt, Jacob},
  journal={Advances in neural information processing systems},
  volume={36},
  pages={80079--80110},
  year={2023}
}

@article{wornow2023shaky,
  title={The shaky foundations of large language models and foundation models for electronic health records},
  author={Wornow, Michael and Xu, Yizhe and Thapa, Rahul and Patel, Birju and Steinberg, Ethan and Fleming, Scott and Pfeffer, Michael A and Fries, Jason and Shah, Nigam H},
  journal={npj digital medicine},
  volume={6},
  number={1},
  pages={135},
  year={2023},
  publisher={Nature Publishing Group UK London}
}

@article{goetz2024generalization,
  title={Generalization—a key challenge for responsible AI in patient-facing clinical applications},
  author={Goetz, Lea and Seedat, Nabeel and Vandersluis, Robert and van der Schaar, Mihaela},
  journal={NPJ Digital Medicine},
  volume={7},
  number={1},
  pages={126},
  year={2024},
  publisher={Nature Publishing Group UK London}
}

@article{lavin2022technology,
  title={Technology readiness levels for machine learning systems},
  author={Lavin, Alexander and Gilligan-Lee, Ciar{\'a}n M and Visnjic, Alessya and Ganju, Siddha and Newman, Dava and Ganguly, Sujoy and Lange, Danny and Baydin, At{\'\i}l{\'\i}m G{\"u}ne{\c{s}} and Sharma, Amit and Gibson, Adam and others},
  journal={Nature Communications},
  volume={13},
  number={1},
  pages={6039},
  year={2022},
  publisher={Nature Publishing Group UK London}
}

@article{you2025clinical,
  title={Clinical trials informed framework for real world clinical implementation and deployment of artificial intelligence applications},
  author={You, Jacqueline G and Hernandez-Boussard, Tina and Pfeffer, Michael A and Landman, Adam and Mishuris, Rebecca G},
  journal={NPJ Digital Medicine},
  volume={8},
  number={1},
  pages={107},
  year={2025},
  publisher={Nature Publishing Group UK London}
}

@article{schrouff2022diagnosing,
  title={Diagnosing failures of fairness transfer across distribution shift in real-world medical settings},
  author={Schrouff, Jessica and Harris, Natalie and Koyejo, Sanmi and Alabdulmohsin, Ibrahim M and Schnider, Eva and Opsahl-Ong, Krista and Brown, Alexander and Roy, Subhrajit and Mincu, Diana and Chen, Christina and others},
  journal={Advances in Neural Information Processing Systems},
  volume={35},
  pages={19304--19318},
  year={2022}
}

@article{al2023machine,
  title={Machine learning for ECG diagnosis and risk stratification of occlusion myocardial infarction},
  author={Al-Zaiti, Salah S and Martin-Gill, Christian and Z{\`e}gre-Hemsey, Jessica K and Bouzid, Zeineb and Faramand, Ziad and Alrawashdeh, Mohammad O and Gregg, Richard E and Helman, Stephanie and Riek, Nathan T and Kraevsky-Phillips, Karina and others},
  journal={Nature Medicine},
  volume={29},
  number={7},
  pages={1804--1813},
  year={2023},
  publisher={Nature Publishing Group US New York}
}

@article{johnson2025artificial,
  title={Artificial intelligence for direct-to-physician reporting of ambulatory electrocardiography},
  author={Johnson, LS and Zadrozniak, P and Jasina, G and Grotek-Cuprjak, A and Andrade, JG and Svennberg, E and Diederichsen, SZ and McIntyre, WF and Stavrakis, S and Benezet-Mazuecos, J and others},
  journal={Nature medicine},
  volume={31},
  number={3},
  pages={925--931},
  year={2025},
  publisher={Nature Publishing Group US New York}
}

@article{thapa2026multimodal,
  title={A multimodal sleep foundation model for disease prediction},
  author={Thapa, Rahul and Kjaer, Magnus Ruud and He, Bryan and Covert, Ian and Moore IV, Hyatt and Hanif, Umaer and Ganjoo, Gauri and Westover, M Brandon and Jennum, Poul and Brink-Kjaer, Andreas and others},
  journal={Nature Medicine},
  pages={1--11},
  year={2026},
  publisher={Nature Publishing Group US New York}
}

@inproceedings{he2023domain,
  title={Domain adaptation for time series under feature and label shifts},
  author={He, Huan and Queen, Owen and Koker, Teddy and Cuevas, Consuelo and Tsiligkaridis, Theodoros and Zitnik, Marinka},
  booktitle={International conference on machine learning},
  pages={12746--12774},
  year={2023},
  organization={PMLR}
}

@article{poterucha2025detecting,
  title={Detecting structural heart disease from electrocardiograms using AI},
  author={Poterucha, Timothy J and Jing, Linyuan and Ricart, Ramon Pimentel and Adjei-Mosi, Michael and Finer, Joshua and Hartzel, Dustin and Kelsey, Christopher and Long, Aaron and Rocha, Daniel and Ruhl, Jeffrey A and others},
  journal={Nature},
  volume={644},
  number={8075},
  pages={221--230},
  year={2025},
  publisher={Nature Publishing Group UK London}
}

@article{sun2024adaptive,
  title={Adaptive spatiotemporal encoding network for cognitive assessment using resting state EEG},
  author={Sun, Jingnan and Shen, Anruo and Sun, Yike and Chen, Xiaogang and Li, Yunxia and Gao, Xiaorong and Lu, Bai},
  journal={npj Digital Medicine},
  volume={7},
  number={1},
  pages={375},
  year={2024},
  publisher={Nature Publishing Group UK London}
}

@article{mahajan2025wearable,
  title={Wearable AI to enhance patient safety and clinical decision-making},
  author={Mahajan, Arjun and Heydari, Kimia and Powell, Dylan},
  journal={npj Digital Medicine},
  volume={8},
  number={1},
  pages={176},
  year={2025},
  publisher={Nature Publishing Group UK London}
}

@article{rossi2025sleepyland,
  title={SLEEPYLAND: trust begins with fair evaluation of automatic sleep staging models},
  author={Rossi, Alvise Dei and Metaldi, Matteo and Bechny, Michal and Filchenko, Irina and Meer, Julia van der and Schmidt, Markus H and Bassetti, Claudio LA and Tzovara, Athina and Faraci, Francesca D and Fiorillo, Luigi},
  journal={npj Digital Medicine},
  year={2025},
  publisher={Nature Publishing Group UK London}
}

@article{wu2025towards,
  title={Towards evaluating and building versatile large language models for medicine},
  author={Wu, Chaoyi and Qiu, Pengcheng and Liu, Jinxin and Gu, Hongfei and Li, Na and Zhang, Ya and Wang, Yanfeng and Xie, Weidi},
  journal={npj Digital Medicine},
  volume={8},
  number={1},
  pages={58},
  year={2025},
  publisher={Nature Publishing Group UK London}
}

@article{khandekar2024medcalc,
  title={Medcalc-bench: Evaluating large language models for medical calculations},
  author={Khandekar, Nikhil and Jin, Qiao and Xiong, Guangzhi and Dunn, Soren and Applebaum, Serina S and Anwar, Zain and Sarfo-Gyamfi, Maame and Safranek, Conrad W and Anwar, Abid A and Zhang, Andrew and others},
  journal={Advances in Neural Information Processing Systems},
  volume={37},
  pages={84730--84745},
  year={2024}
}

@article{wu2024medjourney,
  title={Medjourney: Benchmark and evaluation of large language models over patient clinical journey},
  author={Wu, Xian and Zhao, Yutian and Zhang, Yunyan and Wu, Jiageng and Zhu, Zhihong and Zhang, Yingying and Ouyang, Yi and Zhang, Ziheng and Wang, Huimin and Lin, Zhenxi and others},
  journal={Advances in Neural Information Processing Systems},
  volume={37},
  pages={87621--87646},
  year={2024}
}

@article{karargyris2023federated,
  title={Federated benchmarking of medical artificial intelligence with MedPerf},
  author={Karargyris, Alexandros and Umeton, Renato and Sheller, Micah J and Aristizabal, Alejandro and George, Johnu and Wuest, Anna and Pati, Sarthak and Kassem, Hasan and Zenk, Maximilian and Baid, Ujjwal and others},
  journal={Nature machine intelligence},
  volume={5},
  number={7},
  pages={799--810},
  year={2023},
  publisher={Nature Publishing Group UK London}
}

@article{vasey2022reporting,
  title={Reporting guideline for the early stage clinical evaluation of decision support systems driven by artificial intelligence: DECIDE-AI},
  author={Vasey, Baptiste and Nagendran, Myura and Campbell, Bruce and Clifton, David A and Collins, Gary S and Denaxas, Spiros and Denniston, Alastair K and Faes, Livia and Geerts, Bart and Ibrahim, Mudathir and others},
  journal={bmj},
  volume={377},
  year={2022},
  publisher={British Medical Journal Publishing Group}
}

@article{bycroft2018uk,
  title={The UK Biobank resource with deep phenotyping and genomic data},
  author={Bycroft, Clare and Freeman, Colin and Petkova, Desislava and Band, Gavin and Elliott, Lloyd T and Sharp, Kevin and Motyer, Allan and Vukcevic, Damjan and Delaneau, Olivier and O’Connell, Jared and others},
  journal={Nature},
  volume={562},
  number={7726},
  pages={203--209},
  year={2018},
  publisher={Nature Publishing Group UK London}
}

@article{pereira2020photoplethysmography,
  title={Photoplethysmography based atrial fibrillation detection: a review},
  author={Pereira, Tania and Tran, Nate and Gadhoumi, Kais and Pelter, Michele M and Do, Duc H and Lee, Randall J and Colorado, Rene and Meisel, Karl and Hu, Xiao},
  journal={NPJ digital medicine},
  volume={3},
  number={1},
  pages={3},
  year={2020},
  publisher={Nature Publishing Group UK London}
}

@inproceedings{mikolov2010recurrent,
  title={Recurrent neural network based language model.},
  author={Mikolov, Tomas and Karafi{\'a}t, Martin and Burget, Lukas and Cernock{\`y}, Jan and Khudanpur, Sanjeev},
  booktitle={Interspeech},
  volume={2},
  number={3},
  pages={1045--1048},
  year={2010},
  organization={Makuhari}
}

@article{graves2012long,
  title={Long short-term memory},
  author={Graves, Alex},
  journal={Supervised sequence labelling with recurrent neural networks},
  pages={37--45},
  year={2012},
  publisher={Springer}
}

@article{vaswani2017attention,
  title={Attention is all you need},
  author={Vaswani, Ashish and Shazeer, Noam and Parmar, Niki and Uszkoreit, Jakob and Jones, Llion and Gomez, Aidan N and Kaiser, {\L}ukasz and Polosukhin, Illia},
  journal={Advances in neural information processing systems},
  volume={30},
  year={2017}
}

@article{kaplan2020scaling,
  title={Scaling laws for neural language models},
  author={Kaplan, Jared and McCandlish, Sam and Henighan, Tom and Brown, Tom B and Chess, Benjamin and Child, Rewon and Gray, Scott and Radford, Alec and Wu, Jeffrey and Amodei, Dario},
  journal={arXiv preprint arXiv:2001.08361},
  year={2020}
}

@article{hoffmann2022training,
  title={Training compute-optimal large language models},
  author={Hoffmann, Jordan and Borgeaud, Sebastian and Mensch, Arthur and Buchatskaya, Elena and Cai, Trevor and Rutherford, Eliza and Casas, DDL and Hendricks, Lisa Anne and Welbl, Johannes and Clark, Aidan and others},
  journal={arXiv preprint arXiv:2203.15556},
  volume={10},
  year={2022}
}

@article{floridi2020gpt,
  title={GPT-3: Its nature, scope, limits, and consequences},
  author={Floridi, Luciano and Chiriatti, Massimo},
  journal={Minds and machines},
  volume={30},
  number={4},
  pages={681--694},
  year={2020},
  publisher={Springer}
}

@article{radford2019language,
  title={Language models are unsupervised multitask learners},
  author={Radford, Alec and Wu, Jeffrey and Child, Rewon and Luan, David and Amodei, Dario and Sutskever, Ilya and others},
  journal={OpenAI blog},
  volume={1},
  number={8},
  pages={9},
  year={2019}
}

@article{abbaspourazad2023large,
  title={Large-scale training of foundation models for wearable biosignals},
  author={Abbaspourazad, Salar and Elachqar, Oussama and Miller, Andrew C and Emrani, Saba and Nallasamy, Udhyakumar and Shapiro, Ian},
  journal={arXiv preprint arXiv:2312.05409},
  year={2023}
}

@article{liu2026teaching,
  title={Teaching multimodal LLMs to comprehend 12-lead electrocardiographic images},
  author={Liu, Ruoqi and Bai, Yuelin and Yue, Xiang and Zhang, Ping},
  journal={npj Digital Medicine},
  year={2026},
  publisher={Nature Publishing Group UK London}
}

@article{mathew2024foundation,
  title={Foundation models for cardiovascular disease detection via biosignals from digital stethoscopes},
  author={Mathew, George and Barbosa, Daniel and Prince, John and Venkatraman, Subramaniam},
  journal={NPJ cardiovascular health},
  volume={1},
  number={1},
  pages={25},
  year={2024},
  publisher={Nature Publishing Group UK London}
}

@article{bengio2003neural,
  title={A neural probabilistic language model},
  author={Bengio, Yoshua and Ducharme, R{\'e}jean and Vincent, Pascal and Jauvin, Christian},
  journal={Journal of machine learning research},
  volume={3},
  number={Feb},
  pages={1137--1155},
  year={2003}
}

@article{sundermeyer2015feedforward,
  title={From feedforward to recurrent LSTM neural networks for language modeling},
  author={Sundermeyer, Martin and Ney, Hermann and Schl{\"u}ter, Ralf},
  journal={IEEE/ACM transactions on audio, speech, and language processing},
  volume={23},
  number={3},
  pages={517--529},
  year={2015},
  publisher={IEEE}
}

@inproceedings{kaplan2025tokens,
  title={From tokens to words: On the inner lexicon of LLMs},
  author={Kaplan, Guy and Oren, Matanel and Reif, Yuval and Schwartz, Roy},
  booktitle={International Conference on Learning Representations},
  volume={2025},
  pages={70068--70092},
  year={2025}
}

@inproceedings{jin2024time,
  title={Time-llm: Time series forecasting by reprogramming large language models},
  author={Jin, Ming and Wang, Shiyu and Ma, Lintao and Chu, Zhixuan and Zhang, James and Shi, Xiaoming and Chen, Pin-Yu and Liang, Yuxuan and Li, Yuan-Fang and Pan, Shirui and others},
  booktitle={International conference on learning representations},
  volume={2024},
  pages={23857--23880},
  year={2024}
}

@article{luo2026toward,
  title={Toward foundation model for multivariate wearable sensing of physiological signals},
  author={Luo, Yunfei and Chen, Yuliang and Salekin, Asif and Rahman, Tauhidur},
  journal={ACM Transactions on Computing for Healthcare},
  volume={7},
  number={3},
  pages={1--43},
  year={2026},
  publisher={ACM New York, NY}
}

@article{makarov2025large,
  title={Large language models forecast patient health trajectories enabling digital twins},
  author={Makarov, Nikita and Bordukova, Maria and Quengdaeng, Papichaya and Garger, Daniel and Rodriguez-Esteban, Raul and Schmich, Fabian and Menden, Michael P},
  journal={npj Digital Medicine},
  volume={8},
  number={1},
  pages={588},
  year={2025},
  publisher={Nature Publishing Group UK London}
}

@inproceedings{narayanswamy2025scaling,
  title={Scaling wearable foundation models},
  author={Narayanswamy, Girish and Liu, Xin and Ayush, Kumar and Yang, Yuzhe and Xu, Xuhai and Liao, Shun and Garrison, Jake and Tailor, Shyam and Sunshine, Jacob and Liu, Yun and others},
  booktitle={International Conference on Learning Representations},
  volume={2025},
  pages={59382--59413},
  year={2025}
}

@article{yu2026scaling,
  title={Scaling embedding layers in language models},
  author={Yu, Da and Cohen, Edith and Ghazi, Badih and Huang, Yangsibo and Kamath, Pritish and Kumar, Ravi and Liu, Daogao and Zhang, Chiyuan},
  journal={Advances in Neural Information Processing Systems},
  volume={38},
  pages={31969--31995},
  year={2026}
}

@article{zhang2024found,
  title={Found in the middle: How language models use long contexts better via plug-and-play positional encoding},
  author={Zhang, Zhenyu and Chen, Runjin and Liu, Shiwei and Yao, Zhewei and Ruwase, Olatunji and Chen, Beidi and Wu, Xiaoxia and Wang, Zhangyang},
  journal={Advances in Neural Information Processing Systems},
  volume={37},
  pages={60755--60775},
  year={2024}
}

@article{wan2024teach,
  title={Teach better or show smarter? on instructions and exemplars in automatic prompt optimization},
  author={Wan, Xingchen and Sun, Ruoxi and Nakhost, Hootan and Ar{\i}k, Sercan {\"O}},
  journal={Advances in Neural Information Processing Systems},
  volume={37},
  pages={58174--58244},
  year={2024}
}

@article{wu2024prompt,
  title={Prompt optimization with ease? efficient ordering-aware automated selection of exemplars},
  author={Wu, Zhaoxuan and Lin, Xiaoqiang and Dai, Zhongxiang and Hu, Wenyang and Shu, Yao and Ng, See-Kiong and Jaillet, Patrick and Low, Bryan Kian Hsiang},
  journal={Advances in Neural Information Processing Systems},
  volume={37},
  pages={122706--122740},
  year={2024}
}

@article{zhang2025leveraging,
  title={Leveraging long context in retrieval augmented language models for medical question answering},
  author={Zhang, Gongbo and Xu, Zihan and Jin, Qiao and Chen, Fangyi and Fang, Yilu and Liu, Yi and Rousseau, Justin F and Xu, Ziyang and Lu, Zhiyong and Weng, Chunhua and others},
  journal={npj Digital Medicine},
  volume={8},
  number={1},
  pages={239},
  year={2025},
  publisher={Nature Publishing Group UK London}
}

@article{goswami2024moment,
  title={Moment: A family of open time-series foundation models},
  author={Goswami, Mononito and Szafer, Konrad and Choudhry, Arjun and Cai, Yifu and Li, Shuo and Dubrawski, Artur},
  journal={arXiv preprint arXiv:2402.03885},
  year={2024}
}

@article{zoph2020rethinking,
  title={Rethinking pre-training and self-training},
  author={Zoph, Barret and Ghiasi, Golnaz and Lin, Tsung-Yi and Cui, Yin and Liu, Hanxiao and Cubuk, Ekin Dogus and Le, Quoc},
  journal={Advances in neural information processing systems},
  volume={33},
  pages={3833--3845},
  year={2020}
}

@article{du2024stacking,
  title={Stacking your transformers: A closer look at model growth for efficient llm pre-training},
  author={Du, Wenyu and Luo, Tongxu and Qiu, Zihan and Huang, Zeyu and Shen, Yikang and Cheng, Reynold and Guo, Yike and Fu, Jie},
  journal={Advances in Neural Information Processing Systems},
  volume={37},
  pages={10491--10540},
  year={2024}
}

@article{toma2023clinical,
  title={Clinical camel: An open expert-level medical language model with dialogue-based knowledge encoding},
  author={Toma, Augustin and Lawler, Patrick R and Ba, Jimmy and Krishnan, Rahul G and Rubin, Barry B and Wang, Bo},
  journal={arXiv preprint arXiv:2305.12031},
  year={2023}
}

@article{zheng2025learning,
  title={Learning from models beyond fine-tuning},
  author={Zheng, Hongling and Shen, Li and Tang, Anke and Luo, Yong and Hu, Han and Du, Bo and Wen, Yonggang and Tao, Dacheng},
  journal={Nature Machine Intelligence},
  volume={7},
  number={1},
  pages={6--17},
  year={2025},
  publisher={Nature Publishing Group UK London}
}

@article{peng2023study,
  title={A study of generative large language model for medical research and healthcare},
  author={Peng, Cheng and Yang, Xi and Chen, Aokun and Smith, Kaleb E and PourNejatian, Nima and Costa, Anthony B and Martin, Cheryl and Flores, Mona G and Zhang, Ying and Magoc, Tanja and others},
  journal={NPJ digital medicine},
  volume={6},
  number={1},
  pages={210},
  year={2023},
  publisher={Nature Publishing Group UK London}
}

@inproceedings{qiang2025biecg,
  title={BiECG-LLM: An Approach to ECG Classification and Report Generation Using a Fine-Tuned LLM with Bi-Modal ECG},
  author={Qiang, Yupeng and Dong, Xunde and Liu, Xiuling and Hu, Fei and Wang, Rongjia},
  booktitle={2025 IEEE International Conference on Bioinformatics and Biomedicine (BIBM)},
  pages={7476--7483},
  year={2025},
  organization={IEEE}
}

@inproceedings{hu2024exploring,
  title={Exploring large-scale language models to evaluate eeg-based multimodal data for mental health},
  author={Hu, Yongquan and Zhang, Shuning and Dang, Ting and Jia, Hong and Salim, Flora D and Hu, Wen and Quigley, Aaron J},
  booktitle={Companion of the 2024 on ACM International Joint Conference on Pervasive and Ubiquitous Computing},
  pages={412--417},
  year={2024}
}

@article{thapa2024sleepfm,
  title={SleepFM: multi-modal representation learning for sleep across brain activity, ECG and respiratory signals},
  author={Thapa, Rahul and He, Bryan and Kj{\ae}r, Magnus Ruud and Moore, Hyatt and Ganjoo, Gauri and Mignot, Emmanuel and Zou, James},
  journal={arXiv preprint arXiv:2405.17766},
  year={2024}
}

@article{ding2025ai,
  title={AI modeling photoplethysmography to electrocardiography useful for predicting cardiovascular disease},
  author={Ding, Zhengyao and Hu, Yujian and Li, Ziyu and Mao, Yiheng and Li, Haitao and Zhou, Dongchen and Chu, Xuesen and Yu, Long and Liu, Ziyi and Wu, Fei and others},
  journal={npj Digital Medicine},
  year={2025},
  publisher={Nature Publishing Group UK London}
}

@article{wang2023contrast,
  title={Contrast everything: A hierarchical contrastive framework for medical time-series},
  author={Wang, Yihe and Han, Yu and Wang, Haishuai and Zhang, Xiang},
  journal={Advances in Neural Information Processing Systems},
  volume={36},
  pages={55694--55717},
  year={2023}
}

@article{wang2024medformer,
  title={Medformer: A multi-granularity patching transformer for medical time-series classification},
  author={Wang, Yihe and Huang, Nan and Li, Taida and Yan, Yujun and Zhang, Xiang},
  journal={Advances in Neural Information Processing Systems},
  volume={37},
  pages={36314--36341},
  year={2024}
}

@article{hogan2025scaling,
  title={Scaling convolutional neural networks achieves expert level seizure detection in neonatal EEG},
  author={Hogan, Robert and Mathieson, Sean R and Luca, Aurel and Ventura, Soraia and Griffin, Sean and Boylan, Geraldine B and O’Toole, John M},
  journal={npj Digital Medicine},
  volume={8},
  number={1},
  pages={17},
  year={2025},
  publisher={Nature Publishing Group UK London}
}

@article{fraser2025integration,
  title={Integration of artificial intelligence and wearable technology in the management of diabetes and prediabetes},
  author={Fraser, Raphael A and Walker, Rebekah J and Campbell, Jennifer A and Ekwunife, Obinna and Egede, Leonard E},
  journal={npj Digital Medicine},
  volume={8},
  number={1},
  pages={687},
  year={2025},
  publisher={Nature Publishing Group UK London}
}

@article{pedroso2025leveraging,
  title={Leveraging AI-enhanced digital health with consumer devices for scalable cardiovascular screening, prediction, and monitoring},
  author={Pedroso, Aline F and Khera, Rohan},
  journal={Npj Cardiovascular Health},
  volume={2},
  number={1},
  pages={34},
  year={2025},
  publisher={Nature Publishing Group UK London}
}

@article{sun2025automated,
  title={An Automated Classifier of Harmful Brain Activities for Clinical Usage Based on a Vision-Inspired Pre-trained Framework},
  author={Sun, Yulin and Si, Xiaopeng and He, Runnan and Hu, Xiao and Smielewski, Peter and Wang, Wenlong and Tong, Xiaoguang and Yue, Wei and Pang, Meijun and Zhang, Kuo and others},
  journal={npj Digital Medicine},
  volume={8},
  number={1},
  pages={768},
  year={2025},
  publisher={Nature Publishing Group UK London}
}

@article{partamian2025machine,
  title={Machine learning on interictal intracranial EEG predicts surgical outcome in drug resistant epilepsy},
  author={Partamian, Hmayag and Jahromi, Saeed and Corona, Ludovica and Perry, M Scott and Tamilia, Eleonora and Madsen, Joseph R and Bolton, Jeffrey and Stone, Scellig SD and Pearl, Phillip L and Papadelis, Christos},
  journal={NPJ digital medicine},
  volume={8},
  number={1},
  pages={138},
  year={2025},
  publisher={Nature Publishing Group UK London}
}

@article{zhang2026instruction,
  title={Instruction tuning for large language models: A survey},
  author={Zhang, Shengyu and Dong, Linfeng and Li, Xiaoya and Zhang, Sen and Sun, Xiaofei and Wang, Shuhe and Li, Jiwei and Hu, Runyi and Zhang, Tianwei and Wang, Guoyin and others},
  journal={ACM Computing Surveys},
  volume={58},
  number={7},
  pages={1--36},
  year={2026},
  publisher={ACM New York, NY}
}

@article{shi2024instruction,
  title={Instruction tuning with loss over instructions},
  author={Shi, Zhengyan and Yang, Adam X and Wu, Bin and Aitchison, Laurence and Yilmaz, Emine and Lipani, Aldo},
  journal={Advances in Neural Information Processing Systems},
  volume={37},
  pages={69176--69205},
  year={2024}
}

@article{dettmers2023qlora,
  title={Qlora: Efficient finetuning of quantized llms},
  author={Dettmers, Tim and Pagnoni, Artidoro and Holtzman, Ari and Zettlemoyer, Luke},
  journal={Advances in neural information processing systems},
  volume={36},
  pages={10088--10115},
  year={2023}
}

@inproceedings{liu2024dora,
  title={Dora: Weight-decomposed low-rank adaptation},
  author={Liu, Shih-Yang and Wang, Chien-Yi and Yin, Hongxu and Molchanov, Pavlo and Wang, Yu-Chiang Frank and Cheng, Kwang-Ting and Chen, Min-Hung},
  booktitle={Forty-first International Conference on Machine Learning},
  year={2024}
}

@article{gao2025enhancing,
  title={Enhancing privacy-preserving deployable large language models for perioperative complication detection: a targeted strategy with LoRA fine-tuning},
  author={Gao, Shaowei and Zhao, Xu and Chen, Lihui and Yu, Junrong and Tian, Shuning and Zhou, Huaqiang and Chen, Jingru and Long, Sizhe and He, Qiulan and Feng, Xia},
  journal={NPJ Digital Medicine},
  year={2025},
  publisher={Nature Publishing Group UK London}
}

@article{liu2024deepseek,
  title={Deepseek-v3 technical report},
  author={Liu, Aixin and Feng, Bei and Xue, Bing and Wang, Bingxuan and Wu, Bochao and Lu, Chengda and Zhao, Chenggang and Deng, Chengqi and Zhang, Chenyu and Ruan, Chong and others},
  journal={arXiv preprint arXiv:2412.19437},
  year={2024}
}

@article{comanici2025gemini,
  title={Gemini 2.5: Pushing the frontier with advanced reasoning, multimodality, long context, and next generation agentic capabilities},
  author={Comanici, Gheorghe and Bieber, Eric and Schaekermann, Mike and Pasupat, Ice and Sachdeva, Noveen and Dhillon, Inderjit and Blistein, Marcel and Ram, Ori and Zhang, Dan and Rosen, Evan and others},
  journal={arXiv preprint arXiv:2507.06261},
  year={2025}
}

@article{khan2026ecg,
  title={ECG Foundation Models and Medical LLMs for Agentic Cardiovascular Intelligence at the Edge: A Review and Outlook},
  author={Khan, Mudassir Hasan and Nayfeh, Ahmad and Masood, Mudassir and Al-Shaikhi, Ali Ahmad and Rahman, Muhammad Mahboob Ur and Al-Naffouri, Tareq Y},
  journal={arXiv preprint arXiv:2604.02501},
  year={2026}
}

@inproceedings{arava2025large,
  title={Large Language Models in Wearable Devices: Applications, Challenges, and Future Directions},
  author={Arava, Poojareddy and Singh, Aditi and Ehtesham, Abul and Kumar, Saket},
  booktitle={2025 IEEE World AI IoT Congress (AIIoT)},
  pages={1003--1009},
  year={2025},
  organization={IEEE}
}

@article{ding2024survey,
  title={A Survey of LLMs on Biosignal Applications},
  author={Ding, Cheng and Wu, Chenwei and Chen, Zhaoliang and Ni, Jianyuan},
  journal={Authorea Preprints},
  year={2024},
  publisher={Authorea}
}

@article{moor2023foundation,
  title={Foundation models for generalist medical artificial intelligence},
  author={Moor, Michael and Banerjee, Oishi and Abad, Zahra Shakeri Hossein and Krumholz, Harlan M and Leskovec, Jure and Topol, Eric J and Rajpurkar, Pranav},
  journal={Nature},
  volume={616},
  number={7956},
  pages={259--265},
  year={2023},
  publisher={Nature Publishing Group UK London}
}

@inproceedings{chan2024leveraging,
  title={Leveraging llms for multimodal medical time series analysis},
  author={Chan, Nimeesha and Parker, Felix and Bennett, William and Wu, Tianyi and Jia, Mung Yao and Fackler, James and Ghobadi, Kimia},
  booktitle={Machine Learning for Healthcare Conference. PMLR},
  year={2024}
}

@article{pham2025q,
  title={Q-heart: Ecg question answering via knowledge-informed multimodal llms},
  author={Pham, Hung Manh and Tang, Jialu and Saeed, Aaqib and Ma, Dong},
  journal={arXiv preprint arXiv:2505.06296},
  year={2025}
}

@article{li2026mira,
  title={Mira: Medical time series foundation model for real-world health data},
  author={Li, Hao and Deng, Bowen and Xu, Chang and Feng, Zhiyuan and Schlegel, Viktor and Huang, Yu-Hao and Sun, Yizheng and Sun, Jingyuan and Yang, Kailai and Yu, Yiyao and others},
  journal={Advances in neural information processing systems},
  volume={38},
  pages={98657--98685},
  year={2026}
}

@article{langer2025opentslm,
  title={Opentslm: Time-series language models for reasoning over multivariate medical text-and time-series data},
  author={Langer, Patrick and Kaar, Thomas and Rosenblattl, Max and Xu, Maxwell A and Chow, Winnie and Maritsch, Martin and Jakob, Robert and Wang, Ning and Liu, Juncheng and Verma, Aradhana and others},
  journal={arXiv preprint arXiv:2510.02410},
  year={2025}
}

@inproceedings{li2026anyecg,
  title={Anyecg-chat: A generalist ECG-MLLM for flexible ECG input and multi-task understanding},
  author={Li, Haitao and Li, Ziyu and Mao, Yiheng and Liu, Ziyi and Sun, Zhoujian and Huang, Zhengxing},
  booktitle={Proceedings of the AAAI Conference on Artificial Intelligence},
  volume={40},
  number={1},
  pages={597--605},
  year={2026}
}

@inproceedings{joseph2023multilingual,
  title={Multilingual simplification of medical texts},
  author={Joseph, Sebastian and Kazanas, Kathryn and Reina, Keziah and Ramanathan, Vishnesh and Xu, Wei and Wallace, Byron C and Li, Junyi Jessy},
  booktitle={Proceedings of the 2023 conference on empirical methods in natural language processing},
  pages={16662--16692},
  year={2023}
}

@article{chung2024scaling,
  title={Scaling instruction-finetuned language models},
  author={Chung, Hyung Won and Hou, Le and Longpre, Shayne and Zoph, Barret and Tay, Yi and Fedus, William and Li, Yunxuan and Wang, Xuezhi and Dehghani, Mostafa and Brahma, Siddhartha and others},
  journal={Journal of Machine Learning Research},
  volume={25},
  number={70},
  pages={1--53},
  year={2024}
}

@inproceedings{lewis2020bart,
  title={BART: Denoising sequence-to-sequence pre-training for natural language generation, translation, and comprehension},
  author={Lewis, Mike and Liu, Yinhan and Goyal, Naman and Ghazvininejad, Marjan and Mohamed, Abdelrahman and Levy, Omer and Stoyanov, Veselin and Zettlemoyer, Luke},
  booktitle={Proceedings of the 58th annual meeting of the association for computational linguistics},
  pages={7871--7880},
  year={2020}
}

@article{grattafiori2024llama,
  title={The llama 3 herd of models},
  author={Grattafiori, Aaron and Dubey, Abhimanyu and Jauhri, Abhinav and Pandey, Abhinav and Kadian, Abhishek and Al-Dahle, Ahmad and Letman, Aiesha and Mathur, Akhil and Schelten, Alan and Vaughan, Alex and others},
  journal={arXiv preprint arXiv:2407.21783},
  year={2024}
}

@article{bai2023qwen,
  title={Qwen technical report},
  author={Bai, Jinze and Bai, Shuai and Chu, Yunfei and Cui, Zeyu and Dang, Kai and Deng, Xiaodong and Fan, Yang and Ge, Wenbin and Han, Yu and Huang, Fei and others},
  journal={arXiv preprint arXiv:2309.16609},
  year={2023}
}

@article{hui2024qwen2,
  title={Qwen2. 5-coder technical report},
  author={Hui, Binyuan and Yang, Jian and Cui, Zeyu and Yang, Jiaxi and Liu, Dayiheng and Zhang, Lei and Liu, Tianyu and Zhang, Jiajun and Yu, Bowen and Lu, Keming and others},
  journal={arXiv preprint arXiv:2409.12186},
  year={2024}
}

@article{wang2024qwen2,
  title={Qwen2-vl: Enhancing vision-language model's perception of the world at any resolution},
  author={Wang, Peng and Bai, Shuai and Tan, Sinan and Wang, Shijie and Fan, Zhihao and Bai, Jinze and Chen, Keqin and Liu, Xuejing and Wang, Jialin and Ge, Wenbin and others},
  journal={arXiv preprint arXiv:2409.12191},
  year={2024}
}

@article{ouyang2022training,
  title={Training language models to follow instructions with human feedback},
  author={Ouyang, Long and Wu, Jeffrey and Jiang, Xu and Almeida, Diogo and Wainwright, Carroll and Mishkin, Pamela and Zhang, Chong and Agarwal, Sandhini and Slama, Katarina and Ray, Alex and others},
  journal={Advances in neural information processing systems},
  volume={35},
  pages={27730--27744},
  year={2022}
}

@article{chiang2023vicuna,
  title={Vicuna: An open-source chatbot impressing gpt-4 with 90\%* chatgpt quality},
  author={Chiang, Wei-Lin and Li, Zhuohan and Lin, Ziqing and Sheng, Ying and Wu, Zhanghao and Zhang, Hao and Zheng, Lianmin and Zhuang, Siyuan and Zhuang, Yonghao and Gonzalez, Joseph E and others},
  journal={See https://vicuna. lmsys. org (accessed 14 April 2023)},
  volume={2},
  number={3},
  pages={6},
  year={2023}
}

@misc{taori2023stanford,
  title={Stanford alpaca: An instruction-following llama model},
  author={Taori, Rohan and Gulrajani, Ishaan and Zhang, Tianyi and Dubois, Yann and Li, Xuechen and Guestrin, Carlos and Liang, Percy and Hashimoto, Tatsunori B},
  year={2023},
  publisher={Stanford, CA, USA}
}

@misc{jiang2023mistral7b,
      title={Mistral 7B}, 
      author={Albert Q. Jiang and Alexandre Sablayrolles and Arthur Mensch and Chris Bamford and Devendra Singh Chaplot and Diego de las Casas and Florian Bressand and Gianna Lengyel and Guillaume Lample and Lucile Saulnier and Lélio Renard Lavaud and Marie-Anne Lachaux and Pierre Stock and Teven Le Scao and Thibaut Lavril and Thomas Wang and Timothée Lacroix and William El Sayed},
      year={2023},
      eprint={2310.06825},
      archivePrefix={arXiv},
      primaryClass={cs.CL},
      url={https://arxiv.org/abs/2310.06825}, 
}

@article{guo2025deepseek,
  title={DeepSeek-R1 incentivizes reasoning in LLMs through reinforcement learning},
  author={Guo, Daya and Yang, Dejian and Zhang, Haowei and Song, Junxiao and Wang, Peiyi and Zhu, Qihao and Xu, Runxin and Zhang, Ruoyu and Ma, Shirong and Bi, Xiao and others},
  journal={Nature},
  volume={645},
  number={8081},
  pages={633--638},
  year={2025},
  publisher={Nature Publishing Group UK London}
}

@article{bao2021beit,
  title={Beit: Bert pre-training of image transformers},
  author={Bao, Hangbo and Dong, Li and Piao, Songhao and Wei, Furu},
  journal={arXiv preprint arXiv:2106.08254},
  year={2021}
}

@inproceedings{kim2021vilt,
  title={Vilt: Vision-and-language transformer without convolution or region supervision},
  author={Kim, Wonjae and Son, Bokyung and Kim, Ildoo},
  booktitle={International conference on machine learning},
  pages={5583--5594},
  year={2021},
  organization={PMLR}
}

@inproceedings{zhang2020pegasus,
  title={Pegasus: Pre-training with extracted gap-sentences for abstractive summarization},
  author={Zhang, Jingqing and Zhao, Yao and Saleh, Mohammad and Liu, Peter},
  booktitle={International conference on machine learning},
  pages={11328--11339},
  year={2020},
  organization={PMLR}
}

@article{yu2022coca,
  title={Coca: Contrastive captioners are image-text foundation models},
  author={Yu, Jiahui and Wang, Zirui and Vasudevan, Vijay and Yeung, Legg and Seyedhosseini, Mojtaba and Wu, Yonghui},
  journal={arXiv preprint arXiv:2205.01917},
  year={2022}
}

@article{team2024gemini,
  title={Gemini 1.5: Unlocking multimodal understanding across millions of tokens of context},
  author={Team, Gemini and Georgiev, Petko and Lei, Ving Ian and Burnell, Ryan and Bai, Libin and Gulati, Anmol and Tanzer, Garrett and Vincent, Damien and Pan, Zhufeng and Wang, Shibo and others},
  journal={arXiv preprint arXiv:2403.05530},
  year={2024}
}

@article{team2023gemini,
  title={Gemini: a family of highly capable multimodal models},
  author={Team, Gemini and Anil, Rohan and Borgeaud, Sebastian and Alayrac, Jean-Baptiste and Yu, Jiahui and Soricut, Radu and Schalkwyk, Johan and Dai, Andrew M and Hauth, Anja and Millican, Katie and others},
  journal={arXiv preprint arXiv:2312.11805},
  year={2023}
}

@article{liu2023visual,
  title={Visual instruction tuning},
  author={Liu, Haotian and Li, Chunyuan and Wu, Qingyang and Lee, Yong Jae},
  journal={Advances in neural information processing systems},
  volume={36},
  pages={34892--34916},
  year={2023}
}

@misc{liu2024llavanext,
  title={Llavanext: Improved reasoning, ocr, and world knowledge},
  author={Liu, Haotian and Li, Chunyuan and Li, Yuheng and Li, Bo and Zhang, Yuanhan and Shen, Sheng and Lee, Yong Jae},
  year={2024}
}

@inproceedings{yasunaga2022linkbert,
  title={Linkbert: Pretraining language models with document links},
  author={Yasunaga, Michihiro and Leskovec, Jure and Liang, Percy},
  booktitle={Proceedings of the 60th Annual Meeting of the Association for Computational Linguistics (Volume 1: Long Papers)},
  pages={8003--8016},
  year={2022}
}

@article{zeng2022glm,
  title={Glm-130b: An open bilingual pre-trained model},
  author={Zeng, Aohan and Liu, Xiao and Du, Zhengxiao and Wang, Zihan and Lai, Hanyu and Ding, Ming and Yang, Zhuoyi and Xu, Yifan and Zheng, Wendi and Xia, Xiao and others},
  journal={arXiv preprint arXiv:2210.02414},
  year={2022}
}

@article{glm2024chatglm,
  title={Chatglm: A family of large language models from glm-130b to glm-4 all tools},
  author={Glm, Team and Zeng, Aohan and Xu, Bin and Wang, Bowen and Zhang, Chenhui and Yin, Da and Zhang, Dan and Rojas, Diego and Feng, Guanyu and Zhao, Hanlin and others},
  journal={arXiv preprint arXiv:2406.12793},
  year={2024}
}

@article{zhang2022opt,
  title={Opt: Open pre-trained transformer language models},
  author={Zhang, Susan and Roller, Stephen and Goyal, Naman and Artetxe, Mikel and Chen, Moya and Chen, Shuohui and Dewan, Christopher and Diab, Mona and Li, Xian and Lin, Xi Victoria and others},
  journal={arXiv preprint arXiv:2205.01068},
  year={2022}
}

@article{workshop2022bloom,
  title={Bloom: A 176b-parameter open-access multilingual language model},
  author={Workshop, BigScience and Scao, Teven Le and Fan, Angela and Akiki, Christopher and Pavlick, Ellie and Ili{\'c}, Suzana and Hesslow, Daniel and Castagn{\'e}, Roman and Luccioni, Alexandra Sasha and Yvon, Fran{\c{c}}ois and others},
  journal={arXiv preprint arXiv:2211.05100},
  year={2022}
}

@article{cai2024internlm2,
  title={Internlm2 technical report},
  author={Cai, Zheng and Cao, Maosong and Chen, Haojiong and Chen, Kai and Chen, Keyu and Chen, Xin and Chen, Xun and Chen, Zehui and Chen, Zhi and Chu, Pei and others},
  journal={arXiv preprint arXiv:2403.17297},
  year={2024}
}

@misc{mistralai2024large2,
  title={Large Enough: Introducing Mistral Large 2},
  author={{Mistral AI Team}},
  year={2024},
  howpublished={Mistral AI Blog},
  url={https://mistral.ai/news/mistral-large-2407/},
}

@misc{laion2023leolm,
  title={LeoLM: Igniting German-Language LLM Research},
  author={{LAION AI}},
  year={2023},
  howpublished={LAION Blog},
  url={https://laion.ai/blog/leo-lm/},
}

@misc{gong2025minimind2,
  title={MiniMind2},
  author={Gong, Jingyao},
  year={2025},
  howpublished={Hugging Face model card and GitHub repository},
  url={https://huggingface.co/jingyaogong/MiniMind2},
}

@article{thoppilan2022lamda,
  title={Lamda: Language models for dialog applications},
  author={Thoppilan, Romal and De Freitas, Daniel and Hall, Jamie and Shazeer, Noam and Kulshreshtha, Apoorv and Cheng, Heng-Tze and Jin, Alicia and Bos, Taylor and Baker, Leslie and Du, Yu and others},
  journal={arXiv preprint arXiv:2201.08239},
  year={2022}
}

@article{hurst2024gpt,
  title={Gpt-4o system card},
  author={Hurst, Aaron and Lerer, Adam and Goucher, Adam P and Perelman, Adam and Ramesh, Aditya and Clark, Aidan and Ostrow, AJ and Welihinda, Akila and Hayes, Alan and Radford, Alec and others},
  journal={arXiv preprint arXiv:2410.21276},
  year={2024}
}

@techreport{anthropic2024claude35sonnet,
  title       = {Claude 3.5 Sonnet Model Card Addendum},
  author      = {{Anthropic}},
  institution = {Anthropic},
  year        = {2024},
  type        = {Model Card Addendum},
  url         = {https://www-cdn.anthropic.com/fed9cc193a14b84131812372d8d5857f8f304c52/Model_Card_Claude_3_Addendum.pdf}
}

@misc{openai2022embedding,
  title        = {New and improved embedding model},
  author       = {{OpenAI}},
  year         = {2022},
  month        = dec,
  howpublished = {OpenAI Blog},
  url          = {https://openai.com/index/new-and-improved-embedding-model/},
}

@article{omar2025multi,
  title={Multi-model assurance analysis showing large language models are highly vulnerable to adversarial hallucination attacks during clinical decision support},
  author={Omar, Mahmud and Sorin, Vera and Collins, Jeremy D and Reich, David and Freeman, Robert and Gavin, Nicholas and Charney, Alexander and Stump, Lisa and Bragazzi, Nicola Luigi and Nadkarni, Girish N and others},
  journal={Communications Medicine},
  volume={5},
  number={1},
  pages={330},
  year={2025},
  publisher={Nature Publishing Group UK London}
}

@inproceedings{samsi2023words,
  title={From words to watts: Benchmarking the energy costs of large language model inference},
  author={Samsi, Siddharth and Zhao, Dan and McDonald, Joseph and Li, Baolin and Michaleas, Adam and Jones, Michael and Bergeron, William and Kepner, Jeremy and Tiwari, Devesh and Gadepally, Vijay},
  booktitle={2023 IEEE high performance extreme computing conference (HPEC)},
  pages={1--9},
  year={2023},
  organization={IEEE}
}

@article{wang2024model,
  title={Model compression and efficient inference for large language models: A survey},
  author={Wang, Wenxiao and Chen, Wei and Luo, Yicong and Long, Yongliu and Lin, Zhengkai and Zhang, Liye and Lin, Binbin and Cai, Deng and He, Xiaofei},
  journal={arXiv preprint arXiv:2402.09748},
  year={2024}
}

@article{chan2024medtsllm,
  title={Medtsllm: Leveraging llms for multimodal medical time series analysis},
  author={Chan, Nimeesha and Parker, Felix and Bennett, William and Wu, Tianyi and Jia, Mung Yao and Fackler, James and Ghobadi, Kimia},
  journal={arXiv preprint arXiv:2408.07773},
  year={2024}
}

@article{mckeen2025ecg,
  title={Ecg-fm: An open electrocardiogram foundation model},
  author={McKeen, Kaden and Masood, Sameer and Toma, Augustin and Rubin, Barry and Wang, Bo},
  journal={Jamia Open},
  volume={8},
  number={5},
  pages={ooaf122},
  year={2025},
  publisher={Oxford University Press}
}

@article{jiang2024neurolm,
  title={NeuroLM: A universal multi-task foundation model for bridging the gap between language and EEG signals},
  author={Jiang, Wei-Bang and Wang, Yansen and Lu, Bao-Liang and Li, Dongsheng},
  journal={arXiv preprint arXiv:2409.00101},
  year={2024}
}

@article{xie2025physllm,
  title={Physllm: Harnessing large language models for cross-modal remote physiological sensing},
  author={Xie, Yiping and Zhao, Bo and Dai, Mingtong and Zhou, Jian-Ping and Sun, Yue and Tan, Tao and Xie, Weicheng and Shen, Linlin and Yu, Zitong},
  journal={arXiv preprint arXiv:2505.03621},
  year={2025}
}

@article{khasentino2025personal,
  title={A personal health large language model for sleep and fitness coaching},
  author={Khasentino, Justin and Belyaeva, Anastasiya and Liu, Xin and Yang, Zhun and Furlotte, Nicholas A and Lee, Chace and Schenck, Erik and Patel, Yojan and Cui, Jian and Schneider, Logan Douglas and others},
  journal={Nature Medicine},
  volume={31},
  number={10},
  pages={3394--3403},
  year={2025},
  publisher={Nature Publishing Group US New York}
}

@article{yang2023biot,
  title={Biot: Biosignal transformer for cross-data learning in the wild},
  author={Yang, Chaoqi and Westover, M and Sun, Jimeng},
  journal={Advances in Neural Information Processing Systems},
  volume={36},
  pages={78240--78260},
  year={2023}
}

@article{xu2026deepseek,
  title={Deepseek-v4: Towards highly efficient million-token context intelligence},
  author={Xu, Anyi and Lin, Bangcai and Xue, Bing and Wang, Bingxuan and Xu, Bingzheng and Wu, Bochao and Zhang, Bowei and Lin, Chaofan and Dong, Chen and Ling, Chenchen and others},
  journal={arXiv preprint arXiv:2606.19348},
  year={2026}
}

@article{singh2025openai,
  title={Openai gpt-5 system card},
  author={Singh, Aaditya and Fry, Adam and Perelman, Adam and Tart, Adam and Ganesh, Adi and El-Kishky, Ahmed and McLaughlin, Aidan and Low, Aiden and Ostrow, AJ and Ananthram, Akhila and others},
  journal={arXiv preprint arXiv:2601.03267},
  year={2025}
}

@article{team2025gemma,
  title={Gemma 3 technical report},
  author={Team, Gemma and Kamath, Aishwarya and Ferret, Johan and Pathak, Shreya and Vieillard, Nino and Merhej, Ramona and Perrin, Sarah and Matejovicova, Tatiana and Ram{\'e}, Alexandre and Rivi{\`e}re, Morgane and others},
  journal={arXiv preprint arXiv:2503.19786},
  year={2025}
}

@article{chen2024sparse,
  title={Sparse learned kernels for interpretable and efficient medical time series processing},
  author={Chen, Sully F and Guo, Zhicheng and Ding, Cheng and Hu, Xiao and Rudin, Cynthia},
  journal={Nature machine intelligence},
  volume={6},
  number={10},
  pages={1132--1144},
  year={2024},
  publisher={Nature Publishing Group UK London}
}

@article{yao2025efficient,
  title={Efficient GPT-4V level multimodal large language model for deployment on edge devices},
  author={Yao, Yuan and Yu, Tianyu and Zhang, Ao and Wang, Chongyi and Cui, Junbo and Zhu, Hongji and Cai, Tianchi and Chen, Chi and Li, Haoyu and Zhao, Weilin and others},
  journal={Nature Communications},
  volume={16},
  number={1},
  pages={5509},
  year={2025},
  publisher={Nature Publishing Group UK London}
}

@article{leroux2025analog,
  title={Analog in-memory computing attention mechanism for fast and energy-efficient large language models},
  author={Leroux, Nathan and Manea, Paul-Philipp and Sudarshan, Chirag and Finkbeiner, Jan and Siegel, Sebastian and Strachan, John Paul and Neftci, Emre},
  journal={Nature computational science},
  volume={5},
  number={9},
  pages={813--824},
  year={2025},
  publisher={Nature Publishing Group US New York}
}

@article{kasper2025wearable,
  title={Wearable AI for on-device frailty assessment},
  author={Kasper, Kevin Albert and Thien, Ryan and Stuart, Tucker and Kim, John and Bhatia, Aman and Gutruf, Philipp},
  journal={Nature Communications},
  volume={17},
  number={1},
  pages={991},
  year={2025},
  publisher={Nature Publishing Group UK London}
}

@article{woo2025synthetic,
  title={Synthetic data distillation enables the extraction of clinical information at scale},
  author={Woo, Elizabeth Geena and Burkhart, Michael C and Alsentzer, Emily and Beaulieu-Jones, Brett K},
  journal={npj Digital Medicine},
  volume={8},
  number={1},
  pages={267},
  year={2025},
  publisher={Nature Publishing Group UK London}
}

@article{liu2026survey,
  title={A survey on large language models for medical time series},
  author={Liu, Xingyue and Zhou, Feizhong and Xiao, Hanguang and Li, Zhipeng and Liu, Shuai and Qian, Lingling},
  journal={Expert Systems with Applications},
  pages={131364},
  year={2026},
  publisher={Elsevier}
}

@article{jamieson2025guide,
  title={A guide to consumer-grade wearables in cardiovascular clinical care and population health for non-experts},
  author={Jamieson, Alexandra and Chico, Timothy JA and Jones, Siana and Chaturvedi, Nishi and Hughes, Alun D and Orini, Michele},
  journal={NPJ cardiovascular health},
  volume={2},
  number={1},
  pages={44},
  year={2025},
  publisher={Nature Publishing Group UK London}
}

@article{bedi2026holistic,
  title={Holistic evaluation of large language models for medical tasks with MedHELM},
  author={Bedi, Suhana and Cui, Hejie and Fuentes, Miguel and Unell, Alyssa and Wornow, Michael and Banda, Juan M and Kotecha, Nikesh and Keyes, Timothy and Mai, Yifan and Oez, Mert and others},
  journal={Nature medicine},
  volume={32},
  number={3},
  pages={943--951},
  year={2026},
  publisher={Nature Publishing Group US New York}
}

\end{document}